\documentstyle[dina4p,12pt]{article}
\newcommand{\hc}{h_{v_1}^{(c)}}                
\newcommand{\hcbar}{\bar h_{v_2}^{(c)}}        
\newcommand{\hcbarv}{\bar h_{v_1}^{(c)}}        
\newcommand{\hb}{h_{v_1}^{(b)}}                 
\newcommand{\hbbar}{\bar h_{v_1}^{(b)}}         
\newcommand{\dslash}{\mbox{$\not{\hspace{-1.03mm}D}$}}        
\newcommand{\pslash}{\mbox{$\not{\hspace{-1.03mm}p}$}}        
\newcommand{\vslash}{\mbox{$\not{\hspace{-1.03mm}v}$}}        
\newcommand{\bea}{\begin{eqnarray}}
\newcommand{\eea}{\end{eqnarray}}
\newcommand{\beq}{\begin{equation}}
\newcommand{\eeq}{\end{equation}}
\newcommand{\bay}{\begin{array}}
\newcommand{\eay}{\end{array}}

\catcode`\@=11
\def\@citex[#1]#2{\if@filesw\immediate\write\@auxout{\string\citation{#2}}\fi
  \def\@citea{}\@cite{\@for\@citeb:=#2\do
    {\@citea\def\@citea{,\penalty\@m}\@ifundefined
       {b@\@citeb}{{\bf ?}\@warning
       {Citation `\@citeb' on page \thepage \space undefined}}%
\hbox{\csname b@\@citeb\endcsname}}}{#1}}
\def\citer{\@ifnextchar [{\@tempswatrue\@citexr}{\@tempswafalse\@citexr[]}}
\def\@citexr[#1]#2{\if@filesw\immediate\write\@auxout{\string\citation{#2}}\fi
  \def\@citea{}\@cite{\@for\@citeb:=#2\do
    {\@citea\def\@citea{--\penalty\@m}\@ifundefined
       {b@\@citeb}{{\bf ?}\@warning
       {Citation `\@citeb' on page \thepage \space undefined}}%
\hbox{\csname b@\@citeb\endcsname}}}{#1}}
\relax
\begin{document}
\begin{titlepage}
\hfill\vbox{\hbox{\bf DESY 94--095}
            \hbox{\bf MZ--THEP--94--08}
            \hbox{\bf hep-ph/9406359}}
\begin{flushleft}
\vspace*{3cm}
\hspace*{2cm}{\Large {\bf Heavy Baryons}$\vphantom{X}^{\; *}$}\\[1cm]
\sc
\hspace*{2cm}J.G.K\"orner$^{1}$, M.Kr\"amer$^{1,2}$
{\rm and} D.Pirjol$^{1}$\\[0.5cm]
\em\small
\hspace*{2cm}$^{1}$Johannes Gutenberg--Universit\"at,
Institut f\"ur Physik, D--55099 Mainz, Germany\\
\hspace*{2cm}$^{2}$Deutsches Elektronen--Synchrotron DESY,
D--22603 Hamburg, Germany
\end{flushleft}
\vspace*{1.5cm}
{\sc Abstract}

\rm
\noindent
We review the experimental and theoretical status of baryons containing
one heavy quark. The charm and bottom baryon states are classified and
their mass spectra are listed. The appropriate theoretical framework for
the description of heavy baryons is the Heavy Quark Effective Theory, whose
general ideas and methods are introduced and illustrated in specific
examples. We present simple covariant expressions for the spin wave
functions of heavy baryons including p--wave baryons. The covariant spin
wave functions are used to determine the Heavy Quark Symmetry structure
of flavour--changing current--induced transitions between heavy baryons as
well as one--pion and one--photon transitions between heavy baryons of the
same flavour. We discuss $1/m_Q$ corrections to the current--induced
transitions as well as the structure of heavy to light baryon transitions.
Whenever possible we attempt to present numbers to compare with experiment
by making use of further model--dependent assumptions as e.g. the
constituent picture for light quarks. We highlight recent advances in
the theoretical understanding of the inclusive decays of hadrons containing
one heavy quark including polarization. For exclusive semileptonic decays
we discuss rates, angular decay distributions and polarization effects.
We provide an update of the experimental and theoretical status of
lifetimes of heavy baryons and of exclusive nonleptonic two body decays
of charm baryons.\\[1.5cm]
{\sc Keywords}

\noindent
Angular distributions, heavy baryon decays, heavy quark effective theory,
lifetimes, nonleptonic decays, polarization effects, semileptonic decays.
\\[1.0cm]
$^{*}$ to appear in "Progress in Particle and Nuclear Physics"
\end{titlepage}
\tableofcontents
\newpage

\vspace*{1cm}\hspace*{3cm}
\section{Introduction and Motivation}
\vspace*{2cm}

  This year marks the $20^{th}$ anniversary of the discovery in 1974 of the
$J/\psi$, a narrow meson resonance of mass 3.1 GeV \cite{a,ab}. With the
discovery of the $J/\Psi $ a new era in particle physics began.
In subsequent
 years this state was successfully interpreted as
 a bound state composed of the heavy charm quark
 with mass $m_{c}\approx $1.5 GeV and charge 2/3,
 and its antiparticle. Soon after the discovery of
 the so--called hidden charm state $J/\Psi $ further
 so--called open charm hadrons composed of a charm
 quark and light (u:up , d:down, s:strange)
 quarks/antiquarks were found. The first candidate
 charm baryon states were detected in 1975
 in neutrino interactions \cite{ac} soon
 to be followed by the identification of charm meson
 states at the SPEAR $e^{+}e^{-}$--ring in 1976 \cite{ad,ae}.
 In retrospect, charm hadrons had probably
 made their appearance several years earlier in
 cosmic ray interactions \cite{af}.

The discovery \cite{ag} in 1977 of the $\Upsilon $ family
 of mesons was the first indication of the existence
 of a fifth quark, the bottom quark b, with
 mass $m_{b}\approx 5$ GeV and a charge $-1/3$. Again,
 open bottom meson states composed of a heavy
 bottom quark and a light antiquark were
 identified at a somewhat later stage \cite{ah,ai}. Concerning bottom
baryons the
experimental situation is not yet quite conclusive. There have been reports
on low statistics direct evidence for $\Lambda_b$ in the channels
$\Lambda_b\to\Lambda\psi$ \cite{UA1} and $\Lambda_b\to\Lambda_c\pi^-$
\cite{OPAL93}.
These results need confirmation from other experiments. Some indirect
evidence for semileptonic $\Lambda_b$ and $\Xi_b$ decays exists in the form
of the detection of enhanced $\Lambda_c\ell^-$ (high $p_\perp$)
\cite{HIGHPTLAM} and
$\Xi_c\ell^-$ (high $p_\perp$) \cite{HIGHPTXI} correlations from $Z$--decays
at LEP.
An early 1981 claim to the first observation of the $\Lambda_b$ (at a
rather low mass of $\simeq$ 5425 MeV) in an ISR experiment has not been
upheld and probably was due to a statistical fluctuation \cite{BAS81}.
 Finally, a third species of heavy flavour
 quarks is anticipated but not yet identified in the top quark
 with a mass $m_{t}\approx $ 140 GeV and charge 2/3.

Recent measurements and theoretical calculations
 have substantially enhanced our understanding of
 charm meson states, their spectroscopy and decays.
 Experimental results on charm baryons and their decays are
 beginning to be good enough to apply and test what has been
 learned in the charm meson sector to the charm
 baryon sector. Furthermore, there is a very active ongoing experimental
program at various laboratories to study charm and bottom baryons, their
masses, lifetimes and weak decays. The present experiments and further
planned future experiments will produce a wealth of data on heavy baryons.
 It is therefore timely to review what we know now
 about these states and what we can expect to
 learn from these experiments. How can we extrapolate
 theoretical calculations from mesons to baryons,
 and how do they translate to heavy flavour
 baryons with bottom quantum numbers?

The heavy charm and bottom quarks and the
 heavy hadrons composed of them are quite
 distinct in their properties from the light
 flavoured hadrons composed of u, d and s
 quarks. The large mass of the heavy
 flavoured quarks introduces a mass scale
 much larger than the confinement scale
 $\Lambda \approx $ 400 MeV which governs the physics of
 the light hadrons.

Although heavy hadrons with different
 heavy flavours have distinctly different masses,
 they are in some sense quite similar to
 one another once the appropriate mass scale
 including possible anomalous dimension factors have been
 taken care of. Recently this notion has been given a more
 precise meaning in the Heavy Quark Effective Theory (HQET).
 The HQET provides a systematic expansion of QCD in terms of inverse
powers of the heavy quark mass. The leading term in this expansion gives
rise to a new spin-- and flavour--symmetry, termed Heavy Quark Symmetry.

   Nature has been very accommodating in that it has divided its six
quarks into a heavy and a light quark sector. The ``heavy'' $c,b,t$ quarks
are much heavier than the QCD scale $\Lambda_{QCD}=400$ MeV whereas the
``light'' $u,d,s$ quarks are much (except for the $s$ quark) lighter than
$\Lambda_{QCD}$, i.e. one has
\bea
m_c,m_b,m_t \gg \Lambda_{QCD} \gg m_u,m_d,m_s\,.
\eea

  In the heavy quark sector it then makes sense to first consider QCD in
the limit where the heavy quark masses become very large and then, in the
second stage, to consider power corrections to this limit in terms of a
systematic $1/m_Q$ expansion. Likewise, one can profitably first study the
light quark sector in the zero mass limit, i.e. in the chiral symmetry
limit, and then add corrections to the chiral limit at a later stage.

  It is quite important to realize that the Heavy Quark Symmetry is not a
spectrum symmetry but
it is a new type of equal velocity symmetry. That one cannot expect a
spectrum symmetry to hold in the heavy quark sector should be quite clear
from the fact that there are two orders of magnitude difference between the
masses of the $c$ and $t$ quarks! The new type of
symmetry at equal velocities takes a little bit of getting used to. But
once one has gotten into the habit of thinking in terms of quark and
particle velocities the Heavy Quark Symmetry will in fact look quite natural.

   The basic physics leading to the new spin and flavour symmetries at
equal velocity can easily be appreciated in nontechnical terms by
considering a bottom and charm baryon at rest as shown in Fig.1.

\begin{figure}
\vspace{8cm}
\caption[dummy1]{Portrayal of bottom and charm baryon wave functions at rest.
Upper right corner: wave functions of the hydrogen, deuterium and tritium
atoms.}
\end{figure}

 The heavy
bottom quark and the charm quark at the center are surrounded by a cloud
corresponding to a light diquark system. The only communication between
the cloud and the center is via gluons. But since gluons are flavour blind
the light cloud knows nothing about the flavour at the center. Also,
for infinitely heavy quarks, there is no spin communication between the
cloud and the center. Thus one
concludes that, in the heavy mass limit, a bottom baryon at rest is
identical to a charm baryon at rest regardless of the spin orientation of
the heavy quarks, i.e. one has
\beq
\mbox{Bottom baryon at rest} (\downarrow\uparrow)=
\mbox{Charm baryon at rest} (\uparrow\downarrow)\,.
\eeq
One then just needs to boost the rest configuration by a Lorentz boost from
velocity zero to velocity $v$ to conclude
\beq
\mbox{Bottom baryon at velocity } v (\downarrow\uparrow)=
\mbox{Charm baryon at velocity } v (\uparrow\downarrow)\,,
\eeq
remembering that a Lorentz boost depends only on relative velocities.
Eq.(3) exposes the existence of a new spin and flavour symmetry of QCD at
equal velocities which holds true in the large mass limit. This is nothing
but the advertised Heavy Quark Symmetry.

  In fact, everyone should be quite familiar with the existence of such a
symmetry in the context of QED. Take a hydrogen, deuterium and tritium atom
at rest as also shown in Fig.1. When hyperfine interactions are neglected
they possess identical wave functions and thus identical atomic properties.
The Coulombic interaction between the electron cloud and the nucleus at the
center is sensitive only to the total charge of the nucleus which is the
same for all three isotopes.

   It is quite intriguing that many of the ideas of the HQET date back as
far as 1937, then of course in the context of QED \cite{aq,BOGUL59}. In the
Bloch--Nordsieck
approach to soft photon radiation it was the electron that was
``infinitely'' heavy (on the scale of the soft photons) so it could be
treated as a classical source of radiation. In fact the Bloch--Nordsieck
model was already formulated in terms of an effective theory with the
electron degrees of freedom removed from the field theory. The quantum
mechanical Foldy--Wouthuysen transformation has turned into the
field--theoretical $1/m$ expansion. What used to be
called the eikonal approximation is now referred to as on--mass shell
propagation of heavy quarks with no velocity change (``velocity
superselection rule'').

  We begin in Sec.2 with a discussion of the ground--state charm and bottom
baryons. We list experimental mass values whenever they have been measured.
For the missing mass values we give theoretical extrapolations. In Sec.3
we give a brief outline of HQET where we focus on heavy baryon
applications. In Sec.4 we write down covariant forms of the heavy baryon
spin wave functions including those of the p--wave excited heavy baryon
states. Using the covariant spin wave functions we calculate
current--induced transitions between heavy baryons of different flavours
and discuss their contributions to the sum rule of Bjorken. We also compute
one--pion and one--photon transitions between heavy baryons. In Sec.5 we
review recent advances in the field of inclusive semileptonic decays of
heavy hadrons and hope to convey some of the excitement that has been
spawned by these recent developments. Sec.5 also contains a treatment of
exclusive $c\to s$ and $b\to c$ semileptonic heavy baryon decays including
a discussion of polarization effects where there have been some recent
experimental and theoretical advances. Sec.6 deals with inclusive
nonleptonic decays where we discuss lifetime hierarchies suggested by the
interplay of various theoretical mechanisms contributing to the decays.
Sec.7 treats exclusive nonleptonic charm baryon decays where there has been
a wealth of recent data to compare with theoretical modelling. Sec.8
contains our theoretical summary and an outlook on the heavy baryon physics
that lies ahead of us.

   From the point of view of phenomenological applications, the main
emphasis in this review is on charm baryons and their decays. The obvious
reason is that there already exists enough data on charm baryon decays
to make their study worthwhile, while we are just entering the decade of
experimental bottom baryon physics.

The charm baryons, being the lightest of the
 heavy baryons, may not be the best candidates
 to test and apply the predictions of HQET formulated for
 infinitely heavy quarks. But certainly charm
 hadrons and their decays will be
 the best studied experimentally in the next few years, at least
 what concerns baryons.
 Also they are an ideal laboratory to study the influence of
 preasymptotic 1/m effects to the heavy
 quark limit. And last, but not least, the quality of the
 b$\to$ c physics to be extracted from bottom
 baryon to charm baryon transitions depends
 on the detailed knowledge of the decay properties of
 charm baryons.

   Space limitations preclude an exhaustive treatment of charm and bottom
baryon physics and of the many fascinating aspects of heavy hadron physics
in general. In particular we do not discuss the physics of charm and bottom
baryon production. Instead we focus on the properties of charm and bottom
baryons as revealed in their decays. We refer the reader to earlier reviews
on heavy baryon physics and on heavy hadron physics in general
\citer{ar,SLAC}.

\newpage
\vspace*{1cm}\hspace*{3cm}
\section{Classification of States and Mass Measurements}
\vspace*{2cm}

The ground--state charm baryons are classified
 as usual as members of the SU(4) multiplets $20'$ and 20.
 The $J^P=1/2^+$ ground--state baryons (containing the ordinary
 $C=0$ octet baryons) comprise the $20'$ representation
 and the $J^P=3/2^+$ ground--state baryons (containing the
 ordinary $C=0$ decuplet baryons) make up the 20 representation.
 For the bottom baryons we limit our attention to the lower mass
 $B=1$ and $C=0$ states, which can be classified in analogy to
 the charm baryon states.
 In Tables 1, 2, and 3
 we list the quantum number content and masses of the charm
 baryon members of the $20'$ and 20 representation and of the
 $B=1$, $C=0$ bottom baryon states. We use the same notation
 as the Particle Data Group \cite{pdg}. I and $\mbox{I}_3$ denote
 the isospin; S, C and B refer to the strangeness, charm and
 bottom quantum numbers.

There exist now precise mass measurements for the charm
 $J^P=1/2^+$ baryon states $\Lambda_{c}^{+}$, $\Xi_{c}^{+}$,
 $\Xi_{c}^{0}$, $\Sigma_{c}^{++}$, $\Sigma_{c}^{+}$,
 $\Sigma_{c}^{0}$ \cite{pdg}, $\Omega_{c}^{0}$ \cite{AR,E687} and a first
determination of the mass of the $J^P=3/2^+$ state
 $\Sigma_{c}^{*}$ \cite{SKAT} at 2530 MeV.
 The $\Omega_c^0$ mass listed in Table 1 is an average of the two results
\cite{AR,E687}. Because of the preliminary nature of the $\Sigma_c^*$
mass determination \cite{SKAT} we do not list the experimental mass in
Table 2. Discussion of the recently discovered excited $\Lambda_c$--states
is deferred to Sec.4.

 For the bottom baryons so far only
 the lowest lying state $\Lambda_{b}$ has been observed
 \cite{UA1}. The theoretical predictions of the
 $\Lambda_{b}$ mass range from 5547 to 5660~MeV \cite{kwro}
 and have to be compared with the experimental mass value of
 $5641 \pm 50$~MeV quoted in \cite{pdg}. Using the symmetry
 properties of the static theory for heavy quarks simple relations
 between heavy hadron masses can be derived
 (see e.g. \cite{aglietti}): from the relation
 $M_{\Lambda_c} - 1/4(M_D + 3 M_{D^*}) = M_{\Lambda_b}
  - 1/4(M_B + 3 M_{B^*})$ the $\Lambda_{b}$ mass is predicted to
 be $\cong$ 5630~MeV in good agreement with the experimental
 value. The remaining mass entries in Tables 1, 2, and 3
                                  have been estimated
 in the framework of the one--gluon--exchange model of \cite{rugega} where
isospin splitting effects are not taken into account.
 In the non--relativistic Breit--Fermi reduction the one--gluon--exchange
 contribution leads to a spin--spin interaction of the form
 \begin{eqnarray}\label{spinspin}
  H_{ss}&=& \sum_{i<j}
  \frac{16 \pi \alpha_{s}}{9m_{i}m_{j}}
  \vec s_{i} \cdot \vec s_{j} \delta^{3}(\vec r_{i}-\vec r_{j} )
  \quad .
 \end{eqnarray}
 Starting with the seminal work of \cite{rugega} many authors have
 emphasized the fact that the hyperfine splitting resulting from
 (\ref{spinspin}) is crucial in understanding the mass breaking
 pattern of both heavy and light hadrons \cite{as,kworo}. As long
 as the spin--spin interaction term is taken into account a variety
 of models with differing degrees of sophistication will basically
 reproduce the heavy baryon mass pattern in Tables 1, 2, and 3.
                                   However, for our estimates of
 charm and bottom baryon masses in Tables 1, 2, and 3,
                                  we have retained the original
 version of the one--gluon--exchange model as detailed in \cite{rugega}.

\vspace{1cm}
\begin{table}[h,t]
\begin{center}
\caption[dummy2]{ Charm $1/2^{+}$ baryon states.
 $[ab]$ and $\{ab\}$ denote antisymmetric and symmetric flavour index
 combinations.}\vspace*{5mm}
\renewcommand{\arraystretch}{1.33}
\begin{tabular}{llllrrl}
\hline\hline
 Notation & Quark & SU(3) & $(I, I_{3}) $ & S & C & Mass \\
          & content &     &               &   &   &      \\
\hline
 $\Lambda_{c}^{+}$ & $c[ud]$ & $3^{\ast }$ & (0, 0) &
 0 & 1 & $2285.0\pm 0.6$~MeV \\
 $\Xi_{c}^{+}$ & $c[su]$ & $3^{\ast }$ & (1/2, 1/2) &
 -1 & 1 & $2466.2\pm 2.2$~MeV \\
 $\Xi_{c}^{0}$ & $c[sd]$ & $3^{\ast }$ & (1/2, -1/2) &
 -1 & 1 & $2472.8\pm 1.7$~MeV \\ \hline
 $\Sigma_{c}^{++}$ & $cuu$ & $6$ & (1, 1) &
 0 & 1 & $2453\pm 0.7$~MeV \\
 $\Sigma_{c}^{+}$ & $c\{ud\}$ & $6$ & (1, 0) &
 0 & 1 & $2453 \pm 3.0$~MeV \\
 $\Sigma_{c}^{0}$ & $cdd$ & $6$ & (1, -1) &
 0 & 1 & $2452.5\pm 0.9$~MeV \\
 $\Xi_{c}^{+'}$ & $c\{su\}$ & $6$ & (1/2, 1/2) &
 -1 & 1 & $2.57$~GeV \\
 $\Xi_{c}^{0'}$ & $c\{sd\}$ & $6$ & (1/2, -1/2) &
 -1 & 1 & $2.57$~GeV \\
 $\Omega_{c}^{0}$ & $css$ & $6$ & (0, 0) &
 -2 & 1 & $2719.0 \pm 7.0 \pm 2.5$~MeV \\ \hline
 $\Xi_{cc}^{++}$ & $ccu$ & $3$ & (1/2, 1/2) &
 0 & 2 & $3.61$~GeV \\
 $\Xi_{cc}^{+}$ & $ccd$ & $3$ & (1/2, -1/2) &
 0 & 2 & $3.61$~GeV \\
 $\Omega_{cc}^{+}$ & $ccs$ & $3$ & (0, 0) &
 -1 & 2 & $3.71$~GeV \\ \hline
\end{tabular}
\end{center}
\renewcommand{\arraystretch}{1}
\end{table}


\begin{table}[h]\label{tablec32}
\begin{center}
\caption[dummy3]{ Charm $3/2^{+}$
 baryon states.}\vspace*{5mm}
\renewcommand{\arraystretch}{1.33}
\begin{tabular}{llllrrl}
\hline\hline
 Notation & Quark & SU(3) & $(I, I_{3}) $ & S & C & Mass \\
          & content &     &               &   &   &      \\
\hline
 $\Sigma_{c}^{\ast ++}$ & $cuu$ & $6$ & (1, 1) &
 0 & 1 & $2.51$~GeV \\
 $\Sigma_{c}^{\ast +}$ & $cud$ & $6$ & (1, 0) &
 0 & 1 & $2.51$~GeV \\
 $\Sigma_{c}^{\ast 0}$ & $cdd$ & $6$ & (1, -1) &
 0 & 1 & $2.51$~GeV \\
 $\Xi^{\ast +}$ & $cus$ & $6$ & (1/2, 1/2) &
 -1 & 1 & $2.63$~GeV \\
 $\Xi^{\ast 0}$ & $cds$ & $6$ & (1/2, -1/2) &
 -1 & 1 & $2.63$~GeV \\
 $\Omega_{c}^{\ast 0}$ & $css$ & $6$ & (0, 0) &
 -2 & 1 & $2.74$~GeV \\ \hline
 $\Xi_{cc}^{\ast ++}$ & $ccu$ & $3$ & (1/2, 1/2) &
 0 & 2 & $3.68$~GeV \\
 $\Xi_{cc}^{\ast +}$ & $ccd$ & $3$ & (1/2, -1/2) &
 0 & 2 & $3.68$~GeV \\
 $\Omega_{cc}^{\ast +}$ & $ccs$ & $3$ & (0, 0) &
 -1 & 2 & $3.76$~GeV \\ \hline
 $\Omega_{ccc}^{++}$ & $ccc$ & $1$ & (0, 0) &
 0 & 3 & $4.73$~GeV \\ \hline
\end{tabular}
\renewcommand{\arraystretch}{1}
\end{center}
\end{table}

\begin{table}[t]\label{tableb}
\begin{center}
\caption[dummy4]{ Bottom baryon states with $B=1$,
 $C=0$ and $J^P=1/2^+,3/2^+$ quantum numbers.}\vspace*{5mm}
\renewcommand{\arraystretch}{1.33}
\begin{tabular}{lllllrrl}
\hline\hline
 Notation & Quark   & $J^P$ & SU(3) & $(I, I_{3})$ & S & B & Mass \\
          & content &       &       &              &   &   &      \\
\hline
 $\Lambda_{b}$ & $b[ud]$ & $1/2^+$ & $3^{\ast }$ & (0, 0) &
 0 & 1 & $ 5641 \pm 50$~MeV \\
 $\Xi_{b}^{0}$ & $b[su]$ & $1/2^+$ & $3^{\ast }$ & (1/2, 1/2) &
 -1 & 1 & $ 5.80 $~GeV \\
 $\Xi_{b}^{-}$ & $b[sd]$ & $1/2^+$ & $3^{\ast }$ & (1/2, -1/2) &
 -1 & 1 & $ 5.80 $~GeV \\ \hline
 $\Sigma_{b}^{+}$ & $buu$ & $1/2^+$ & $6$ & (1, 1) &
 0 & 1 & $ 5.82 $~GeV \\
 $\Sigma_{b}^{0}$ & $b\{ud\}$ & $1/2^+$ & $6$ & (1, 0) &
 0 & 1 & $ 5.82 $~GeV \\
 $\Sigma_{b}^{-}$ & $bdd$ & $1/2^+$ & $6$ & (1, -1) &
 0 & 1 & $ 5.82 $~GeV \\
 $\Xi_{b}^{0'}$ & $b\{su\}$ & $1/2^+$ & $6$ & (1/2, 1/2) &
 -1 & 1 & $ 5.94 $~GeV \\
 $\Xi_{b}^{-'}$ & $b\{sd\}$ & $1/2^+$ & $6$ & (1/2, -1/2) &
 -1 & 1 & $ 5.94 $~GeV \\
 $\Omega_{b}^{-}$ & $bss$ & $1/2^+$ & $6$ & (0, 0) &
 -2 & 1 & $ 6.04 $~GeV\\ \hline
 $\Sigma_{b}^{\ast +}$ & $buu$ & $3/2^+$ & $6$ & (1, 1) &
 0 & 1 & $ 5.84 $~GeV \\
 $\Sigma_{b}^{\ast 0}$ & $bud$ & $3/2^+$ & $6$ & (1, 0) &
 0 & 1 & $ 5.84 $~GeV \\
 $\Sigma_{b}^{\ast -}$ & $bdd$ & $3/2^+$ & $6$ & (1, -1) &
 0 & 1 & $ 5.84 $~GeV \\
 $\Xi_{b}^{\ast 0}$ & $bus$ & $3/2^+$ & $6$ & (1/2, 1/2) &
 -1 & 1 & $ 5.94 $~GeV \\
 $\Xi_{b}^{\ast -}$ & $bds$ & $3/2^+$ & $6$ & (1/2, -1/2) &
 -1 & 1 & $ 5.94 $~GeV \\
 $\Omega_{b}^{\ast -}$ & $bss$ & $3/2^+$ & $6$ & (0, 0) &
 -2 & 1 & $ 6.06 $~GeV \\ \hline
\end{tabular}
\renewcommand{\arraystretch}{1}
\end{center}
\end{table}

Of the observed charm and bottom baryons, the $\Lambda_{c}$,
 $\Xi_{c}$, $\Omega_{c}$, and $\Lambda_{b}$ states are weakly
 decaying. According to theoretical expectations, the unobserved
 $\Xi_{cc}$, $\Omega_{cc}$ and $\Omega_{ccc}$ as well as the
 $\Xi_{b}$ and $\Omega_{b}$ states in Tables 1 and 3 are also anticipated
to be weakly decaying.

Because of the spatial $\delta $--function in Equation (\ref{spinspin})
 the matrix elements of the spin--spin interaction term are
 proportional to the square of the baryon wave function at
 the origin. The experimental hyperfine splittings thus provide
 a reliable measure of the wave function at the origin of the
 ground--state baryons, the value of which is needed in lifetime
 estimates (cf.~Sec.6.2).

\newpage
\vspace*{1cm}\hspace*{3cm}
\section{Outline of Heavy Quark Effective Theory}
\vspace*{2cm}

  Over the last past years it has become widely recognized that the
fundamental theory of the strong interactions, quantum chromodynamics
(QCD), simplifies enormously in the presence of a very heavy quark
\cite{pioneers}.
By heavy it is understood that the quark mass must be much larger than the
typical scale of the strong interactions $\Lambda_{QCD}\simeq 400$ MeV.
The Heavy Quark Effective Theory (HQET) \cite{HQET} is a set of rules which
embodies
in a natural way the new symmetries appearing in this limit and describes
in a systematic manner the deviations from the symmetry limit.

   There exist two quarks in nature to which the ideas of HQET can
be applied: the charmed ($m_c\simeq 1.5$ GeV) and the bottom quark ($m_b
\simeq 4.8$
GeV). In this review we will mainly be concerned with applications of the
HQET to the study of baryons made up of one of these two heavy quarks and
two light quarks (denoted as $Qqq$, with $Q$ = c,b and $q$ = u,d,s). The
scope of the method is, however, not limited to this situation. It is
possible to regard a baryon of the type $QQq$ as a bound state of the
heavy pair $QQ$ looking like a pointlike heavy object and the light quark
$q$ \cite{JS}.

   The idea of the Heavy Quark Symmetry is very simple and can be
best explained using a quantum mechanical analogy. A bound state $Qqq$
can be looked upon as consisting of the heavy quark $Q$ surrounded by the
two light valence quarks, gluons and vacuum pairs, which will be
collectively referred to as the ``light diquark system'' or, more
generally, as the ``light degrees of freedom''. The Heavy Quark Symmetry
 expresses
the fact that the state of the ``light degrees of freedom'' is
independent of that of the heavy quark, in the
limit when the mass of the latter goes to infinity. In particular this
means that the ``light degrees of freedom'' will look the same regardless of
the flavour
type and the spin orientation of the heavy quark. Thus, there are actually
two distinct heavy quark symmetries, the flavour symmetry and the spin
symmetry, and we will turn to a separate discussion of their properties and
consequences.

  Before doing this, it is convenient to introduce a formal
field--theoretical description for the heavy quark in a hadron of the
type $Qqq$ or $Q\bar q$. The heavy quark in such a bound state
continually exchanges momentum with the ``light degrees of freedom'', of
the order $\Lambda_{QCD}$, and therefore its change in velocity is of the
order $\Lambda/m_Q$,
which vanishes when the quark is infinitely heavy. Let us consider for
simplicity the rest frame of the hadron. Then the heavy quark will be also
at rest and we can disregard all its dynamical degrees of freedom, except for
colour. It can be then described in terms of the Lagrangian
\beq\label{1}
L_{HQET}=\bar Q(iD_0-m_Q)Q\,,
\eeq
where $D_\mu=\partial_\mu+igA_\mu$. This results from the QCD Lagrangian
\beq\label{2}
L_{QCD}=\bar Q(i\dslash -m_Q)Q\,,
\eeq
when the condition
\beq\label{3}
Q=\frac{1+\gamma_0}{2}Q
\eeq
is imposed. This condition expresses the fact that the heavy quark is at rest
and is equivalent to saying that such a quark is only described by the upper
two components of the Dirac spinor, the Pauli spinor. The generalization
of (\ref{1}) to a heavy quark moving with velocity $v$ is
\beq\label{4}
L_{HQET}=\bar Q_v(iv\cdot D-m_Q)Q_v\,,
\eeq
where the field $Q_v$ now satisfies the condition
\beq\label{5}
Q_v=\frac{1+\vslash}{2}Q_v\,.
\eeq
It is easy to see that in this limit the quark mass $m_Q$ becomes irrelevant,
as it can be completely removed from the Lagrangian through a simple field
transformation
\beq\label{6}
h_v=e^{im_Qv\cdot x}Q_v\,.
\eeq
In terms of the new field $h_v$, the HQET Lagrangian becomes
\beq\label{7}
L_{HQET}=\bar h_v(iv\cdot D)h_v\,.
\eeq

  The reader will have noticed that we labeled the heavy quark field $h_v$
with the heavy quark velocity $v$. This is to say that for each possible
velocity we introduce a distinct field, which duplicates the initial one,
and for each such field we have one term in the Lagrangian similar to
(\ref{7}).
There is no term in this Lagrangian which connects heavy quark fields of
different velocities, so the quark velocity is a good quantum number.
This statement is sometimes called the ``velocity superselection rule''.

  Having developed the formalism of the HQET at zero$^{th}$ order in
$1/m_Q$, we are in a position to discuss the two symmetries specific
to the infinite mass limit. As discussed before, the flavour symmetry
relates heavy hadrons containing different (heavy) quarks. Let us denote
the two heavy quark species by $b$ and $c$. Then the total Lagrangian is
\beq\label{8}
L_{HQET}=\bar h_{v}^{(b)}(iv\cdot D)h_{v}^{(b)}
 + \bar h_{v}^{(c)}(iv\cdot D)h_{v}^{(c)}\,.
\eeq
Note that the two heavy quarks must have the same velocity. It is easy
to see that this Lagrangian is invariant under the transformation
\begin{displaymath}
{h_v^{(c)}\choose h_{v}^{(b)}} \to U{h_v^{(c)}\choose h_{v}^{(b)}}
\end{displaymath}
where $U$ is an arbitrary SU(2) matrix. This is the formal statement of
the flavour symmetry. Any symmetry in quantum mechanics has in general two
consequences: degeneracies and an associated conservation law, and this one
is no exception to the rule. The degeneracy implied by the flavour
symmetry can be expressed as
\beq\label{9}
m_{bqq}-m_b=m_{cqq}-m_c\,,
\eeq
that is, the mass of the light degrees of freedom (for given quantum
numbers) is independent of the
type of the heavy quark. The other consequence of this symmetry is the
existence of a conserved operator, a kind of isospin, which we will denote
by $\vec\tau$. It has the following properties
\bea\label{10}
&&\tau_3|bqq\rangle=|bqq\rangle\,\qquad\tau_3|cqq\rangle=-|cqq\rangle\,\\
&&\tau_-|bqq\rangle=|cqq\rangle\,\qquad\tau_+|cqq\rangle=|bqq\rangle\,
\eea
where $\tau_\pm=\tau_1\pm i\tau_2$. An explicit representation for
$\vec\tau$ is
\bea\label{12}
\vec\tau=\frac{1}{2}\int\mbox{d}^3x(\bar h_{v}^{(b)}(x)\,\,\bar h_{v}^{(c)}
(x)) \gamma_0\vec\sigma{h_{v}^{(b)}(x)\choose h_v^{(c)}(x)}
\eea
where $\vec\sigma$ are the Pauli matrices. Note that they act in the flavour
space, not on the Dirac indices of the fields. The conservation of $\vec\tau$
can be used to relate amplitudes for processes involving $b$ quarks to
those involving $c$ quarks, much in the same way as the conservation of
the usual isospin can be used to relate processes involving protons to
those involving neutrons.

   Another symmetry appearing in the infinite mass limit is the spin
symmetry: the ``light degrees of freedom'' are insensitive to the spin
orientation of the heavy quark. This can be seen by noting that the HQET
Lagrangian (\ref{7}) is invariant under an arbitrary spin rotation of the
heavy quark (for a heavy quark at rest)
\beq\label{13}
h_v\to \exp(\frac{i}{2}\vec\Sigma\cdot\vec n\theta)h_v
\eeq
where $\vec n$ and $\theta$ are respectively, the rotation axis and the
rotation angle, and $\vec\Sigma$ is the spin operator.
This symmetry implies a degeneracy between the two states of
spin $J=j\pm\frac{1}{2}$ obtained by coupling the heavy quark spin
$s=\frac{1}{2}$ with the angular momentum of the ``light degrees of freedom''
$j$. Such states are for example $B$ and $B^*$ for mesons and $\Sigma_b$
and
$\Sigma^*_b$ for baryons. The corresponding conserved quantity is, of course,
the heavy quark spin
\beq\label{14}
\vec s=\frac{1}{2}\int\mbox{d}^3x\bar h(x)\gamma_0\vec\Sigma h(x)\,.
\eeq
The conservation of this operator can be used to relate amplitudes for
processes involving the two partners of a multiplet described above.

  As an example of how the HQET can relate various transition
amplitudes, we consider the calculation of the matrix element
\cite{B,HKKT91}
\beq\label{15}
M_\mu=
\langle\Lambda_c(v_2,s_2)|\bar c\gamma_\mu b|\Lambda_b(v_1,s_1)\rangle
\eeq
appearing in the description of the semileptonic decay $\Lambda_b\to
\Lambda_ce\bar\nu_e$.
It will be shown that, in the limit of infinitely heavy $b$ and $c$ quarks,
it can be completely described in terms of $\xi(\omega)$ ($\omega=v_1\cdot
v_2$), which is one of
the elastic form--factors of the $\Lambda_c$ baryon, defined through
\beq\label{16}
\langle\Lambda_c(v_2,s_2)|\bar c\gamma_\mu c|\Lambda_c(v_1,s_1)\rangle =
\bar u(v_2,s_2)[\xi(\omega)\gamma_\mu+\zeta(\omega)(v_1+v_2)_\mu ]
u(v_1,s_1)
\,.
\eeq
This is the most general form which is allowed for this matrix element from
current conservation.
Furthermore, in the infinite mass limit, the form--factor $\zeta(\omega)$
will be shown to vanish.

  The states in (\ref{16}) are normalized according to
\bea
\langle \Lambda_Q(p_1,s_1)|\Lambda_Q(p_2,s_2)\rangle =
2E_1 (2\pi)^3\delta^{(3)}(\vec p_1-\vec p_2)\nonumber
\eea
and the spinors $u(v,s)$ satisfy $\bar u(v,s)u(v,s)=2m_{\Lambda_Q}$.

  Let us write the matrix element (\ref{16}) in the HQET using the
Lagrangian (\ref{4}) and go to the reference
frame where $v_1=(1,\vec 0)$. Also, take $s_1=+\frac{1}{2}$ and consider the
$(1+i2)$ component of the vector current. This gives
\bea\label{17}
\langle\Lambda_c(v_2,s_2)|\hcbar\gamma_{1+i2}\hc|\Lambda_c(\vec 0,\uparrow)
\rangle = \zeta(\omega)\bar u(v_2,s_2)u(\vec 0,\uparrow)v_{2_{1+i2}}\,,
\eea
since $\gamma_{1+i2}u(\vec 0,\uparrow)=0$. We now use the commutation
relation
\beq\label{18}
[s_3^{(v_1)},\hcbar\gamma_{1+i2}\hc] =
{\textstyle \frac{1}{2}}\hcbar\gamma_{1+i2}\hc
\eeq
which can be obtained from the defining relation for $\vec s\,^{(v_1)}$
(\ref{14}).
Here the fields $\hc$ and $\hcbar$ must be considered as distinct,
according to the discussion above, and therefore they anticommute. Thus we
can write
\bea\label{19}
& &\frac{1}{2}\langle\Lambda_c(v_2,s_2)|\hcbar\gamma_{1+i2}\hc|\Lambda_c
(\vec 0,
\uparrow)\rangle =\nonumber\\
& & \langle\Lambda_c(v_2,s_2)|s_3^{(v_1)}\hcbar\gamma_{1+i2}\hc
-\hcbar\gamma_{1+i2}\hc s_3^{(v_1)}|\Lambda_c(\vec 0,\uparrow)\rangle=
\\
& & -\frac{1}{2}\langle\Lambda_c(v_2,s_2)|\hcbar\gamma_{1+i2}\hc|\Lambda_c(
\vec 0,\uparrow)\rangle = 0\,,\nonumber
\eea
where we have used that $s_3^{(v_1)}|\Lambda_c(v_2,s_2)\rangle=0$ because
this state contains no $v_1$--type heavy quarks. Comparing with (\ref{17}),
this gives that
\beq\label{20}
\zeta(\omega)=0\,.
\eeq

  We have seen in the above calculation the heavy quark spin symmetry
in action. Next we make use of the flavour symmetry to relate the transition
matrix element for $c\to c$ to the matrix element for $b\to c$.
The proof is entirely analogous to the preceding one, and makes use of
the following commutation relation
\beq\label{21}
[\tau_-,\,\hcbar\gamma_\mu\hc ]=-\hcbar\gamma_\mu\hb\,.
\eeq
This gives
\bea\label{22}
& &\langle\Lambda_c(v_2,s_2)|\hcbar\gamma_\mu\hb|\Lambda_b(v_1,s_1)\rangle=
-\langle\Lambda_c(v_2,s_2)|\tau_-\hcbar\gamma_\mu\hc - \hcbar\gamma_\mu\hc
\tau_-|\Lambda_b(v_1,s_1)\rangle\nonumber\\
& &=\langle\Lambda_c(v_2,s_2)|\hcbar\gamma_\mu\hc|\Lambda_c(v_1,s_1)\rangle
=\bar u(v_2,s_2)\gamma_\mu u(v_1,s_1)\xi(\omega)\,.
\eea
This proves the promised result. Actually it is much more general: any
matrix element between any $\Lambda_Q$ baryon states can be expressed,
in the infinite mass limit, in terms of the same function $\xi(\omega)$
\beq\label{23}
\langle\Lambda_c(v_2,s_2)|\hcbar\Gamma\hb|\Lambda_b(v_1,s_1)\rangle=
\bar u(v_2,s_2)\Gamma u(v_1,s_1)\xi(\omega)\,,
\eeq
with $\Gamma$ an arbitrary gamma matrix.
The universal function $\xi(\omega)$ is called the Isgur--Wise function
or the reduced form--factor.
It encodes nonperturbative, long--distance properties of QCD, which are
at present noncalculable (except on a lattice or from QCD sum rules) and
has therefore to be extracted from experiment. There is one thing which can
be said about this function, its value for $\omega=1$ is known
\beq\label{24}
\xi(1)=1\,.
\eeq
This may be seen by taking $v_1=v_2=v$ in (\ref{16}) and noting that the
value of the
matrix element on the l.h.s. is equal to $2m_{\Lambda_c}v_\mu\delta_{s_1,
s_2}$
because the vector current is conserved. Also, on the r.h.s. $\bar u(v,s_2)
\gamma_\mu u(v,s_1)=2m_{\Lambda_c}v_\mu \delta_{s_1,s_2}$ (because $\vslash
u(v,s)=u(v,s)$), which gives (\ref{24}).

  In principle, the relation (\ref{23}) and its generalization to other
baryon states could be proved by making use of commutation relations like the
ones above. This procedure is rather tedious and, fortunately, there exists
a more elegant method which allows one to do the same with considerably
less effort. This is the covariant wave--function method, which will be
presented later on in Sec.4.

   In the infinite mass limit, the matrix element in (\ref{23}) becomes
independent of the heavy quark mass. This is to be expected, because
it is a measure of the overlap between the ``light degrees of freedom'' of
the initial
and final baryons, which only depends on the velocity change $\omega$
but not on the masses of the heavy quarks. There are two sorts of
corrections to this result which induce a mass--dependence: i) radiative
corrections and ii) power--suppressed mass corrections.

   The first type of corrections are due to hard gluon exchange between
the initial and final heavy quarks and bring about a logarithmic mass
dependence of the matrix elements. Because of lack of space, we only
quote the result in the leading logarithm approximation \cite{LLA}
and refer the reader to the literature for its derivation:
\bea\label{25}
\lefteqn{\langle\Lambda_c(v_2,s_2)|\hcbar\Gamma\hb|\Lambda_b(v_1,s_1)
\rangle =}\\
& &\left(\frac{\alpha_s(m_b)}{\alpha_s(m_c)}\right)^{-\frac{6}{25}}
\left(\frac{\alpha_s(m_c)}{\alpha_s(\mu)}\right)^{-\frac{8}{27}
  [\omega r(\omega)-1]} [\bar u(v_2,s_2)\Gamma u(v_1,s_1)]
  \xi(\omega,\mu)\nonumber
\eea
with
\beq\label{26}
r(x)=\frac{1}{\sqrt{x^2-1}}\log (x+\sqrt{x^2-1})\,.
\eeq
The Isgur--Wise function $\xi(\omega,\mu)$ acquires a dependence on the
renormalization scale $\mu$, which is exactly compensated by the
$\mu$--dependence in the anomalous dimension factor, so that the matrix
element is $\mu$--independent.

   In contrast to the radiative corrections discussed above, which have
left unchanged the prediction (\ref{23}) (apart from multiplying it with a
correction factor), the corrections suppressed by one or more powers of
$1/m_Q$ change its form. This will be seen to happen because these
corrections break in general both the spin and the flavour heavy quark
symmetries. To order $1/m_Q$ they can be obtained in a simple way as
follows. Consider the Dirac equation for the heavy quark field
\beq\label{27}
(i\dslash - m_Q)Q=0\,.
\eeq
Decompose $Q(x)$ into an ``upper'' and a ``lower'' component
\beq\label{28}
Q(x)=e^{-im_Qv\cdot x}(h^{(+)}(x)+h^{(-)}(x))\,, \qquad
\vslash h^{(\pm)}=\pm h^{(\pm)}\,.
\eeq
Introducing this into the Dirac equation (\ref{27}) and projecting the
resulting relation onto the ``upper'' and ``lower''--spaces gives two
equations
\bea\label{29}
& &i\dslash_\perp h^{(+)} - (iv\cdot D+2m_Q)h^{(-)} =0\\
& &iv\cdot Dh^{(+)} + i\dslash_\perp h^{(-)} =0\,.
\eea
The first relation can be used to solve for $h^{(-)}$ in terms of
$h^{(+)}$:
\beq\label{31}
h^{(-)}=\frac{i\dslash_\perp}{2m_Q}h^{(+)}+{\cal O}(1/m_Q^2)
\eeq
which, when inserted into the second relation, gives the equation of
motion for the $h^{(+)}$ field
\bea\label{32}
\left(iv\cdot D+\frac{(i\dslash_\perp)^2}{2m_Q}+\cdots\right)h^{(+)}=0\,.
\eea
This can be considered to have arisen from the Lagrangian \citer{1m,MRR}
\bea\label{33}
{\cal L}_{HQET}=\bar h^{(+)}iv\cdot D h^{(+)} + \bar h^{(+)}
\frac{(i\dslash_\perp)^2}{2m_Q}h^{(+)} + \cdots
\eea
which gives the generalization of (\ref{7}) by including corrections of order
$1/m_Q$ (the derivation presented here has been taken from \cite{SLAC}).
The relation of the QCD field $Q(x)$ to the HQET field $h^{(+)}$ can be found
from (\ref{28},\ref{31}) to be given by
\bea\label{34}
Q(x)=e^{-im_Qv\cdot x}\left(1+\frac{i\dslash_\perp}{2m_Q}+\cdots\right)
h^{(+)}(x)\,.
\eea

  We mention that the extension of this method to obtain the higher order
corrections to the HQET Lagrangian (see also \cite{MRR}) is not without risk
\cite{BKP,DAS}. This is due to the fact that the $h^{(+)}(x)$ field has a
mass--dependent normalization. The correct heavy quark
field must be determined from the condition of its having the same
normalization as the QCD field $Q(x)$ \cite{KT}.

  We proceed now with investigating the effects of the new terms which
appear in the HQET Lagrangian at order $1/m_Q$. They can be written
in a slightly different form as
\bea\label{35}
{\cal L}_{HQET}=\bar h^{(+)}iv\cdot D h^{(+)} + \frac{1}{2m_Q}\bar h^{(+)}
\left((iD)^2-(iv\cdot D)^2-\frac{g}{2}\sigma_{\mu\nu}
F^{\mu\nu}\right)h^{(+)} + \cdots
\eea
with
\beq\label{36}
F_{\mu\nu}^a=\partial_\mu A_\nu^a - \partial_\nu A_\mu^a - gf_{abc}
A_{\mu}^bA_{\nu}^c\,.
\eeq
All three terms at order $1/m_Q$ break the flavour symmetry. Indeed, one
cannot rotate anymore the fields of two heavy quarks into each other, as
before, because the mass factors $1/m_Q$ are different for the two quarks.
The last term breaks also the spin symmetry, because it is no longer
invariant under the field transformation (\ref{13}). One therefore expects
corrections to the predictions (\ref{23}), which have been obtained under
the assumption that these two symmetries are valid. There is, however, one
important
kinematical point where these corrections vanish and the leading order
result (\ref{23}) remains valid. This happens for the zero recoil point
$v_1=v_2$. This result is called Luke's theorem
\cite{LUKE,GGW,BB1,KT}.
We shall prove Luke's
theorem for the special case of $\Lambda_b\to\Lambda_c$ transitions.
We will do this by explicitly calculating the form of the $1/m_c$--order
correction to the matrix elements of the vector and axial transition
currents away from the equal--velocity point \cite{GGW}. The correction
proportional to $1/m_b$ can be calculated in a completely analogous manner.

   There are two sources of corrections to the prediction
(\ref{23}): i) corrections due to the new terms in the Lagrangian (\ref{35})
and ii) corrections due to the modified form of the current:
\beq\label{37}
\bar c\Gamma b \to \hcbar\Gamma\hb - \frac{1}{2m_c}\hcbar
(i\stackrel{\leftarrow}{\dslash_\perp})\Gamma\hb\,.
\eeq

\begin{figure}
\vspace{11.5cm}
\caption[dummy5]{ a) ${\cal O}(1)$ matrix element for a $b\to c$ transition;
b) the same for a $c\to c$ transition; c) insertion of a ${\cal O}(1/m_c)$
term in the HQET Lagrangian for a $b\to c$ transition; d) the same for
a $c\to c$ transition; e) vertex correction of order ${\cal O}(1/m_c)$
to a $b\to c$ transition.}
\end{figure}

These corrections are represented into a graphical form in Fig.2. The
insertion of a $1/m_c$ term in the HQET Lagrangian is shown as a
cross on the $c$ propagator and contributes a correction to (\ref{23}) equal
to
\bea\label{38}
& &\frac{i}{2m_c}
\langle\Lambda_c(v_2,s_2)|\int\mbox{d}^4x\mbox{T}(\hcbar(iD)^2\hc)(x)
(\hcbar\Gamma\hb)(0)|\Lambda_b(v_1,s_1)\rangle\nonumber\\
& &-\frac{i}{4m_c}
\langle\Lambda_c(v_2,s_2)|\int\mbox{d}^4x\mbox{T}(\hcbar g\sigma_{\mu\nu}
F^{\mu\nu}\hc)(x)(\hcbar\Gamma\hb)(0)|\Lambda_b(v_1,s_1)\rangle=\nonumber\\
& &\frac{1}{2m_c}\bar u(v_2,s_2)\Gamma u(v_1,s_1)\eta(\omega) +
\frac{1}{2m_c}\bar u(v_2,s_2)\sigma^{\mu\nu}\frac{1+\vslash_2}{2}
\Gamma u(v_1,s_1)\zeta_{\mu\nu}(\omega)
\eea
where $\eta(\omega)$ is an unknown function and $\zeta_{\mu\nu}$ is the
most general antisymmetric tensor which can be built from $v_1,v_2$ and hence
is proportional to $v_{1_\mu} v_{2_\nu}-v_{1_\nu} v_{2_\mu}$ (the combination
$\epsilon_{\mu\nu\lambda\xi}v_1^\lambda v_2^\xi$ is not allowed because it
has the wrong parity, it transforms as a pseudotensor). Actually, this
form for $\zeta_{\mu\nu}$ vanishes when inserted in (\ref{38}) because
$\sigma_{\mu\nu}v_2^\nu =0$ when sandwiched between projectors $(1+\vslash_2)
/2$ and we are only left with the first term.

   The correction to the current (\ref{37}) is shown in Fig.2 as a cross on
the upper vertex and contributes to the matrix element (\ref{23}) the term
\bea\label{39}
-\frac{1}{2m_c}
\langle\Lambda_c(v_2,s_2)|(\hcbar(i\stackrel{\leftarrow}{\dslash_\perp})
\Gamma\hb)(0)|\Lambda_b(v_1,s_1)\rangle\,.
\eea
Let us first calculate the matrix element
\bea\label{40}
\langle\Lambda_c(v_2,s_2)|\hcbar(i\stackrel{\leftarrow}{D_\mu})
\Gamma\hb|\Lambda_b(v_1,s_1)\rangle=\bar u(v_2,s_2)\Gamma u(v_1,s_1)
[Av_{1_\mu}+Bv_{2_\mu}]\,,
\eea
in terms of which one can readily express (\ref{39}). Here
$A,B$ are functions of $v_1\cdot v_2$ which we will compute now. Multiplying
(\ref{40}) with $v_2{_\mu}$ and using the equation of motion for the $\hc$
field $\hcbar
iv_2\cdot\stackrel{\leftarrow}{D}=0$, one obtains
\beq\label{41}
B=-A v_1\cdot v_2\,.
\eeq
$A$ can be obtained by noting that
\bea\label{42}
\langle\Lambda_c(v_2,s_2)|i\partial_\mu(\hcbar\Gamma\hb)|\Lambda_b(v_1,s_1)
\rangle &=& \bar\Lambda(v_{1_\mu}-v_{2_\mu})\langle\Lambda_c(v_2,s_2)|\hcbar
\Gamma\hb|\Lambda_b(v_1,s_1)\rangle\nonumber\\
&=&\bar\Lambda(v_{1_\mu}-v_{2_\mu})\bar u(v_2,s_2)
\Gamma u(v_1,s_1)\xi(\omega)
\eea
with $\bar\Lambda=m_{\Lambda_Q}-m_Q$ the binding energy of the $\Lambda_Q$
baryon, which is independent of $Q$ in the infinite mass limit. On the other
hand, the first matrix element in this equation can be written as
\bea\label{43}
\langle\Lambda_c(v_2,s_2)|\hcbar(i\stackrel{\leftarrow}{D_\mu})\Gamma\hb|
\Lambda_b(v_1,s_1)\rangle + \langle\Lambda_c(v_2,s_2)|\hcbar\Gamma
(i\stackrel{\to}{D_\mu})\hb|\Lambda_b(v_1,s_1)\rangle
\eea
By contracting this with $v_{1_\mu}$ and comparing with (\ref{40},\ref{42}),
the following result emerges
\beq\label{44}
A=\frac{\bar\Lambda\xi(\omega)}{1+\omega}\,.
\eeq

  Thus, for example the correction to (\ref{23}) for $\Gamma=\gamma_\mu$ can
be obtained to be of the form
\bea\label{45}
-\frac{1}{2m_c}\bar u(v_2,s_2)[2v_{1_\mu}-\gamma_\mu(1+\omega)] u(v_1,s_1)
\frac{\bar\Lambda\xi(\omega)}{1+\omega}+\frac{1}{2m_c}[\bar u(v_2,s_2)
\gamma_\mu u(v_1,s_1)]\eta(\omega)\,.
\eea
One can obtain a condition on $\eta(1)$ by considering the same correction to
the matrix element
\bea
\langle\Lambda_c(v,s_2)|\hcbarv\gamma_\mu\hc|\Lambda_c(v,s_1)\rangle\nonumber
\eea
which, by current conservation, is known to vanish. This can be calculated
in a completely analogous way, with the result
\beq\label{46}
\frac{1}{m_c}[\bar u(v_2,s_2)\gamma_\mu u(v_1,s_1)](\eta(\omega)+\bar\Lambda
\xi(\omega))-\frac{1}{m_c}\frac{\bar\Lambda\xi(\omega)}{1+\omega}
[\bar u(v_2,s_2)u(v_1,s_1)](v_{1_\mu}+v_{2_\mu})\,.
\eeq
This should vanish for $v_1=v_2$, whence $\eta(1)=0$.

  The correction to (\ref{23}) for the other case of physical significance
$\Gamma=\gamma_\mu\gamma_5$, is equal to
\bea\label{47}
-\frac{1}{2m_c}\bar u(v_2,s_2)[2v_{1_\mu}+\gamma_\mu(1-\omega)]\gamma_5
u(v_1,s_1)
\frac{\bar\Lambda\xi(\omega)}{1+\omega}+\frac{1}{2m_c}[\bar u(v_2,s_2)
\gamma_\mu\gamma_5 u(v_1,s_1)]\eta(\omega)\,.
\eea

  We can summarize our findings by introducing six form--factors $f_{1,2,
3}^V(\omega)$ and $f_{1,2,3}^A(\omega)$, defined by
\bea\label{48}
\langle\Lambda_c(v_2,s_2)|\bar c\gamma_\mu b|\Lambda_b(v_1,s_1)\rangle
&=&\bar u(v_2,s_2)[f_1^V\gamma_\mu+f_2^Vv_{1_\mu}+f_3^Vv_{2_\mu}]u(v_1,s_1)\\
\label{49}
\langle\Lambda_c(v_2,s_2)|\bar c\gamma_\mu\gamma_5 b|\Lambda_b(v_1,s_1)
\rangle
&=&\bar u(v_2,s_2)[f_1^A\gamma_\mu+f_2^Av_{1_\mu}+f_3^Av_{2_\mu}]\gamma_5
u(v_1,s_1)\,.
\eea
In terms of these form--factors, the relations (\ref{23},\ref{45},\ref{47})
can be compactly expressed as
\bea\label{50}
f_1^V(\omega)&=&\xi(\omega)+\left(\frac{1}{2m_c}+\frac{1}{2m_b}\right)
 (\eta(\omega)+\bar\Lambda\xi(\omega))\\
f_2^V(\omega)&=&-\frac{1}{m_c}\frac{\bar\Lambda
 \xi(\omega)}{1+\omega}\\
f_3^V(\omega)&=&-\frac{1}{m_b}\frac{\bar\Lambda
 \xi(\omega)}{1+\omega}\\
f_1^A(\omega)&=&\xi(\omega)+\left(\frac{1}{2m_c}+\frac{1}{2m_b}\right)
 (\eta(\omega)+\frac{\bar\Lambda\xi(\omega)(\omega-1)}{1+\omega})\\
f_2^A(\omega)&=&-\frac{1}{m_c}\frac{\bar\Lambda
 \xi(\omega)}{1+\omega}\\\label{55}
f_3^A(\omega)&=&\frac{1}{m_b}\frac{\bar\Lambda
 \xi(\omega)}{1+\omega}\,.
\eea
We have included here also the corrections proportional to $1/m_b$, which
can be obtained from those of order $1/m_c$ by taking the complex
conjugate followed by the interchange of the quark labels $c\leftrightarrow
b$.

   For $v_1=v_2$ one can see, by making use of $\xi(1)=1$ and $\eta(1)=0$,
that
\bea\label{56}
& &f_1^V(1)+f_2^V(1)+f_3^V(1)=1\\\label{57}
& &f_1^A(1)=1\,
\eea
which is just the content of Luke's theorem for the $\Lambda_{Q_1} \to
\Lambda_{Q_2}$ transitions
(the original derivation of the theorem \cite{LUKE} was
 given for the transition matrix elements between meson states
$B\to  D^{(*)}$). The linear combination of vector current amplitudes
in Eq.(\ref{56}) and the axial vector current amplitude $f_{1}^{A}$
in Eq.(\ref{57}) are the partial wave s--wave amplitudes that survive
in the limit $\omega \to 1$ (see the discussion in Sec.5.2.2). We
emphasize that Luke's theorem and the ${\cal O}(1/m_Q)$ normalization
condition applies to the s--wave amplitudes as written down in
(\ref{56},\ref{57}).
We mention that the $1/m_Q$ corrections have been worked
out also for other baryonic transitions of interest, like $\Omega_b\to
\Omega_c^{(*)}$ \cite{BB2}, for which simplifications similar to
(\ref{56},\ref{57}) hold true at the zero--recoil point $v_1=v_2$.

We mention that Luke's theorem can be proved in a very general context
using a diagrammatic language \cite{KT}. Referring to the $b\to c$ and
$c\to c$ transitions drawn in Fig.2 one notes that there is a doubling
up of the contributions of both the Lagrangian and the current insertion
when going from the inelastic $b\to c$ to the elastic $c\to c$ case. Since
one knows from current conservation that the elastic $1/m_c$ corrections
have to vanish at the zero recoil point (or, equivalently, at $q^2=0$) one
concludes that also the inelastic $1/m_c$ corrections have to vanish at
the zero recoil point --- they are exactly one--half of the vanishing
elastic $1/m_c$ corrections. This way of proving Luke's theorem is
independent of what happens on the light side and thus immediately applies
to any heavy particle transition, be it baryons, mesons or even
supersymmetric heavy particles. Using the diagrammatic language one can
also immediately appreciate that Luke's theorem breaks down at
${\cal O}(1/m_{c}^{2})$. The elastic ${\cal O}(1/m_{c}^{2})$ transition
shown in diagram 2f has no analogue in the inelastic $b\to c$ case and thus
there is no longer a doubling up argument as was used in the
${\cal O}(1/m_c)$ proof.

   This concludes our brief presentation of the basic ideas and methods
of the HQET. The reader can find more details and applications in the many
good existing reviews on this subject \cite{review,SLAC}.

\newpage
\vspace*{1cm}\hspace*{3cm}
\section{\label{wf and trans}Heavy Baryon Spin Wave Functions and Heavy
Baryon Transitions}
\vspace*{2cm}
\newcommand{\kslash}{\mbox{$\not{\hspace{-1mm}k}$}}             
\newcommand{\eslash}{\mbox{$\not{\hspace{-0.8mm} \varepsilon}$}}
\newcommand{\matheins}{\mbox{$\rule{2.5mm}{0.1mm}
                          {\hspace{-2.7mm}1}
                          {\hspace{-0.2mm}\rule{0.07mm}{2.7mm}}$}}

\newcommand{\ersetze}{\rule[-3.5mm]{.2mm}{7mm}}
\newcommand{\young}{\mbox{$\Box{\hspace{-1mm} \Box} $}}

   The heavy baryons that we are mainly concerned with in this review are
bound states formed from a heavy quark and a light diquark system. The
spin--parity quantum numbers $j^{P}$ of the light diquark system are
determined from the spin and orbital degree of freedom of the two light
quarks that make up the diquark system. From the spin degrees of freedom
of the two light quarks one obtains a spin 0 and a spin 1 state. The total
orbital state of the diquark system is characterized by two angular degrees
of freedom which
we take to be the two independent relative momenta $k=\frac{1}{2}(p_{1}-
p_{2})$ and $K=\frac{1}{2}(p_{1}+p_{2}-2p_{3})$ that can be formed from the
two light quark momenta $p_{1}$ and $p_{2}$ and the heavy quark momentum
$p_{3}$. The k--orbital momentum describes relative orbital excitations of
the two light quarks, and the K--orbital momentum describes orbital
excitations of the center of mass of the two light quarks relative to the
heavy quark as drawn in Fig.3.
\begin{figure}[h]
\vspace{7cm}
\caption[dummy6]{ Orbital angular momenta of the light diquark system.
$l_{k}$ describes relative orbital momentum of the two light quarks and
$l_{K}$ describes orbital momentum of the center of mass of the light quarks
relative to the heavy quark.}
\end{figure}

In this review we limit our discussion to the ground state baryons with
$l_{k}=l_{K}=0$ and the p--wave baryons with ($l_{k}=0$, $l_{K}=1$) or
($l_{k}=1$, $l_{K}=0$). A treatment of higher orbital excitations can be
found in \cite{HKT}. The flavour symmetry nature of the light diquark
state can then be determined from the generalized Pauli principle as
applied to the light quark sector. Totally antisymmetric (symmetric)
spatial configurations are antisymmetric (symmetric) in flavour space. The
antisymmetric flavour configurations $\Lambda_{[q_{1}q_{2}]Q}$ will be
generically
referred to as $\Lambda$--type states and the symmetric flavour
configurations $\Sigma_{\{q_{1}q_{2}\}Q}$ as $\Sigma$--type states. In SU(3)
$(q=u,d,s)$ the $\Lambda$--type states form an antitriplet $3^{*}$ and the
$\Sigma$--type states a sextet 6 according to the decomposition $3\otimes3=
3^{*}\oplus 6$.

Mass values of the ground state charm and bottom baryons have been listed in
Tables 1,2 and 3. As concerns p--wave levels there
are altogether seven $\Lambda$--type and seven $\Sigma$--type p--wave states
for a given flavour configuration.
According to a quark model calculation \cite{CIK79} done in the charm baryon
sector the p--wave levels are well separated from the ground states. For the
$\Lambda_{c}$--type p--wave states the two $(l_{k}=0, \; l_K=1)$ states are
lowest because orbital and spin--spin splitting effects work in the same
direction to lower these two states while the five $(l_{k}=1, \; l_K=0)$
states are raised. The total orbital and spin--spin splitting effect amounts
to $\cong 350$~MeV. For the $\Sigma_{c}$--type states, however, the orbital
and spin--spin splitting effects work in opposite directions leading to a
close level spacing of the seven $\Sigma_{c}$--type states.

These qualitative features clearly show up in the $\Lambda_{c}$-- and $\Sigma
_{c}$--type charm baryon level plot in Fig.4 taken from the calculation
of \cite{CIK79}. Copley et al. \cite{CIK79} used a constituent quark model
based on harmonic oscillator interquark forces. The two recently found
excited
$\Lambda_{c}$--states at $\cong \,2593$~MeV \cite{CLEOLAMBDA93} and at
$\cong\,
2627$~MeV \citer{CLEOLAMBDA93,E687LAMBDA93}
lie almost on top of the two $(l_{k}=0,\;
l_{K}=1)$ levels predicted by \cite{CIK79} thus inviting an interpretation
of these two new states as forming the $1/2^{-}$ and $3/2^{-}$ members of the
$(l_{k}=0, \; l_{K}=1)$ Heavy Quark Symmetry spin doublet. The details of the
closely spaced level ordering of the remaining five $\Lambda_{c}$--type and
seven $\Sigma_{c}$--type p--wave states awaits to be unravelled by further
experimental and theoretical effort.

There exists no universal agreement on how to label the excited heavy baryon
states. In a spectroscopic notation one would write ${}^{2s+1}_{\quad j}(l;
l_{1},l_{2})_{J}$ where $l_{1}$ and $l_{2}$ are the two light--side orbital
degrees of freedom coupling to a total orbital momentum $l$. The total
orbital momentum then couples with the spin singlet or triplet state $2s+1$
to form a light diquark state with spin j. The total spin J  of the heavy
baryon is then obtained by coupling j with the heavy quark spin $S_{Q}=1/2$
to form $J=j \pm 1/2$. In order to avoid the cumbersome spectroscopic
notation we use a more concise notation in this review which is i) tailored
to the p--wave states and ii) uses the $l_{1}=l_{k}$ and $l_{2}=l_{K}$ basis
which diagonalizes the Hamiltonian, at least in the harmonic oscillator
approximation \cite{CIK79}. The $(l_{k}=0, l_{K}=1)$ and the $(l_{k}=1, l_{K}
=0)$ states will be referred to as the K-- and k--states, respectively. Heavy
Quark Symmetry doublets will be denoted by $\{ B_{QKj} \}$ or $\{ B_{Qkj} \}$
(j=1,2) and the singlets (j=0) by $B_{QK0}$ and $B_{Qk0}$ ($B=\Lambda$-- or
$\Sigma$--type). The two degenerate members of the doublets are denoted by
$\{B_{QKj}\}:=\{ B_{QKj}, B_{QKj}^{*} \}$ (and the same for $K \rightarrow
k$) for total heavy baryon spins $\{ J=j-1/2, J=j+1/2 \}$. When summarily
referring
to excited heavy baryon states these will be called $B_{Q}^{**}$ as in
$\Lambda_{Q}^{**}$ or $\Sigma_{Q}^{**}$.

\subsection{\label{ground state wf}Ground State Spin Wave Function}
\begin{figure}
\vspace{12cm}
\caption[dummy7]{ Spectrum of strangeness zero  $\Lambda_{c}$-- and
$\Sigma_{c}$--type s--wave and p--wave charm baryon states. Mass values for
s--wave states from Tables 1 and 2.
Mass values of the two lowest
$\Lambda_{c}$--type p--wave states are taken from
\protect{\citer{CLEOLAMBDA93,E687LAMBDA93}}. Masses of
remaining p--wave states are taken from Copley et al.\protect{\cite{CIK79}}.
They are adjusted upward to be in agreement with the measured $\Lambda_{c}$
mass.}
\end{figure}
The ground state heavy baryons ($l_{k}=l_{K}=0$) are made from the heavy
quark Q with spin--parity $J^{P}=\frac{1}{2}^{+}$ and a light diquark system
with spin--parity $0^{+}$ ($\Lambda$--type) and $1^{+}$ ($\Sigma$--type)
moving
in a s--wave state relative to the heavy quark. The spin wave functions of
the light diquark system will be denoted by $\chi^{0}$ and $\chi^{1,\mu}$ for
the spin 0 and spin 1 diquark, respectively. When one combines the diquark
spin with the heavy quark's spin one obtains the ground state heavy baryons
$\Lambda_{Q}$ and $\{\Sigma_{Q},\Sigma_{Q}^{*}\}$ according to the coupling
scheme
\begin{eqnarray}
0^{+}&\otimes&{\textstyle
\frac{1}{2}^{+} \to \; \frac{1}{2}^{+}}\qquad\qquad\Lambda_Q \\
&\nonumber\\
1^{+}&\otimes &{\textstyle \frac{1}{2}^{+}} {\nearrow \atop \searrow }
  \begin{array}{l}
  \frac{1}{2}^{+} \\
  \\
  \frac{3}{2}^{+}
  \end{array}
\qquad
\left \{  \begin{array}{l}
  \Sigma_{Q} \\
  \\
  \Sigma_{Q}^{*}
\end{array} \right \}
\end{eqnarray}
The two states $\Sigma_{Q}$ and $\Sigma_{Q}^{*}$ are  exactly degenerate
in the heavy quark limit since the heavy quark possesses no spin interaction
with the light--side diquark system as $m_{Q}\rightarrow \infty$.

For the purposes of deriving the consequences of the Heavy Quark Symmetry
the only information needed about the light diquark system is its spin and
its
parity. This entails the transversality condition on the $1^{+}$ state
$v^{\mu} \chi^{1}_{\mu}=0$. Nevertheless it is convenient (but not necessary)
to regard the spin 0 and spin 1 diquark system as being composed of two
light quarks according to
\begin{equation}
{\textstyle \frac{1}{2}^{+}}
\otimes {\textstyle \frac{1}{2}^{+}} = 0^{+} \oplus 1^{+} \quad .
\end{equation}
The explicit forms of the covariant bispinor spin wave functions in the
constituent quark model read (see \cite{HKT})
\beq\label{SWF}
\renewcommand{\baselinestretch}{1.5}
\begin{array}{lll}
0^{+} : \qquad \hat \chi^{0}_{ \alpha \beta }& =& \frac{1}{2\sqrt{2} }{ [(
\vslash + 1)\gamma_{5} C ]}_{\alpha  \beta}\\[1mm]
1^{+} : \qquad \hat \chi^{1, \mu}_{\alpha \beta} &=& \frac{1}{2\sqrt{2} }{ [(
\vslash +1 )\gamma^{\mu}_{\bot}C]}_{\alpha \beta}
\end{array}
\eeq
where  $\gamma^{\mu}_{\perp}$ is the (four--) transverse $\gamma$--matrix
defined by $\gamma^{\mu}_{\bot} =\gamma^{\mu} -\vslash v^{\mu} $, and
$v^{\mu}$ is the four--velocity of the diquark system (equal to the heavy
baryon's four velocity $v^{\mu}=\frac{P^{\mu}}{M}$). C is the $4\times 4 $
charge conjugation matrix $ C=i \gamma_{0} \gamma_{2} $ and serves to
"pull down"
the antispinor index in the remaining spinor--antispinor $\gamma$--matrix
combination. In the following we shall drop explicit reference to the spin
of the bispinor state and write $\hat \chi$ and $ \hat \chi^{\mu} $ for
$ \hat \chi^{0}$ and $ \hat \chi^{1,\mu}$, respectively,
where this does not lead
to confusion.

The spin wave functions $\hat \chi$ and $\hat \chi^{\mu}$ satisfy the
so--called Bargmann--Wigner equations on both labels, i.e.
\begin{equation}\label{BW}
\vslash_{\alpha \alpha^{'}}  \hat \chi_{\alpha^{'} \beta} =\vslash_{
\alpha \beta^{'} } \hat \chi_{\beta \beta^{'} } = \hat \chi_{\alpha
\beta}
\end{equation}
and similarly for ${\hat\chi}^{\mu} $. They further possess the symmetry
properties
\beq
\begin{array}{ccc}
\hat \chi_{\alpha \beta}&=&-\hat \chi_{\beta \alpha}\\
{\hat\chi_{\alpha \beta}}^{\mu}&=& {\hat\chi_{\beta \alpha}}^{\mu}
\end{array}
\eeq
The transverse $\gamma_{\mu}$--matrix is used in the spin 1 part of
Eq.(\ref{SWF}) in order to ensure that the spin 1 wave function is
transverse to the
four--velocity $v^\mu$, i.e.
\beq\label{trans}
v_{\mu} \hat \chi^{\mu} =0
\eeq
The transversality condition (\ref{trans}) insures that $\hat \chi^{\mu}$
reduces to a three--component object in the particle's rest frame (r.f.)
$v_{\mu}=(1,0,0,0)$.
The Bargmann--Wigner condition (\ref{BW}) in turn implies that the bispinor
wave functions reduce to an upper--left two by two matrix in the rest frame
which, from the transversality condition (\ref{trans}), has the appropriate
r.f. spin transformation behaviour. In fact one has
\beq \label{bispin}
\begin{array}{ccc}
\hat \chi_{\alpha \beta}\ersetze_{\: r.f.} & = & -\frac{1}{\sqrt{2}}\left(
  \begin{array}{cc}
  i\sigma^{2} & 0 \\
  0 & 0
  \end{array}
\right )\\
&&\\
\hat \chi^{\mu}_{\alpha \beta}\ersetze_{\: r.f.}&=&-\frac{1}{\sqrt{2}}\left(
  \begin{array}{cc}
  \sigma^{k} i \sigma^{2}& 0\\
  0&0
  \end{array}
\right )
\end{array}
\eeq
where the $\sigma^{k}\hspace{1mm} (k=1,2,3)$ are the usual Pauli matrices
and $i\sigma^{2}$ is the $2\times2$ charge conjugation matrix \cite{JONES}.
When
reading Eq.(\ref{bispin}) component--wise in the spherical basis (see
Eq.(\ref{4-73})) one recovers the familiar spin wave functions
$\hat \chi \hspace{0.7mm}\ersetze_{\: r.f.}=
\frac{1}{\sqrt{2}}(\uparrow \downarrow-\downarrow \uparrow)$ etc. The
representations (\ref{SWF}) can be seen to form a covariant way of writing
the Clebsch--Gordan coupling in a moving frame
$v^{\mu}$ with $\vec v \not= 0$ or,
put differently, the covariant spin wave functions are just boosted
rest frame spin wave functions.

Let us briefly have a look at the normalization of the bispinor diquark
spin wave functions Eq.(\ref{SWF}). The conjugate spin wave functions are
given by (see e.g. \cite{GU})
\beq\label{conj1}
\bar {\hat \chi} (v)=C^{-1} C^{-1} \hat \chi(-v)
\eeq
such that
\beq\label{conj2}
\begin{array}{lll}
\bar {\hat \chi} &=& -\frac{1}{2\sqrt{2}}[C^{-1} \gamma_{5} (\vslash+1)]
\\[3mm]
\bar {\hat \chi}^{\mu} &=& -\frac{1}{2\sqrt{2}}[C^{-1} \gamma^{\mu}_{\perp}
(\vslash+1)]
\end{array}
\eeq
The normalization can then be calculated to be
\beq\label{norm}
\begin{array}{lll}
{\bar {\hat \chi}}_{\alpha \beta} \hat \chi_{\alpha \beta }&=&1\\[2mm]
{\bar {\hat \chi}}^{\mu}_{\alpha \beta} \hat\chi^{\nu}_{\alpha \beta}&=&
-g^{\mu\nu}_{\perp}
\end{array}
\eeq
where $g^{\mu \nu}_{\perp}=g^{\mu \nu}-v^{\mu}v^{\nu}$.

It is sometimes convenient to transform to the spherical basis for the
spin 1
diquarks which can be done with the help of the spin 1 polarization vectors.
One has ($\lambda=\pm 1,0$)
\beq\label{4-73}
\hat \chi (1,\lambda)=\varepsilon_{\mu}(\lambda) \hat \chi^{\mu}
\eeq
and the inverse
\beq
\hat \chi^{\mu}=\sum_{\lambda}{}\varepsilon^{*\mu}(\lambda) \hat
\chi(1,\lambda)
\eeq
where
\beq
\begin{array}{lll}
\varepsilon_{\mu}(\pm)&=&\mp \frac{1}{\sqrt{2}}(0,1,\pm i,o)\\[2mm]
\varepsilon_{\mu}(0)&=&(\mid \vec v \mid,0,0,v_{0})
\end{array}
\eeq

In as much as the spin wave functions $\hat \chi $ and ${\hat \chi}^{\mu}$
satisfy the Bargmann--Wigner equation on both spinor labels they are
spin wave functions built from constituent on--mass shell quarks. While
these do not adequately describe the actual physical situation of the light
diquark system the on--mass shell spin wave functions are nevertheless quite
useful
when one wants to construct non--constituent spin wave functions with the
correct spin and parity of the diquark system. To obtain the full light
side spin wave functions one just multiplies the on--shell spin wave
functions ${\hat \chi}_{\alpha \beta}$ with a spinor valued matrix $A_{
\alpha \beta}^{\alpha' \beta'}$ such that the spin and parity of the light
diquark system remain untouched. The resulting off--shell spin wave functions
will be denoted by an unhatted object and reads
\beq\label{off-shell}
\chi_{\alpha \beta} = A_{\alpha \beta}^{\alpha' \beta'} {\hat \chi}_{\alpha'
\beta'}
\eeq
where the wave function matrix $A_{\alpha \beta}^{\alpha' \beta'}$ in general
depends on the nature of the diquark state as well as on the velocity
$v_{\mu}$ and the relative momenta $k$ and $K$. The multiplication with the
spin and parity neutral matrix $A_{\alpha \beta}^{\alpha' \beta'}$ serves to
soften the light--side spinor structure. This would e.g. be achieved by the
replacement $(\vslash +1)\rightarrow (A \vslash + B +C \kslash + D \kslash
\cdot \vslash)$ for the positive energy projection in the spin wave functions
(\ref{SWF}). The off--shell spin wave functions $\chi_{\alpha \beta}$ no
longer satisfy the Bargmann--Wigner equations but still represent a diquark
system transforming as $ J^{P}=0^{+} $ and $J^{P}=1^{+} $ under parity
transformations and SO(3) rest frame rotations. The constituent quark model
can be seen to be a special case of Eq.(\ref{off-shell}) where the matrix
$A_{\alpha \beta}^{\alpha' \beta'}$ takes the special form $A_{\alpha
\beta}^{\alpha'\beta'}=\delta_{\alpha}^{\alpha'} \delta_{\beta}^{\beta'} A$.

The normalization of the wave function matrix $A_{\alpha \beta}^{\alpha'
\beta'}$ must be such that the normalization conditions Eq.(\ref{norm})
generalize to
\begin{eqnarray}
(\chi,\chi)&=&1\label{gnorm1} \\
(\chi^{\mu},\chi^{\nu})&=&-g^{\mu \nu}_{\perp}\label{gnorm2}
\end{eqnarray}
where the inner product ( , ) is defined with regard to integrations and
traces over the internal degrees of freedom of the diquark state. The exact
form of the phase space integral, the spinor trace and the form of the
integrand need not concern us here since we are only interested in the rest
frame transformation properties of the spinor tensors $\chi$ and
$\chi^{\mu}$,
and their correct normalization which we define through Eqs.(\ref{gnorm1})
and (\ref{gnorm2}).

We are now in the position to write down the spin wave functions  $\Psi_
{\alpha \beta \gamma}$ of the ground state heavy baryons by writing down
invariant couplings between the light--side spinor tensors $\chi$ and
$\chi^{\mu}$and the heavy--side spinor tensors $\psi$ and $\psi^{\mu}$ of the
ground state baryons according to the coupling scheme
\beq\label{coupling}
\underbrace{{\textstyle\frac{1}{2}^{+}} \otimes
{\textstyle \frac{1}{2}^{+}}}_{light \hspace{1mm} side}
 \otimes\underbrace{{\textstyle\frac{1}{2}^{+}} \otimes
{J}^{P}}_{heavy \hspace{1mm} side}
\Longrightarrow 0^{+}.
\eeq
The heavy--side spinor tensors $\psi$ and $\psi^{\mu}$ involve the heavy
baryon spinor $u$ (for $J^{P}=\frac{1}{2}^{+}$) and the Rarita--Schwinger
spinor--vector $u^{\mu}$ (for $J^{P}=\frac{3}{2}^{+}$), and their couplings
to the heavy quark spinor label. The rule is that if additional tensor
structure is required on the heavy side one brings in a factor of $\gamma_{
\perp}^{\mu}$ (remember that $v^{\mu}$ annihilates on the light side tensor).
One then has
\beq\label{gr.state1}
 \hspace{-4.5mm}\Lambda_{Q} \hspace{0.5mm} : \qquad \Psi_{\alpha \beta
 \gamma}\hspace{3mm}
 = \chi_{\alpha \beta}
\psi_{\gamma} \equiv \chi u
\eeq
\beq\label{gr.state2}
  \begin{array}{ll}
  \{ \Sigma_{Q} \}: \qquad \Psi_{\alpha \beta \gamma}&
 =\chi^{\mu}_{\alpha \beta} \psi_{\mu,\gamma}\\[4mm]
 &
 \equiv \chi^{\mu} \left\{
     \begin{array}{c}
     \frac{1}{\sqrt{3}} \gamma^{\perp}_{\mu} \gamma_{5} u\\
     u_{\mu}
     \end{array}
  \right\} \end{array}
\eeq
where the curly bracket notation always implies a two--fold Heavy Quark
Symmetry degeneracy. The Lorentz contraction on the r.h.s. of
Eq.(\ref{gr.state2}) is required because the total heavy baryon
wave function $\Psi_{\alpha \beta \gamma}$ on the l.h.s. of
Eq.(\ref{gr.state2}) transforms as a scalar in Lorentz space (but
not in spinor space).\footnote{For added emphasis we keep the
transversality labels in scalar products such as $\chi^\mu\gamma_\mu^\perp$
most of the time even though the the transversality label could be
dropped since $\chi^\mu\gamma_\mu^\perp = \chi^\mu\gamma_\mu$ since
$v^\mu\chi_\mu=0$.}

The total spin wave functions $\Psi_{\alpha \beta \gamma}$ satisfy the
Bargmann--Wigner (or mass--shell) condition on the heavy quark spinor label
$\gamma$, i.e.
\beq\label{on-mass}
\vslash_{\gamma \gamma'} \Psi_{\alpha \beta \gamma'}= \Psi_{\alpha \beta
\gamma}
\eeq
showing that the heavy quarks appear in the theory as freely propagating
 on--mass shell quarks as is required in the Heavy Quark Symmetry limit.
Note that a factor of $\gamma_{5}$ is needed in the $\Sigma_{Q}$ spin wave
function in order to satisfy the mass--shell condition (\ref{on-mass}).

The normalization of the heavy--side spin wave functions $\psi$ and
$\psi^{\mu}$ can be seen to follow from the overall normalization condition
\beq
(\Psi,\Psi)=2M
\eeq
where the inner product ( , ) is defined as in Eqs.(\ref{gnorm1}) and
(\ref{gnorm2}). Using the fact that the light--side and heavy--side spin wave
functions factorize in the sense of Eqs.(\ref{gr.state1}) and
(\ref{gr.state2}) one then obtains the appropriate normalization conditions
for the heavy--side spin wave functions $\psi$ and $\psi^{\mu}$. In fact one
has
\begin{eqnarray}
 \Lambda_{Q}  &:&2M=(\Psi,\Psi)=(\chi,\chi){\bar\psi}\psi\label{normal1}\\
 \{ \Sigma_{Q} \} &:&2M=(\Psi,\Psi)=(\chi^{\mu},\chi^{\nu}){\bar\psi}_{\mu}
                 \psi_{\nu}\label{normal2}
\end{eqnarray}
Using the normalization conditions Eqs.(\ref{gnorm1}) and (\ref{gnorm2})
for the light--side wave functions one obtains
\beq\label{lnorm1}
\bar\psi \psi = 2M
\eeq
\beq\label{lnorm2}
-g^{\mu \nu}_{\perp} {\bar\psi}_{\mu} \psi_{\nu}=2M.
\eeq
The $\Lambda$-- and $\Sigma$--type heavy--side spin wave functions $\psi$ and
$\psi_{\mu}$ in Eqs.(\ref{gr.state1}) and (\ref{gr.state2}) can be seen to
satisfy the normalization conditions (\ref{lnorm1}) and (\ref{lnorm2})
using ${\bar u}u=2M$ and ${\bar u}^{\mu} u_{\mu} =-2M$.

\subsection{\label{excited wf}Excited Heavy Baryon States}
The spin wave function formalism introduced in Sec.(\ref{ground state wf})
for
the ground state baryons can easily be extended to describe excited baryon
states. The coupling scheme (\ref{coupling}) now involves also orbital
angular momentum and reads
\beq
\underbrace{{\textstyle \frac{1}{2}^{+}} \otimes
{\textstyle \frac{1}{2}^{+}} \otimes l_{k}^{P=(-)
^{l_{k}}} \otimes l_{K}^{P=(-)^{l_{K}}}}_{light \hspace{1mm} side }
\otimes\;
\underbrace{
{\textstyle \frac{1}{2}^{+}} \otimes J^{P} }_{heavy \hspace{1mm}side}
\Longrightarrow 0^{+}.
\eeq

In the tensor formalism the orbital excitations are represented by tensor
products of the relative momenta $k_{\mu}^{\perp}=k_{\mu} - k \hspace{-0.7mm}
\cdot \hspace{-0.7mm} v \, v_{\mu}$ and $K_{\mu}^{\perp}=K_{\mu} - K
\hspace{-0.7mm} \cdot \hspace{-0.7mm} v \, v_{\mu}$ where the
transversality again reduces the relative four--momenta to relative
three--momenta in the rest frame $ \vec v = 0 $. Here we shall only discuss
p--wave orbital excitations.

Combining the p--wave negative parity orbital angular momentum state $j^{P}=
1^{-}$ with the two $j^{P}=0^{+},1^{+}$ spin states one has the
following spin--parity content for the total light--side diquark states:
\begin{eqnarray}
0^{+} \otimes 1^{-}&=& 1^{-}\label{s.-p.content1}\\
1^{+} \otimes 1^{-}&=& 0^{-} \oplus 1^{-} \oplus 2^{-}\label{s.-p.content2}
\end{eqnarray}
For example, in the tensor formalism the decomposition (\ref{s.-p.content2})
 is achieved by writing e.g.
\beq\label{decomp}
{\hat \chi }^{\mu_{1}} k_{\perp}^{\mu_{2}}=
{\textstyle \frac{1}{3}} {\hat \chi}^{\mu}
k_{\perp\mu} g^{\mu_{1} \mu_{2}}_{\perp} +
{\textstyle \frac{1}{2}} [\hat \chi^
{\mu_{1}}k_{\perp}^{\mu_{2}}] +
{\textstyle \frac{1}{2}} \{ {\hat \chi}^{\mu_{1}}
k_{\perp}^{\mu_{2}} \}_{0}
\eeq
where $\{\hspace{1mm}\}_{0}$ stands for the traceless symmetric tensor
product. Without any loss of generality we have taken $k^{\perp}_{\mu} $ to
represent the orbital excitation in the above example. For our purposes it is
more convenient to represent the spin one piece of (\ref{decomp}) by an
one--index tensor according to
\begin{eqnarray}
{\textstyle \frac{1}{2}}
[\hat\chi^{\mu_{1}}k_{\perp}^{\mu_{2}}]&\approx &-{\textstyle \frac{1}{2}}
\varepsilon_{
\mu \mu_{1} \mu_{2} \alpha } {\hat \chi}^{\mu_{1}} k^{\mu_{2}} v^{\alpha}
\label{one ind}\\
&:=& -{\textstyle \frac{1}{2}} \varepsilon
(\mu\, {\hat \chi}\, k\, v) \nonumber
\end{eqnarray}
(we define $\varepsilon_{0123} = 1 $ as in Bjorken--Drell).
In the following we use a
concise notation for $\varepsilon$--tensor contractions, cf.
\beq
\varepsilon (\mu {\hat \chi } k v ) := \varepsilon_{\mu \mu_{1}
\mu_{2} \alpha } {\hat \chi }^{\mu_{1}} k^{\mu_{2}} v^{\alpha} \quad .
\eeq

As in the case of the ground--state baryons we can now use the on--shell
diquark
states $\hat \chi$ and  ${\hat \chi}^{\mu}$ together with the orbital momenta
represented by $k^{\mu}_{\perp}$ and $K^{\mu}_{\perp}$ to build up the
light--side states ${\hat \phi}^{\mu_{1} \dots  \mu_{j}} $ with the desired
$j^{P}$ quantum numbers. For illustrative purposes such explicit
constructions
have been listed in Tables 4 and 5
(where the spin zero metric contraction $g_{\perp} ^{\mu_{1}\mu}$
has been already absorbed into the heavy--side spin wave functions). The
construction has to be such that the resulting spinor--tensors ${\hat
\phi}^{\mu_{1} \dots \mu_{j}}$ representing a spin j diquark state have to be
i) transverse on all indices ii) totally symmetric in all indices and iii)
traceless w.r.t. any pair of indices. This would be the approach that one
would take in a constituent type quark model approximation.

%
\begin{table}

\caption[dummy8]{ Spin wave functions (s.w.f.) of $\Lambda$--type s-- and
          p--wave
          heavy baryons. Light--side spin wave functions are constituent
          spin wave functions.}
\vspace{5mm}
\renewcommand{\baselinestretch}{1.2}
\small \normalsize
\begin{center}
\begin{tabular}{*{5}c}
\hline \hline
& \begin{tabular}{c}
    light side s.w.f.\\
    ${\hat\phi}^{\mu_{1} \dots \mu_{j}}$
  \end{tabular}
& $j^{P}$
& \begin{tabular}{c}
    heavy side s.w.f.\\
    $\psi_{\mu_{1} \dots \mu_{j}}$
  \end{tabular}
&$J^{P}$
\\ \hline \hline
$\Lambda_{Q}$&$\hat \chi$&$ 0^{+}$&$u$&$\frac{1}{2}^{+}$\\
\hline
$\{ \Lambda_{QK1} \} $&$ {\hat \chi}^{0} K_{\perp}^{\mu_{1}}$&
$1^{-}$&
    $\begin{array}{r}
      \frac{1}{\sqrt{3}} \gamma^{\perp}_{\mu_{1}} \gamma_{5} u\\
      u_{\mu_{1}}
     \end{array} $
   &$\begin{array}{c}
     \frac{1}{2}^{-} \\
     \frac{3}{2}^{-}
     \end{array}$
\\ \hline
$\Lambda_{Qk0}$&${\hat \chi}^{1} \cdot k_{\perp}$&$0^{-}$&$u$&$
\frac{1}{2}^{-}$
\\ \hline
$\{ \Lambda_{Qk1} \}$&$\frac{1}{2} \varepsilon (\mu_{1} {\hat
\chi}^{1} k_{\perp} v)$&$1^{-}$&
    $ \begin{array}{r}
      \frac{1}{\sqrt{3}} \gamma^{\perp}_{\mu_{1}}\gamma_{5}u\\
      u_{\mu_{1}}
      \end{array} $
  &$  \begin{array}{c}
      \frac{1}{2}^{-} \\
      \frac{3}{2}^{-}
      \end{array} $
\\ \hline
$ \{ \Lambda_{Qk2} \} $&$ \frac{1}{2}  {\hat \chi}^{1\{\mu_{1}}
k_{\perp}^{\mu_{2}\}_{0}}$&$2^{-}$&
    $\begin{array}{r}
     \frac{1}{\sqrt{10}} \gamma_{5} \gamma_{\{ \mu_{1}}^{\perp}
                               u_{\mu_{2} \}_{0} }^{\hspace{1mm}}\\
     u_{\mu_{1} \mu_{2} }
     \end{array} $
  &$\begin{array}{c}
    \frac{3}{2}^{-}\\
    \frac{5}{2}^{-}
    \end{array}$
\\ \hline \hline
\end{tabular}
\end{center}
\renewcommand{\baselinestretch}{1}
\small \normalsize
\vspace{1cm}
\caption[dummy9]{ Spin wave functions (s.w.f.) of $\Sigma$--type s-- and
                  p--wave heavy baryons. Light--side spin wave functions
                  are constituent spin wave functions. }
\vspace{5mm}
\renewcommand{\baselinestretch}{1.2}
\small \normalsize
\begin{center}
\begin{tabular}{*{5}c}
\hline \hline
& \begin{tabular}{c}
     light side s.w.f.\\
     ${\hat\phi}^{\mu_{1} \dots \mu_{j}} $
  \end{tabular}
& $j^{P}$
& \begin{tabular}{c}
     heavy side s.w.f.\\
     $\psi_{\mu_{1} \dots \mu_{j}}$
  \end{tabular}
& $J^{P}$
\\ \hline \hline
$\{ \Sigma_{Q} \}$ &${\hat \chi}^{1\mu_{1}}$& $1^{+}$&
        $\begin{array}{r}
             \frac{1}{\sqrt{3}}\gamma^{\perp}_{\mu_{1}} \gamma_{5}u\\
             u_{\mu_{1}}
         \end{array} $
       &$\begin{array}{c}
             \frac{1}{2}^{+}\\
             \frac{3}{2}^{+}
         \end{array} $
\\ \hline
$\{ \Sigma_{Qk1} \}$&${\hat \chi}^{0} k^{\mu_{1}}_{\perp}$&$1^{-}$&
       $\begin{array}{r}
            \frac{1}{\sqrt{3}}\gamma^{\perp}_{\mu_{1}} \gamma_{5} u\\
            u_{\mu_{1}}
        \end{array}$
     &$\begin{array}{c}
            \frac{1}{2}^{-}\\
            \frac{3}{2}^{-}
        \end{array}$
\\ \hline
$\Sigma_{QK0} $&${\hat\chi}^{1} \cdot K_{\perp}$&$0^{-}$&$u$&$
\frac{1}{2}^{-}$
\\ \hline
$\{ \Sigma_{QK1} \} $&$ \frac{1}{2} \varepsilon(\mu_{1} {\hat \chi}
^{1} K_{\perp}v)$&$1^{-}$&
       $\begin{array}{r}
            \frac{1}{\sqrt{3}} \gamma^{\perp}_{\mu_{1}} \gamma_{5} u\\
            u_{\mu_{1}}
       \end{array}$
     &$\begin{array}{c}
            \frac{1}{2}^{-} \\
            \frac{3}{2}^{-}
       \end{array}$
\\ \hline
$ \{ \Sigma_{QK2} \}$&$ \frac{1}{2} {\hat \chi}^{1\{\mu_{1}} K_{
\perp}^{\mu_{2}\}_{0}}$&$2^{-}$&
     $\begin{array}{r}
           \frac{1}{\sqrt{10}}\gamma_{5} \gamma^{\perp}_{ \{ \mu_{1} }
                                     u_{\mu_{2} \}_{0} }^{\hspace{1mm}}\\

           u_{\mu_{1} \mu_{2}}
      \end{array}$
    &$\begin{array}{c}
          \frac{3}{2}^{-} \\
          \frac{5}{2}^{-}
      \end{array}$
\\ \hline \hline
\end{tabular}
\renewcommand{\baselinestretch}{1}
\small \normalsize
\end{center}
\end{table}

The spinor--tensor spin wave functions ${\hat \phi}^{\mu_{1} \dots \mu_{j}}$
listed in Tables 4 and 5 have the correct parity and
spin angular momentum to describe the diquark states. As in the ground--state
case the spinor--tensor has to be multiplied by a wave function matrix $A^
{\alpha' \beta'}_{\alpha \beta}$ in order to obtain the full diquark state
wave functions. One thus has
\beq\label{diquark wf}
[ \phi^{\mu_{1} \dots \mu_{j}}]_{\alpha \beta}= A^{\alpha' \beta' }_{\alpha
\beta} [{\hat \phi}^{\mu_{1} \dots \mu{j}}]_{\alpha' \beta'}
\eeq
where the wave function matrix $A^{\alpha' \beta'}_{\alpha \beta }$ in
general depends on the external and internal degrees of freedom of the
diquark state, i.e. it is different for different diquark states. In the
following we shall mostly suppress spinor labels.

The full diquark wave function $\phi^{\mu_{1} \dots \mu_{j}}$ satisfies the
normalization condition
\beq\label{diquark norm}
( \phi^{\nu_{1} \dots \nu_{j} },\phi^{\mu_{1} \dots \mu_{j}})= G^{\mu_{1}
 \dots \mu_{j}; \nu_{1} \dots \nu_{j}}
\eeq
where the inner product is defined as an integration and trace over the
internal degrees of freedom of the diquark state as in Eqs.(\ref{gnorm1})
and (\ref{gnorm2}). $G^{\mu_{1} \dots \mu_{j}; \nu_{1} \dots \nu_{j}}$ is a
generalized transverse metric tensor which is i) transverse in all indices,
ii) symmetric in the sets of indices $\{ \mu_{i} \}$ and $\{ \nu_{i} \}$ and
iii) traceless w.r.t. to any index pair in $\{ \mu_{i} \}$ or in $\{ \nu_{i}
\}$. Its general explicit form can be found in \cite{HKT}.

We emphasize again that all that is needed for the purposes of Heavy Quark
Symmetry is the $j^{P}$ transformation behaviour of the light--side diquark
states together with the normalization condition (\ref{diquark norm}). The
constituent states listed in Tables 4 and 5 can be viewed
as possible "interpolating fields" of the true diquark states. Furthermore
explicit forms of the constituent states are needed in later
applications if one wants to make reference to the constituent quark
model approach.

Although we are dealing only with diquark spins $j=0,1$ and $2$ in this
review the generic notation introduced in (\ref{diquark wf}) and
(\ref{diquark norm}) turns out to be quite convenient even for
these simple cases. It is also
easily generalized to higher spins \cite{HKT} (see also \cite{FALK}). To be
explicit we list the
subsidiary conditions and the normalization tensors for the light--side
diquark
states with spins $j=0,1,2$. One has
\begin{eqnarray}
i)\quad    j=0&:&  \phi; \qquad G=1 \\[4mm]
ii)\quad   j=1&:& \phi^{\mu_{1}} ;\qquad G^{\mu_{1} \nu_{1}}=-g_{\perp}
                     ^{\mu_{1}\nu_{1}}\nonumber\\[2mm]
&\hphantom{:}&
                     \mbox{transversality: } v^{\mu_{1}} \phi_{\mu_{1}}=0
                                 \\[4mm]
iii)\quad j=2&:&  \phi^{\mu_{1} \mu_{2}} ;\qquad G^{\mu_{1} \mu_{2}; \nu_{1}
                     \nu_{2}}= {\textstyle \frac{1}{2}}
                     ( g_{\perp}^{\mu_{1} \nu_{1}}
                     g_{\perp}^{\mu_{2} \nu_{2}}+g_{\perp} ^{\mu_{1} \nu_{2}}
                     g_{\perp} ^{\mu_{2} \nu_{1}}  - {\textstyle \frac{2}{3}}
                     g^{\mu_{1}\mu_{2}} g^{\nu_{1} \nu_{2}})\nonumber\\[2mm]
&\hphantom{:}&
                     \mbox{symmetry: } \phi^{\mu_{1} \mu_{2}}=\phi^{\mu_{2}
                     \mu_{1}} \nonumber\\[2mm]
&\hphantom{:}&
                   \mbox{transversality: }v^{\mu_{1}} \phi _{\mu_{1} \mu_{2}}
                    =0\nonumber\\[2mm]
&\hphantom{:}&
                    \mbox{tracelessness: } g_{\perp}^{\mu_{1} \mu_{2}} \phi_
                    {\mu_{1} \mu_{2}}=0
\end{eqnarray}

It is a useful and instructive exercise to transform the cartesian tensors
$\phi^{\mu_{1} \dots \mu_{j}}$ to a spherical basis. First note that because
of the normalization condition Eq.(\ref{diquark norm}) the cartesian tensors
$ \phi^{\mu_{1} \dots \mu_{j}}$ can be looked upon as forming a set of
 orthonormal vectors
in a (2j+1) dimensional linear vector space. They can thus be represented
by the state vectors $\mid \mu_{1} \dots \mu_{j}\rangle$ which satisfy
orthonormality and completeness relations. One has
\begin{eqnarray}
\mbox{orthonormality} &:& \quad \langle\nu_{1} \dots \nu_{j} \mid \mu_{1}
                          \dots \mu_{j}\rangle= G^{\nu_{1} \dots \nu_{j}}
                          _{\mu_{1}\dots \mu_{j}} \label{orthonormality}\\
\mbox{completeness}   &:& \quad \mid \mu_{1} \dots \mu_{j}\rangle \langle
                          \mu_{1}\dots \mu_{j} \mid ={\matheins}
                          \label{completeness}
\end{eqnarray}
where the Einstein summation convention is used in Eq.(\ref{completeness}).

One can then transform to a spherical basis $\mid \phi_{j},\lambda\rangle:=
\mid j, \lambda\rangle$
\beq\label{spherical basis}
\mid j,\lambda\rangle = \varepsilon^{\mu_{1} \dots \mu_{j}}(\lambda) \mid
\mu_{1} \dots \mu_{j} \rangle
\eeq
where the $\varepsilon_{\mu_{1} \dots \mu_{j}}$ are the usual spin--j
polarization tensors. The inverse of (\ref{spherical basis}) is given by
\beq
\mid \mu_{1} \dots \mu_{j}\rangle= \sum_{\lambda} \varepsilon^{*}_{\mu_{1}
\dots \mu_{j}} (\lambda) \mid j,\lambda\rangle
\eeq
The spherical basis vectors $\mid j,\lambda\rangle$ in turn satisfy
orthonormality and completeness relations. One has
\begin{eqnarray}
\mbox{orthonormality} &:& \langle j,\lambda \mid j, \lambda' \rangle =
                          \delta_{ \lambda\lambda'}\\
\mbox{completeness}   &:& \sum_{\lambda} \mid j,\lambda \rangle \langle j ,
                          \lambda\mid =\matheins
\end{eqnarray}
 From the above one can then derive orthonormality and completeness relation
for the polarization tensors which read
\beq\label{s-orthonormality}
\varepsilon^{*}_{\mu_{1} \dots \mu_{j}}(\lambda) \varepsilon^{\mu_{1}
\dots \mu_{j}}(\lambda')=(-)^{j} \delta_{\lambda \lambda'}
\eeq
\beq\label{s-completeness}
\sum_{\lambda} \varepsilon^{ \mu_{1} \dots \mu_{j}} (\lambda) \varepsilon^{
*}_{\nu_{1} \dots \nu{j}}(\lambda)= G^{\mu_{1} \dots \mu_{j}}_{\nu_{1}
\dots \nu_{j}}
\eeq

After this technical aside we return to the construction of the excited
heavy baryon states. Using the diquark states $\phi^{\mu_{1} \dots \mu_{j}}$
the heavy baryon spin wave functions can easily be obtained from the
contraction
\beq
\Psi_{\alpha \beta \gamma} = [ \phi ^{\mu_{1} \dots \mu_{j}} ]_{\alpha
\beta} \psi_{\mu_{1} \dots \mu_{j};\gamma}
\eeq
The heavy--side spin wave functions $\psi_{\mu_{1} \dots \mu_{j}}$ are then
uniquely determined in terms of spinor--tensor forms that involve the heavy
baryons Rarita--Schwinger spinor--tensors $u^{\mu_{1} \dots \mu_{j}}$ and
$u^{\mu_{1} \dots \mu_{j-1}}$  for the $j \pm 1/2$ high and low spin
partners, respectively,
in the degenerate Heavy Quark Symmetry baryon doublet. In the latter
case an additional $\gamma_{\perp}^{\mu_{j}}\gamma_{5}$ needs to be
introduced to complete the tensor structure ($v^{\mu_{j}}$ cannot be used
because it annihilates on the light side).
The $\gamma_{5}$ enters since the $\psi^{\mu_{1} \dots \mu_{j}}$ have to
satisfy the heavy quark mass--shell condition
\beq
\vslash \psi^{\mu_{1} \dots \mu_{j}}= \psi^{\mu_{1} \dots \mu_{j}}
\eeq
The normalization of the heavy--side spin wave functions is fixed through the
normalization condition
\beq
{\bar \psi}^{\nu_{1} \dots \nu_{j}} \psi^{\mu_{1} \dots \mu_{j}} G_{\mu_{1}
\dots \mu_{j} ; \nu_{1} \dots \nu_{j}}=2M
\eeq
by the same reasoning as in Eq.(\ref{lnorm1}) and (\ref{lnorm2}). The
generalized Rarita--Schwinger spinor--tensors are normalized according to
\begin{displaymath}
{\bar u}^{\mu_{1} \dots \mu_{j}}u_{\mu_{1} \dots \mu_{j}} = (-)^{j}2M .
\end{displaymath}

In this  way one can then write down all the spin wave functions of the
excited heavy baryons. In Table 4 we have listed the spin wave
functions for the excited $\Lambda$--type states together with the ground
state $\Lambda$ and in Table 5 we have done the same for the
$\Sigma$--type states. The p--wave states are labelled according to the
nature of their orbital state and their light--side diquark spin $j$. Of
interest is also whether the states are in a spin singlet $\chi^{0}$ or
triplet $\chi^{1,\mu}$ state. This can be determined by invoking the
generalized Pauli principle for the light diquark system. One thus finds that
 one has a spin singlet
configuration $\chi^{0}$ for $\{\Lambda_{QK1}\}$ and $\{\Sigma_{Qkj}
(j=0,1,2)\}$ and a spin triplet configuration $\chi^{1, \mu}$ for $\{\Lambda
_{Qkj}; (j=0,1,2)\}$ and $\{\Sigma_{QK1}\}$. In the remaining part of
Sec.4 we shall make repeated use of the covariant spin wave functions
written down in Sec.\ref{ground state wf} and \ref{excited wf}.

A last comment concerns flavour wave functions. The antisymmetric
(antitriplet) and symmetric (sextet) light diquark flavour wave functions
are given by $[q_{i} q_{j}]=\frac{1}{\sqrt{2}}(q_{i}q_{j}-q_{j}q_{i})$ and
$\{ q_{i} q_{j} \}=\frac{1}{\sqrt{2}}(q_{i}q_{j}+q_{j}q_{i})$ with $q_{i},
q_{j}=u,d,s.$ It is sometimes convenient to represent the antisymmetric
flavour wave function by an one--index object by raising indices with the
help of $\varepsilon^{ijk}$, i.e. $T^{k}=\varepsilon^{ijk} [q_{i} q_{j}]$,
as is appropriate for the antitriplet representation.

\subsection{\label{weak trans}Current--Induced Heavy Baryon Transitions}
The exclusive  semileptonic
decays $H_{b} \rightarrow H_{c} +l^{-} +\bar{\nu_{l}}$ have played a central
role in the development of the Heavy Quark Symmetry.
Originally the prime motivation for studying these decays was the
desire to get a handle on the value of the Kobayashi--Maskawa matrix element
$ V_{bc}$. Once HQET was formulated it was noticed that the structure of
these decays is sufficiently rich to put the predictions of the heavy quark
limit and the $1/m_{Q}$ corrections to this limit to a detailed test in these
decays. Quite naturally in the beginning the main emphasis was on the mesonic
$b\rightarrow c$ transitions. But as more and more data is being collected on
heavy baryon decays a new important field for the applications of
Heavy Quark Symmetry has been opening from the investigation of
current--induced heavy baryon transitions.

As explained in Sec.2 the following ground state to ground state weak
semileptonic transitions are expected to be observable
\begin{eqnarray}
\Lambda_{Q}-\mbox{type} \quad &:&\Lambda_{b} \rightarrow \Lambda_{c} +
                                 l^{-}+\bar \nu_{l} \nonumber \\
                               &&\Xi_{b} \rightarrow \Xi_{c} + l^{-} +
                                 \bar \nu_{l} \nonumber \\
\Sigma_{Q}-\mbox{type} \quad  &:&\Omega_{b} \rightarrow \Omega_{c} +
                                 l^{-} +\bar \nu_{l} \nonumber \\
                              &:&\Omega_{b} \rightarrow \Omega_{c}^{*} +
                                 l^{-} +\bar \nu_{l}
\end{eqnarray}

The other ground state bottom baryons also have semileptonic modes but their
semileptonic branching ratios are so small as to make their semileptonic
decays unobservable for all practical purposes.

Further there are transitions to excited charm baryon states as in
$\Lambda_{b} \rightarrow \Lambda_{c}^{**}$. One would then want to know
how big the rate into the inelastic channels $\Lambda_{c}^{**}$ is, in
particular as the elastic rate and the inelastic rates are intimately
linked together by the sum rule of Bjorken. Again one would like to test
the predictions of HQET also for the inelastic contributions. For example,
the excited $\Lambda_{c}^{**}$--states produced in these decays (and for
that matter the $\Lambda_{c}$) will be polarized with definite predictions
for the polarization density matrices from HQET. The polarization of the
$\Lambda_{c}^{**}$'s would reveal itself by the angular decay distribution
of its subsequent decay products. Such considerations can e.g. be used
to pin down the $J^{P}$ quantum numbers of the excited $\Lambda_{Q}^{**}$
states.
\begin{figure}
\vspace{7.5cm}
\caption[dummy10]{ Current--induced transition between heavy
baryons. Heavy--side transition $Q_{1}(v_{1}) \rightarrow Q_{2}(v_{2})$
mediated by (V-A) heavy quark current. Light--side transition $j_{1}^{P_{1}}
(v_{1}) \rightarrow j_{2}^{P_{2}}(v_{2})$ depends only on the invariant
velocity transfer variable $\omega =v_{1} \! \cdot \! v_{2}$.}
\end{figure}

There are two ingredients that go into the Heavy Quark Symmetry description
of the semileptonic transitions as shown in Fig.5. First
there is the $b \rightarrow c$ transition which is mediated through the known
(V-A) structure. Second there is the transition from the initial diquark
system to the final diquark system whose strength and structure is not known.
The lack of knowledge concerning the light--side diquark transition can be
parameterized in terms of independent transition amplitudes which are called
reduced form factors. These can only depend on the one kinematical Lorentz
invariant $\omega=v_{1} \hspace{-1mm} \cdot \hspace{-1mm} v_{2}$ that arises
in the transition. Finally the heavy quarks and the light diquark system in
the initial and final state have to combine to form heavy baryons with the
correct spin--parity quantum numbers. The correct spin coupling factors that
achieve this can be determined from products of C.G. coefficients \cite{K,Z}
(or alternatively from 6--j symbols). Here we use the covariant approach to
determine the correct spin coupling factors as in \cite{HKT}.

For the current--induced transitions we then obtain
\beq\label{v-a matrix}
M^{\lambda}=\langle B_{Q_{2}}(v_{2})\mid J^{\lambda}\mid B_{Q_{1}}(v_{1})
\rangle =
       {\bar\psi}_{2,\alpha}^{\mu_{1} \dots \mu_{j_{1}}}\Gamma_{\alpha\beta}
       ^{\lambda} \psi_{1,\beta}^{\nu_{1} \dots \nu_{j_{2}}} \nonumber\\
        (\sum_{i} f_i(\omega) t^{i}_{\mu_{1} \dots \mu_{j_{1}};\nu_{1}
        \dots \nu_{j_{2}}})
\eeq
where the $\psi^{\mu_{1}\dots\mu_{j}}$ are the heavy--side spin wave
functions. $\Gamma^{\lambda}$ determines the structure of the $Q_{1}
\rightarrow Q_{2}$ current transition (e.g. $\gamma^{\lambda}(1-\gamma^{5})$
for a (V-A) interaction). The tensors $t^{i}_{\mu_{1} \dots \mu_{j_{1}};
\nu_{1} \dots \nu_{j_{2}}}$ describe the diquark transition. They have to be
build from the vectors $v_{1}^{\mu_{i}}$ and $v_{2}^{\nu_{i}}$, the metric
tensors $g_{\mu_{i} \nu_{k}}$ and, depending on parity, from the Levi--Civita
object $\varepsilon (\mu_{i}\nu_{k}v_{1} v_{2})$. The $f_{i}(\omega)$ are
reduced form factors that depend only on the invariant velocity transfer
variable $\omega=v_{1} \hspace{-1mm} \cdot \hspace{-1mm} v_{2}$.

Note that the Heavy Quark Symmetry prediction for the current matrix element
(\ref{v-a matrix}) has some structural similarity to the Wigner--Eckart
theorem. The $f_{i}(\omega)$ can be regarded as "reduced matrix elements" and
the ${\bar\psi}_{2}^{\mu_{1}\dots\mu_{j_{1}}}\Gamma^{\lambda}
\psi_{1}^{\nu_{1}\dots\nu_{j_{2}}}t^{i}_{ \mu_{1} \dots \mu_{j_{1}};
\nu_{1_{2}} \dots \nu_{j_{2}}}$ are the Clebsch--Gordan coefficients that
project onto the reduced matrix elements $f_{i}(\omega)$. The reduced form
factors serve to parameterize our ignorance about the dynamics of the
light--side transitions. Heavy Quark Symmetry can tell us nothing about the
reduced form factors $f_{i}(\omega)$ except for the existence of a
normalization condition at zero recoil $\omega=1$ for the
elastic transitions, as mentioned in Sec.3. The magnitude and the
$\omega$--dependence of the reduced form factors would have to be calculated
using nonperturbative methods such as QCD sum rules, lattice gauge theory or,
more conventionally, explicit quark models.

The number of the independent tensors $t^{i}_{\mu_{1} \dots \mu_{j_{1}},\nu
_{1}\dots \nu_{j_{2}}}$ and their parity depend
of course on the particular transition that is being considered. For the
simple cases considered here they can easily be written down using the
building blocks $ v_{1}^{\mu_{i}}$, $v_{2}^{\nu_{i}},g_{\mu_{i}\nu
_{k}}$ and $\varepsilon(\mu_{i}\nu_{k}v_{1}v_{2})$, as mentioned
before.

For the $\Lambda$--type transitions the tensor structure is particularly
simple since the diquark in the initial state has $j^{P}=0^{+}$. This implies
that there is at most one reduced form factor for the $\Lambda$--type
transitions. One has \cite{HKT,IWY91}
\begin{eqnarray}
i)\quad \Lambda_{Q_{1}} \rightarrow \Lambda_{Q_{2}}\hspace{12.5mm}&:&
           {\textstyle \frac{1}{2}^{+}}\rightarrow
           {\textstyle \frac{1}{2}^{+}} \nonumber\\
       && M^{\lambda}={\bar u}_{2} \Gamma^{\lambda} u_{1}
          f^{(0)}(\omega)\label{cur1}\\
       && \mbox{form factor normalization:}\quad f^{(0)}(\omega =1)=1
          \nonumber\\
       && \nonumber\\
ii)\quad\Lambda_{Q_{1}} \rightarrow \{ \Lambda_{Q_{2}K1} \}\hspace{3mm}
            &:&
{\textstyle \frac{1}{2}^{+}} \rightarrow \left \{ \begin{array}{c}
                                                    \frac{1}{2}^{-}\\
                                                    \frac{3}{2}^{-}
                                                \end{array}\right \}
                          \nonumber\\[3mm]
       &&M^{\lambda}= \left \{
                  \begin{array}{c}
                     -\frac{1}{\sqrt{3}} {\bar u}_{2} \gamma_{5}
                     \gamma^{\mu}_{\perp_{2}} \\
                     \bar u^{\mu}_2
                  \end{array} \right \}
          \Gamma^{\lambda}u_{1} f^{(1)}_{1} (\omega)v_{1\mu}\label{cur2}\\
       && \nonumber\\
iii)\quad\Lambda_{Q_{1}}\rightarrow \Lambda_{Q_{2}k0}\hspace{12mm}
          &:&
{\textstyle \frac{1}{2}^{+}}\rightarrow
{\textstyle \frac{1}{2}^{-}}\nonumber\\
       &&M^{\lambda}=0 \qquad \mbox{(forbidden)}\label{cur3}\\
       && \nonumber\\
iv)\quad \Lambda_{Q_{1}}\rightarrow \{\Lambda_{Q_{2}k1} \}\hspace{8mm}
          &:&{\textstyle \frac{1}{2}^{+}}
                          \rightarrow \left \{
                            \begin{array}{l}
                               \frac{1}{2}^{-}\\
                               \frac{3}{2}^{-}
                            \end{array}
                          \right \} \nonumber\\[3mm]
       &&M^{\lambda}= \left \{  \begin{array}{c}
                                  -\frac{1}{\sqrt{3}}{\bar u}_{2}
                                  \gamma_{5} \gamma^{\mu}_{\perp_{2}}\\
                                  \bar u^{\mu}_2
                                \end{array}\right \}
\Gamma^{\lambda} u_{1} f^{(1)}_{2}(\omega)v_{1\mu}\label{cur4}\\
     && \nonumber\\
v)\quad \Lambda_{Q_{1}}\rightarrow \{ \Lambda_{Q_{2}k2}\} \hspace{9mm}
                    &:&
{\textstyle\frac{1}{2}^{+}}\rightarrow\left \{
                            \begin{array}{c}
                              \frac{3}{2}^{-}\\
                              \frac{5}{2}^{-}
                            \end{array}
                         \right \} \nonumber\\
       &&M^{\lambda}=0\qquad \mbox{(forbidden)}\label{cur5}
\end{eqnarray}

   For the elastic transition there is a change of notation from the one
used in Sec.3 ($f^{(0)}(\omega)=\xi(\omega)$) to allow for the inclusion
of the $p$--wave contributions.

Let us make a few comments about the structure of the current--induced
matrix elements (\ref{cur1})--(\ref{cur5}). An alternative way of
determining the tensors $t^{i}_{\mu_{1} \dots \mu_{j_{1}};\nu_{1} \dots
\nu_{j_{2}}}$ and their associated reduced form factors $f_{i}(\omega)$
consists in considering the structure of the light--side transitions. For
this purpose it is convenient to consider the light--side transition in the
spherical basis (see Eq.(\ref{orthonormality})--(\ref{s-completeness})). One
has
\beq\label{ditrans}
(\phi_{2}(j_{2}^{P_{2}}, \lambda, v_{2}), \phi_{1}(j_1^{P_1},\lambda, v_1))
 = \varepsilon^{*\nu_{1} \dots \nu_{j_{2}}}(\lambda)(\sum_{i} f_{i}
(\omega)t^{i}_{\nu_{1} \dots \nu_{j_{2}};\mu_{1} \dots \mu_{j_{1}}})
\varepsilon^{\mu_{1} \dots \mu_{j_{1}}}(\lambda)
\eeq
where the tensors $t^{i}_{\nu_{1} \dots \nu_{j_{2}};\mu_{1} \dots
\mu_{j_{1}}}$ describe the light side transition. Explicitly one has
\beq
(\phi_{2 \nu_{1} \dots \nu_{j_{2}}}(v_{2}), \phi_{1 \mu_{1} \dots
\mu_{j_{1}}}(v_{1})) =\sum_{i} f_{i}(\omega)t^{i}_{\nu_{1} \dots \nu_
{j_{2}};\mu_{1} \dots \mu_{j_{1}}}
\eeq
In Eq.(\ref{ditrans}) we have made use of the fact that the helicity (or
$j_{z}$) is conserved in the light side diquark transition, i.e. $\lambda_{1}
=\lambda_{2}:=\lambda$.
The inner product is defined as in Eq.(\ref{diquark norm}), but now for the
parity conserving inelastic transitions $j_{1}^{P_{1}}(v_{1}) \rightarrow
j_{2}^{P_{2}}(v_{2}$). It is evident that Eq.(\ref{ditrans}) possesses the
same
tensor structure as Eq.(\ref{v-a matrix}). Using the alternative form
(\ref{ditrans}) it is then easy to understand the absence of $\Lambda_{Q_{1}}
\rightarrow \Lambda_{Q_{2}k0},\{ \Lambda_{Q_{2}k2} \}$ transitions since
there
can be no parity conserving transitions $0^{+} \rightarrow 0^{-}$ and
$0^{+} \rightarrow 2^{-}$.

The form (\ref{ditrans}) also provides for the zero recoil normalization
condition $ \mbox{$f(\omega=1)$}=1$ for the $\Lambda_{Q_{1}} \rightarrow
\Lambda_{Q_{2}}$ transition mentioned in Sec.3 and written down in
(\ref{cur1}). In this case one has an elastic $0^{+} \rightarrow 0^{+}$
transition which is evidently normalized to 1 at $v_{1}=v_{2}$ according to
Eq.(\ref{diquark norm}). Physically speaking, the normalization condition
arises because there is a complete overlap of the wave function of the
diquark system before and after the $Q_{1} \rightarrow Q_{2}$ transition
at zero recoil.

The counting of the number of reduced form factors that describe the
heavy baryon transitions can readily be done by referring to the number of
independent diquark transition amplitudes N in Eq.(\ref{ditrans}).
Defining the normality n of a diquark state with  quantum numbers $j^{P}$
by $n=P(-)^{j}$ one has to differentiate between the two cases
where the product of normalities of the two diquark states is even or odd.
One finds
\beq\label{normality}
\begin{array}{llll}
   i)  &n_{1} \! \cdot \! n_{2} =1   &:&  N=j_{min} +1\\
   ii) &n_{1} \! \cdot \! n_{2} =-1  &:&  N=j_{min}
\end{array}
\eeq
where $j_{min}=Min\{j_{1},j_{2}\}$. In closed form one has $N=j_{min}+
\frac{1}{2}(1+n_{1}\cdot n_{2})$. The above analysis agrees with
Eqs.(\ref{cur1})--(\ref{cur5}) as it must. Eq.(\ref{normality}) can be
derived by counting the number of independent helicity amplitudes in the
transition Eq.(\ref{ditrans}) \cite{POL90}. The difference between the
$n_{1}\! \cdot\! n_{2}$--even and --odd
case comes about because parity invariance forbids helicity zero
transitions when $n_{1}\! \cdot \! n_{2}=-1$. An even more elementary way of
deriving Eq.(\ref{normality}) is by performing a simple LS analysis in which
the transition operator is treated as a $0^{+}$ "spurion" state either in the
initial or final state \cite{K}.

For the $\Sigma$--type transitions the form factor structure predicted by
Heavy Quark Symmetry is a trifle more complex since now the initial diquark
is a $j^{P}=1^{+}$ diquark state. For each of the $\Sigma$--type transitions
there are now at most two reduced form factors.\footnote{As a curious
            byline we would like to remind the
reader that the full Heavy Quark Symmetry spin structure of the
$\Lambda_Q$-- and $\Sigma_Q$--type ground state transitions had already been
written down some 17 years ago in the crossed $e^+e^-$
channel \cite{kkuro77}.} One has
\begin{eqnarray}
i) \{ \Sigma_{Q_{1}}\} \rightarrow \{ \Sigma_{Q_{2}} \}\hspace{6mm}&:&
                 \left \{ \begin{array}{c}
                            \frac{1}{2}^{+} \\
                            \frac{3}{2}^{+}
                          \end{array}  \right \}
                 \rightarrow
                 \left \{ \begin{array}{c}
                            \frac{1}{2}^{+} \\
                            \frac{3}{2}^{+}
                          \end{array}  \right \}
                 \nonumber \\[3mm]
    &&\hspace{-4cm}M^{\lambda}= \left \{
             \begin{array}{c}
               -\frac{1}{\sqrt{3}}{\bar u}_{2} \gamma_{5} \gamma^{
                                       \nu_{1}}_{\perp_{2}} \\
               {\bar u}_{2}^{\nu_{1}}
             \end{array}
        \right \} \Gamma^{\lambda} \left \{
             \begin{array}{c}
              \frac{1}{\sqrt{3}}\gamma^{\mu_{1}}_{\perp_{1}}\gamma_{5}u_{1}\\
               u_{1}^{\mu_{1}}
             \end{array}
        \right \}(-g_{1}^{(0)}(\omega)g_{\mu_{1}\nu_{1}}+g_{2}^{(0)}
             (\omega)v_{1\nu_{1}}v_{2\mu_{2}})\label{curs1} \nonumber\\
 && \mbox{form factor normalization:}\quad g_{1}^{(0)}(\omega=1)=1\nonumber\\
    && \nonumber\\[4mm]
ii)\{ \Sigma_{Q_{1}} \} \rightarrow \{ \Sigma_{Q_{2}k1} \}\hspace{1.5mm}&:&
            \left \{ \begin{array}{c}
                       \frac{1}{2}^{+}\\
                       \frac{3}{2}^{+}
                     \end{array}
            \right \} \rightarrow \left \{
                     \begin{array}{c}
                       \frac{1}{2}^{-}\\
                       \frac{3}{2}^{-}
                     \end{array}
            \right \} \nonumber\\[3mm]
    &&\hspace{-4cm}M^{\lambda}= \left \{
            \begin{array}{c}
               -\frac{1}{\sqrt{3}}{\bar u}_{2} \gamma_{5} \gamma^{\nu_{1}}_
                      {\perp_{2}}\\
               {\bar u}_{2}^{\nu_{1}}
            \end{array}
        \right \} \Gamma^{\lambda} \left \{
            \begin{array}{c}
               \frac{1}{\sqrt{3}} \gamma^{\mu_{1}}_{\perp_{1}}
               \gamma_{5}u_{1}\\
               u_{1}^{\mu_{1}}
            \end{array}
        \right \} ig_{1}^{(1)}(\omega)\varepsilon(\mu_{1}\nu_{1}v_{1}v_{2})
        \nonumber \label{curs2}\\
     && \nonumber\\[4mm]
iii)\{ \Sigma_{Q_{1}} \} \rightarrow \Sigma_{Q_{2}K0} \hspace{3mm}&:&
           \left \{ \begin{array}{c}
                      \frac{1}{2}^{+}\\
                      \frac{3}{2}^{+}
                    \end{array}  \right \}
           \rightarrow  \frac{1}{2}^{-} \nonumber \\[3mm]
   && \hspace{-4cm}M^{\lambda} = {\bar u}_{2} \Gamma^{\lambda} \left \{
           \begin{array}{c}
              \frac{1}{\sqrt{3}}\gamma^{\mu_{1}}_{\perp_{1}}\gamma_{5}u_{1}\\
              u_{1}^{\mu_{1}}
           \end{array}
      \right \} g_{2}^{(1)}(\omega)v_{2\mu_{1}}\label{curs3}\\
   && \nonumber\\[4mm]
iv) \{ \Sigma_{Q_{1}}\} \rightarrow \{ \Sigma_{Q_{2}K1}\} &:&
          \left \{ \begin{array}{c}
                    \frac{1}{2}^{+}\\
                    \frac{3}{2}^{+}
                   \end{array}
          \right \} \rightarrow \left \{
                   \begin{array}{c}
                     \frac{1}{2}^{-}\\
                     \frac{3}{2}^{-}
                   \end{array}
          \right \} \nonumber \\[3mm]
  &&\hspace{-4cm}M^{\lambda}= \left \{
       \begin{array}{c}
         -\frac{1}{\sqrt{3}}{\bar u}_{2}\gamma_{5}\gamma_{\perp_{2}}^
            {\nu_{1}}\\
         {\bar u}^{\nu_{1}}_2
       \end{array}
    \right \} \Gamma^{\lambda} \left \{
       \begin{array}{c}
         \frac{1}{\sqrt{3}} \gamma_{\perp_{1}}^{\mu_{1}} \gamma_{5}u_{1}\\
         u_{1}^{\mu_{1}}
       \end{array}
    \right \}ig_{3}^{(1)}(\omega)\varepsilon(\mu_{1}\nu_{1}v_{1}
     v_{2}) \nonumber \label{curs4}\\
   && \nonumber\\[4mm]
v)\{ \Sigma_{Q_{1}} \} \rightarrow \{\Sigma_{Q_{2}K2}\}&:&
          \left \{ \begin{array}{c}
                     \frac{1}{2}^{+}\\
                     \frac{3}{2}^{+}
                   \end{array}
          \right \} \rightarrow \left \{
                   \begin{array}{c}
                     \frac{3}{2}^{-}\\
                     \frac{5}{2}^{-}
                   \end{array}
          \right \}  \nonumber \\[3mm]
  &&\hspace{-4cm}M^{\lambda}= \left \{
            \begin{array}{c}
               -\frac{1}{\sqrt{10}}{\bar u}_{2}^{\{\nu_{1}}\gamma_{5}
                  \gamma_{\perp_{2}}^{\nu_{2}\}_{0}}\\
               {\bar u}_{2}^{\nu_{1}\nu_{2}}
            \end{array}
     \right \} \Gamma^{\lambda} \left \{
            \begin{array}{c}
               \frac{1}{\sqrt{3}} \gamma_{\perp_{1}}^{\mu_{1}} \gamma_{5}
                  u_{1}\\
               u_{1}^{\mu_{1}}
           \end{array}
     \right \}(-g_{4}^{(1)}(\omega)v_{1\nu_{1}} g_{\nu_{2} \mu_{1}}
      +g_{5}^{(1)}
     (\omega)v_{1\nu_{1}} v_{1\nu_{2}}v_{2\mu_{1}} )\nonumber\\
\label{curs5}
\end{eqnarray}
As discussed before the form factor counting can of course be done equally
well by counting the number of form factors in the diquark transitions $1^{+}
\rightarrow j_{2}^{P_{2}}$ using the general formula Eq.(\ref{normality}).
The normalization condition for the "elastic" transition $\{ \Sigma_{Q_{1}}\}
\rightarrow \{ \Sigma_{Q_{2}} \}$ at zero recoil applies only to the metric
form factor $g_{1}^{(0)}(\omega)$ since the second form factor $g_{2}^{(0)}
(\omega)$ does not contribute when $v_{1} = v_{2}$. The zero recoil
normalization $g_{1}^{(0)} (1) =1$ follows directly from the normalization
of the diquark state, cf. Eq.(\ref{diquark norm}).

Equations (\ref{cur1})--(\ref{cur5}) and (\ref{curs1})--(\ref{curs5})
represent the most general transition form factor structure in the Heavy
Quark Symmetry limit. We shall discuss some possible simplifications at the
end of Sec.\ref{bjorken sum rule} by resolving the diquark transitions into
constituent quark transitions. It is important to keep in mind, though, that
any simplification of the form factor structure predicted by Heavy
Quark Symmetry necessarily involves further model dependent assumptions.

Before closing this subsection we briefly want to discuss the Heavy Quark
Symmetry structure of current--induced transitions from a heavy baryon to a
light baryon. This would be of relevance for e.g. $c\rightarrow s$ or $b
\rightarrow u$ transitions.
Here we limit ourselves to ground state transitions. For heavy to light
transitions one must now allow for spin interactions of the light "active"
quark coming from
the weak interaction vertex with the light diquark system, i.e. now one has
a factorization only in the initial state. The spin interaction can be
introduced by adding $\gamma$--structure to the light "active" fermion in
the final
state. This amounts to the replacements $ f^{(0)} \rightarrow
f' + \vslash_{1} f'', \; g_{1}^{(0)} \rightarrow g'_{1} + \vslash_{1}
g''_{1}$ and $g_{2}^{(0)} \rightarrow g'_{2} + \vslash_{} g''_{2}$ in
Eq.(\ref{cur1}) and (\ref{curs1}). Explicitly one has \cite{HKKT91,HLKKT92}
\begin{eqnarray}\label{4-122}
\Lambda_{Q} \rightarrow \Lambda_{q} &:& \; M^{\lambda}=
      {\bar u }_{2}(p_{2})(f'(p_{2} \! \cdot \! v_{1})+\vslash_{1} f''(p_{2}
      \! \cdot \! v_{1}))\Gamma^{\lambda} u_{1}(v_{1})\\
\{ \Sigma_{Q} \} \rightarrow \Sigma_{q} &:& \; M^{\lambda}=
      {\textstyle \frac{1}{\sqrt{3}}}
      {\bar u}_{2} (p_{2}) \gamma_{5} \gamma^{\nu_{1}}
      _{\perp_{2}} [-g_{\mu_{1} \nu_{1}} (g'_{1}(p_{2}  \! \cdot \! v_{1})+
      \vslash_{1}g''_{1}(p_{2}  \! \cdot \! v_{1}))\nonumber\\
    &&\hspace{1.5cm} +v_{1 \nu_{1}} p_{2\mu_{1}} (g'_{2}(p_{2}  \! \cdot \!
    v_{1}) + \vslash_{1} g''_{2}(p_{2}  \! \cdot \! v_{1})] \nonumber\\[2mm]
    &&\hspace{1.5cm} \Gamma^{\lambda}
      \left \{ \begin{array}{c}
         \frac{1}{\sqrt{3}} \gamma_{\perp_{1}}^{\mu_{1}} \gamma_{5} u_{1}
         (v_{1})\\
         u_{1}^{\mu_{1}}(v_{1})
      \end{array} \right \}
\end{eqnarray}
and similarly for $\{ \Sigma_{Q} \} \rightarrow  \Sigma_{q}^{*}$, albeit with
a new set of form factors. Now  there is no normalization condition for any
of the form factors, and the transitions  $\{ \Sigma_{Q} \} \rightarrow
\Sigma_{q}$ and $\{ \Sigma_{Q} \} \rightarrow  \Sigma_{q}^{*}$ are not
related.

\subsection{\label{bjorken sum rule}Contribution of Transition Form Factors
to the Bjorken Sum Rule}
As mentioned before, Heavy Quark Symmetry says nothing about the $\omega$--
dependence of the reduced form factors except for the normalization
condition at zero recoil in the elastic case. However, Bjorken has
pointed out \cite{BJORKEN} that one may extract useful information on the
reduced form
factors by considering the contribution of the heavy decay baryons
$B_{Q_{2}}, B_{Q_{2}}^{**} \dots $ to the structure functions $H^{i}$
occurring in the semileptonic decay $B_{Q_{1}} \rightarrow (B_{Q_{2}}+
B_{Q_{2}}^{**} \dots )+\nu_{l} + l$ (for a definition of the structure
functions $H_{i}$ see Sec.5.2). Then, by invoking duality, one
equates the sum of particle structure functions to the corresponding
inclusive structure function calculated from the free quark decay $Q_{1}
\rightarrow Q_{2} + \nu_{l} + l$ for any value of $\omega$.

Technically, it is simplest to consider the contribution of the decay baryons
to the longitudinal structure function $H_{L}$. One first calculates
longitudinal helicity transition amplitudes and then squares them in order to
obtain the contribution of a given final baryon to the Bjorken sum rule. This
is an elegant method that avoids the tedium of having to do lengthy spin sums
in squared covariant matrix elements.

In this way it is not difficult to obtain the contribution of the s-- and
p--wave
baryons to the Bjorken sum rule. For the $\Lambda$--type transitions one has
\begin{eqnarray}\label{lamb-sum}
1 &=& \mid f^{(0)}(\omega)\mid^{2} + (\omega^{2}-1)(\mid f_{1}^{(1)}(\omega)
      \mid^{2} + \mid f_{2}^{(1)}(\omega) \mid^{2})\\
  && +\dots \nonumber
\end{eqnarray}
where the ellipsis stand for the contributions of higher radial and orbital
excitations not considered here, and for continuum contributions.

The higher orbital and radial excitations, and the continuum will
contribute to the sum rule with threshold powers $(\omega^{2}-1)^{n}$ at
least as high as the p--wave contributions, i.e. $n\geq1$. For the zero
recoil point $\omega=1$ all but the elastic contribution vanish and one
recovers the normalization condition $f^{(0)}(1)=1$ for the elastic form
factor. As one is moving away from the zero recoil point $\omega=1$ rate is
disappearing from the elastic channel while it appears in the inelastic
channels. Using positivity one obtains bounds for the elastic form factor and
for its derivative at the zero recoil point $\omega=1$. One has
\begin{eqnarray}
&&f^{(0)}(\omega) \leq 1 \label{labounds}\\
&&\frac{df^{(0)}}{d\omega}\ersetze_{\,\omega=1}
 \leq -(\mid f^{(1)}(1) \mid^{2} + \mid f^{(1)}(1)
\mid^{2}) \leq 0 \label{laderbounds}
\end{eqnarray}
For the $\Sigma$--type transitions one finds \cite{HKT,XU}
\begin{eqnarray}
1&=&{\textstyle \frac{2}{3}} \mid g_{1}^{(0)} (\omega) \mid^{2}
+ {\textstyle \frac{1}{3}} \mid \omega
  g_{1}^{(0)}(\omega) -(\omega^{2}-1) g_{2}^{(0)}(\omega) \mid^{2}\nonumber\\
  &&+(\omega^{2}-1) \{{\textstyle \frac{2}{3}}
  \mid g_{1}^{(1)}(\omega) \mid^{2} +
    {\textstyle \frac{1}{3}} \mid g_{2}^{(1)}(\omega)\mid^{2}
+ {\textstyle \frac{4}{3}} \mid g_{3}^{(1)} (\omega) \mid^{2}
  \label{sig-sum}\\
  &&+{\textstyle \frac{2}{3}}
\mid g_{4}^{(1)}(\omega) \mid^{2} + {\textstyle \frac{4}{9}} \mid
    \omega g_{4}^{(1)} (\omega) - (\omega^{2} -1)g_{5}^{(1)} \mid^{2} \}
    + \dots \nonumber
\end{eqnarray}
We have diagonalized the form factor contributions of the
ground state and the $\{ \Sigma_{Q_{2}K2}\}$ multiplet in terms of the
longitudinal and the transverse diquark transitions, $F_{L}=g_{1}^{(0)}$
and $F_{T}=\omega g_{1}^{(0)}- (\omega^{2}-1) g_{2}^{(0)}$ and similarly
for the $1^{+} \rightarrow 2^{-}$ transition.

The bounds on the elastic form factors and their derivatives now read
\beq \label{bounds}
{\textstyle \frac{2}{3}}\mid g_{1}^{(0)}(\omega)\mid^{2}+
{\textstyle \frac{1}{3}} \mid \omega
  g_{1}^{(0)} (\omega) -(\omega^{2}-1) g_{2}^{(0)}(\omega)\mid^{2} \leq 1
\eeq
and
\beq\label{derbounds}
 \frac{dg_{1}^{(0)}(\omega)}{d\omega}\ersetze_{\,\omega=1} \leq -
 {\textstyle \frac{1}{3}}+ {\textstyle \frac{2}{3}}g_{2}^{(0)}(1)
\eeq

The bounds on the form factors (\ref{bounds}) and (\ref{derbounds}) are not
very strong. Still one of the bounds suffices to e.g. rule out the "quark
confinement model (QCM)" of \cite{EIKL92}. In the QCM model one finds
$f^{(0)}
(\omega)=\Phi(\omega)$, $g_{1}^{(0)}=\omega\Phi(\omega)$ and $g_{2}^{(0)}
(\omega)=\Phi(\omega)$ where the form factor function
\beq
\Phi =\frac{\ln (1+\sqrt{\omega^{2}-1})}{\sqrt{\omega^{2}-1}}
\eeq
has a similar origin as the one--loop correction to the
current--quark--quark
vertex discussed in Sec.3. Substituting the QCM results in
the bound (\ref{bounds}) the bound translates into
\beq
\Phi^{2}(\omega) \leq \frac{3}{1+2\omega^{2}}
\eeq
which can be checked to be wrong. One can check that the sum rule bounds
(\ref{labounds}), (\ref{laderbounds}) and (\ref{transoperator}) are indeed
satisfied by the QCM form factors ($\Phi'(1)=-1$). However, in view of the
violation of the bound (\ref{bounds}) one concludes that the QCM model has
to be ruled out as it predicts form factors which are too hard.

Let us consider the simplification that occur when one adopts a
constituent quark model description for the ground state to ground state
transitions $\Lambda_{Q_{1}} \rightarrow \Lambda_{Q_{2}}$ and
$\{\Sigma_{Q_{1}}\} \rightarrow \{ \Sigma_{Q_{2}} \}$. As already mentioned
in Sec.\ref{ground state wf} the constituent approximation for the
spin--wave
functions consists in writing $A(k,K,v)^{\alpha \beta}_{\gamma \delta} =
A(k,K,v)\delta_{\gamma}^{\alpha} \delta_{\delta}^{\beta}$ such that the
$\Lambda$--type and $\Sigma$--type states become related. In the language of
collinear $SU(6)_{W}$ the diquark is the 21--dimensional \young
--representation
of the $SU(6)_{W}$ symmetry, where $21=1_{a} \otimes 3_{a}^{*}+3_{s}\otimes
6_{s}$ is the $SU(2)_{spin} \otimes SU(3)_{flavour}$ decomposition of the
21--dimensional representation. The spin 0 antisymmetric $3^{*}$
representation is made up by the three antisymmetric combinations of u, d
and s while the spin 1 symmetric $6$ representation is made up by the six
symmetric combinations of u, d and s.

The diquark transition will now be resolved into a pair of quark--quark
transitions. A first simplification occurs when the quark--quark transitions
are taken to be superpositions of a scalar and a vector interaction which are
parameterized by the form factors $f(\omega)$ and $g(\omega)$,
respectively, as drawn
in Fig.6. A spin--spin interaction term could be due to
an effective one--gluon exchange force as described in \cite{rugega} or,
alternatively, would show up as a remnant of the fermionic propagator effect
in the Bethe--Salpeter approach of \cite{HKKT91}.
\begin{figure}
\vspace{5.5cm}
\caption[dummy11]{ Light--side diquark transition resolved
into constituent quark transitions. Reduced form factors $f(\omega)$ and
$g(\omega)$ multiply scalar--scalar and vector--vector constituent quark
transitions. Zero recoil normalization condition for elastic case is
$f(1)+g(1)=1$.}
\end{figure}
It is then a simple matter to calculate the resolved diquark transition
using the spin wave functions (\ref{SWF}) and the transition operator
\beq \label{transoperator}
I(\omega)=f(\omega)1\otimes 1+g(\omega)\gamma^{\mu}\otimes\gamma_{\mu}
\eeq
One finds
\begin{eqnarray}
{\bar {\hat \chi}}^{0}(v_{2})I(\omega){\hat \chi}^{0}(v_{1})&=&
    \frac{\omega+1}{2} f(\omega) +(2-\omega)g(\omega)
    \label{transoperator0}\\
{\bar{\hat \chi}}^{1}_{\nu}(v_{2})I(\omega){\hat \chi}^{1}_{\mu} (v_{1})&=&
    -\left(\frac{\omega+1}{2} f(\omega)+g(\omega)\right)g_{\mu \nu}
    +{\textstyle \frac{1}{2}}
    f(\omega)v_{2\mu}v_{1\nu}
    \label{transoperator1}
\end{eqnarray}
with the zero recoil normalization condition
\beq\label{f/g norm}
f(1)+g(1)=1.
\eeq
Eq.(\ref{transoperator1}) is understood to be taken between the spin
polarization vector $\varepsilon^{*\nu}_{2}$ and $\varepsilon^{\mu}_{1}$. In
terms of the elastic form factors defined in Eqs.(\ref{cur1}) and
(\ref{curs1}) one finds
\begin{eqnarray}
\Lambda_{Q}:\quad&&f^{(0)}(\omega)=\frac{\omega+1}{2}f(\omega)+(2-\omega)
                   g(\omega)\label{f-form}\\
\{ \Sigma_{Q} \} :\quad&&g_{1}^{(0)}(\omega)=\frac{\omega+1}{2}f(\omega)
       +g(\omega)
                  \label{g1-form}\\
                &&g_{2}^{(0)}(\omega)= {\textstyle \frac{1}{2}} f(\omega)
                  \label{g2-form}
\end{eqnarray}

It will not be easy to experimentally test the above relation between
the $\Lambda$--type and the $\Sigma$--type form factors in semileptonic
decays. In the test one would have to compare $\Lambda_{b} \rightarrow
\Lambda_{c}$ and $\Omega_{b} \rightarrow \Omega_{c}$ transitions where
there are additional SU(3) breaking effects. However going to the
$e^{+}e^{-}$--production channel, one can predict the relative rates of
heavy baryon pair production from (\ref{f-form}--\ref{g2-form}). For example,
close to the threshold the contribution of $f(\omega)$ is strongly
suppressed and one obtains \cite{HKKT91}
\footnote{The spin coupling factor $(\omega +1)/2$ multiplying the
scalar form factor $f(\omega)$ in \protect{(\ref{f-form})} and
\protect{(\ref{g1-form})}has a simple interpretation in the crossed
$e^{+}e^{-}$--channel where $(\omega+1) \rightarrow -(\omega-1)$ after
crossing. Each factor of $\sqrt{\omega-1}$ in $(\omega -1)=
(\sqrt{\omega -1})^{2}$ accounts for one p--wave suppression for each of the
light quark--antiquark pairs
that are independently produced \protect{\cite{HKKT91}}.}
\beq\label{rates}
\sigma_{\Lambda_{Q} \bar\Lambda_{Q}}:\sigma_{\Sigma_{Q} \bar \Sigma_{Q}}
:\sigma_{\Sigma_{Q} \bar \Sigma_{Q}^{*}}+\sigma_{\Sigma_{Q}^{*} \bar
\Sigma_{Q}}:\sigma_{\Sigma_{Q}^{*} \bar \Sigma_{Q}^{*}}=27:1:16:10
\eeq
In the so called spectator quark model one sets the spin--spin interaction
term to zero. In this case one finds
\beq\label{spec-form}
f^{(0)}(\omega)=g_{1}^{(0)}(\omega)=\frac{\omega+1}{2} f(\omega)
\eeq
\beq\label{spec-form141}
g_{2}^{(0)}(\omega)= {\textstyle \frac{1}{2}} f(\omega)
\eeq

All three reduced form factors are now related to one single form
factor $f(\omega)$. We shall refer to $f(\omega)$ as the residual quark
model form factor since the spin coupling factor $(\omega +1)/2$ has been
factored out. When Eq.(\ref{spec-form}) and (\ref{spec-form141}) is
substituted into either of
the Bjorken sum rules (\ref{lamb-sum}) or (\ref{sig-sum}) one now finds
\beq\label{spec-bounds}
f(\omega)\leq \frac{2}{\omega +1}
\eeq
for the residual form factor $f(\omega)$. Since $2/(\omega +1)$ is the
normalized monopole form factor in the heavy quark limit one finds that the
residual form factors $f(\omega)$ has to fall at least as fast as a
monopole form factor in the heavy quark limit. The form factor $f(\omega)$
used in the spectator quark model calculation of \cite{ct} and \cite{XU}
is consistent with the bound (\ref{spec-bounds}).

Carone, Georgi and Osofsky \cite{CGO94} have recently presented arguments
that the light constituent quarks have no spin interactions to leading
order in $1/N_{C}$ ($N_{C}$ is the number of colours). This would lend
support to the viability of the constituent spectator approach (light
constituent quarks plus absence of spin interactions of light quarks). This
would have dramatic implications for threshold production of heavy baryon
pairs in $e^{+} e^{-}$ --interactions because of the extra $|\vec p|^4$
threshold suppressions present in the spectator model as
argued before. In the absence of spin interactions of the light
constituent quarks there are no spin singlet -- spin triplet transitions
because of the orthogonality of the spin wave functions ${\hat \chi}^{0}$
and ${\hat \chi}^{1}_{\mu}$.
Consequently one would predict e.g. that the s--wave to p--wave transitions
$\Lambda_{Q_{1}} \rightarrow \{ \Lambda_{Qk1} \}$ and  $\Sigma_{Q_{1}}
\rightarrow \{ \Sigma_{Qk1} \}$ vanish (see also \cite{XU} and \cite{CW94}).
This would imply that only two of the seven $\Lambda_{c}$--type p--wave
states contribute to the Bjorken sum rule (\ref{lamb-sum}). If there is only
little rate
into $\Lambda_{c}^{**}$'s much of the inclusive rate must go into the elastic
$\Lambda_{b} \rightarrow \Lambda_{c} $ channel. One would then conclude that
the elastic channel has a large branching ratio, or, in the light of the sum
rule of Bjorken, that the elastic transition form factor $f^{(0)} (\omega)$
should be quite flat.

Finally, the simplest quark model configuration is given by the
``independent quark motion approximation'' where each light quark moves
around the heavy quark source independently with no interaction between the
light  quarks. The analogue of this configuration in atomic physics is the
''unperturbed'' helium atom configuration where the interaction between
the electrons has been switched off.
This approximation corresponds to a totally factorized
form of the diquark wave function \cite{HKKT91}
\beq
A_{\alpha \gamma}^{\beta \delta}(k,K,v)=A(p_{1},v)A(p_{2},v) \delta_{\alpha}
^{\beta} \delta_{\gamma}^{\delta}.
\eeq
In this approximation the baryonic form factor is nothing but the square of
the mesonic reduced form factor $\xi(\omega)$ with $\xi(1)=1$, i.e.
\beq\label{baryonic form}
f(\omega)=\xi^{2}(\omega).
\eeq
Substituting (\ref{baryonic form}) into (\ref{spec-bounds}) one obtains
\beq
\xi(\omega) \leq \sqrt{\frac{2}{\omega+1}}
\eeq
which, not surprisingly, is just the Bjorken bound for the heavy meson
reduced form factor \cite{BJORKEN}.

\subsection{\label{pi-trans}Pion Transitions Between Heavy Baryons}
A look at the spectrum of the s--wave charm baryon states shows that there
is enough phase space for the two members of the $\{ \Sigma_{c} \}$ doublet
to decay into $\Lambda_{c}$ via pion emission. In fact the ${1/2}^{+}
\; \Sigma_{c}$--states revealed themselves as peaks in the $(\Lambda_{c}
\pi)$ invariant mass spectrum. Last year the SCAT
collaboration reported evidence for the corresponding one--pion decay mode
of the $ {3/2}^{+} \; \Sigma_{c}^{*}$ state via the decay $\Sigma_{c}^{*}
\rightarrow \Lambda_{c} \pi$ \cite{SKAT}.

Further interest in pion transitions between heavy charm baryons
has been triggered by the recent observation of two excited
$\Lambda$--type charm baryon states at $\simeq 2593$~MeV and at $\simeq
2627$~MeV by the ARGUS \cite{ARGUSLAMBDA93}, CLEO \cite{CLEOLAMBDA93} and
E687 \cite{E687LAMBDA93}
collaborations. These states show up as peaks in the $(\Lambda\pi\pi)$
invariant mass distribution. The resonances are narrow, in fact too narrow
to be resolved by the experiments.

The lower resonance of the two appears to be dominantly decaying via the
decay chain $\Lambda_{c}(2593) \rightarrow\Sigma_{c}(\rightarrow
\Lambda_{c}\pi)+\pi$. For the higher lying resonance state ARGUS reports on
some evidence for the existence of the decay chain $\Lambda_{c}(2627)
\rightarrow \Sigma_{c}(
\rightarrow\Lambda_{c}\pi)+\pi$ \cite{ARGUSLAMBDA93} which, however, is not
corroborated by the other experiments.
\begin{figure}
\vspace{9.5cm}
\caption[dummy12]{ Allowed pion transitions between p--wave and
s--wave charm baryon states. For the one--pion transitions drawn in the
figure there are no orbital angular momentum selection rules from Heavy Quark
Symmetry. The transitions $\{ \Lambda_{cK1} \} \rightarrow \Lambda_{c} $ are
via two--pion transitions. One--pion transitions are forbidden for these
transitions by isospin and,
in the case $\frac{3}{2}^{-} \rightarrow \frac{1}{2}^{+} + \pi^{-}$,
also by parity in the heavy quark limit.}
\end{figure}

The two new states are very likely the two $J^{P}=1/2^{-}$ and
$3/2^{-}$ members of the p--wave multiplet $\{\Lambda_{cK1} \}$.
In fact, an early quark model calculation \cite{CIK79} predicted mass values
of 2.53 GeV and 2.61 GeV for the $1/2^{-}$ and $3/2^{-}$
members
\footnote{The numerical mass values were adjusted upward such that the
          input $\Lambda_{c}$ mass agrees with its measured value.},
respectively,
of the $\{ \Lambda_{cK1}\}$ multiplet which are quite close to the
experimental mass values. A direct experimental determination of the
$J^{P}$ quantum numbers of the two new states is still outstanding. However,
there is some indirect evidence for the validity of the $J^{P}=1/2^{-}$
and $3/2^{-}$ assignments of the two new states from the
existence or nonexistence of the intermediate state $(\Sigma_{c}
\pi)$ in the $\Lambda_{c}(2593)$ and $\Lambda_{c}(2627)$ decays. The
argument goes
as follows (see Fig.7). Both decays $\Lambda_{c}(2593) \rightarrow
\Sigma_{c}\pi$ and
$\Lambda_{c}(2627) \rightarrow \Sigma_{c} \pi$ are kinematically allowed.
Although the phase space for $\Lambda_{c}(2627) \rightarrow \Sigma_{c}\pi$
is larger than for $\Lambda_{c}(2593) \rightarrow \Sigma_{c}\pi$ the former
channel is not seen by two of the three experiments. This would find a
natural explanation if $\Lambda_{c}(2627) \rightarrow \Sigma_{c} \pi$ is a
very much suppressed $3/2^{-} \rightarrow 1/2^{+} +\pi $
d--wave decay and $\Lambda_{c}(2593) \rightarrow \Sigma_{c} \pi$ is an
unhindered $1/2^{-} \rightarrow 1/2^{+} + \pi\,$ s--wave decay.
\begin{figure}
\vspace{7cm}
\caption[dummy13]{ One--pion (left) and one--photon (right)
transitions between heavy baryons in the heavy quark limit. The pion and the
photon couple only to the light--side diquark which makes a transition from
spin--parity $j_{1}^{P_{1}}$ to $j_{2}^{P_{2}}$ at the same velocity v. The
heavy quark is not affected by the transition. Pion transitions are labelled
by pion's orbital momentum. Photon transitions are labelled in terms of
electric $(EJ_{\gamma})$ and magnetic $(MJ_{\gamma})$ multipoles.}
\end{figure}

The physics of the one--pion transitions between heavy baryons is depicted
in Fig.8. The pion is emitted from the light diquark while the heavy
quark propagates unaffected by the pion emission process. Since the heavy
baryon is infinitely heavy the heavy baryon will not recoil in the pion
emission process, i.e. the velocity of the heavy quark and thereby the
heavy baryon remains unchanged, as indicated in Fig.8.

The number of independent amplitudes describing the one--pion transitions
on the light side can be determined by the same reasoning as in
Sec.\ref{weak trans}. When counting the number N of independent helicity
amplitudes one
has to distinguish again between the two cases that the product of the
normalities of the diquark states is even or odd. One obtains
\beq
\begin{array}{lll}
  i) & n_{1} \! \cdot \! n_{2}=1  & N=j_{min}\\
  ii)& n_{1} \! \cdot \! n_{2}=-1 & N=j_{min} + 1
\end{array}
\eeq
or, in closed form, $N=j_{min}-\frac{1}{2}(n_{1} n_{2} -1)$. The counting
of amplitudes can of course equally well be done using the LS--coupling
scheme.

In fact, in Table 6 we list the relevant forms of the covariant
couplings of the pions in a definite orbital state $l_{\pi}$. In the
heavy quark limit the orbital momenta of the pion
relative to the diquark $l_{\pi}$ and relative to the baryon $L_{\pi}$ are
identical, i.e. $l_{\pi}=L_{\pi}$. As we shall see at the end of
Sec.4.5 the couplings listed in \mbox{Table 6} can easily be
transcribed into chiral invariant couplings.

\begin{table}
\caption[dummy15]{Tensor structure of pion couplings to diquark states.
The pion is in a definite orbital state $l_{\pi}$. Tensor structure of
transitions with $(j_{1}^{P_{1}},j_{2}^{P_{2}})\rightarrow(j_{1}^{-P_{1}}
,j_{2}^{-P_{2}})\rightarrow(j_{2}^{P_{2}},j_{1}^{P_{1}})\rightarrow
(j_{2}^{P_{-2}},j_{1}^{-P_{1}})$ are identical and are not always listed
here. \label{tab3}}
\vspace{5mm}
\renewcommand{\baselinestretch}{1.2}
\small \normalsize
\begin{center}
\begin{tabular}{rlcl}
\hline \hline
\multicolumn{2}{c}{diquark transition}& orbital wave & covariant coupling\\
\multicolumn{2}{c}{$j_{1}^{P_{1}} \rightarrow j_{2}^{P_{2}} + \pi$}&
    $l_{\pi}$& $t^{i}_{\mu_{1} \dots \mu_{j_{1}};\nu_{1} \dots \nu_{j_{2}}}$\\
\hline \hline
$0^{+}\rightarrow $&$ 0^{+}+\pi$&forbidden&-\\
\hline
$1^{+} \rightarrow$&$ 0^{+} +\pi$&1&$p_{\mu_{1}}^{\perp}$\\
 &$ 1^{+}+\pi$&1&$\frac{1}{\sqrt{2}}\varepsilon (\mu_{1} \nu_{1} p v )$\\
\hline
$0^{-} \rightarrow$&$0^{+}+\pi$&0&1 (scalar)\\
  &$ 1^{+}+\pi$&forbidden&-\\
  &$ 0^{-}+\pi$&forbidden&-\\
\hline
$1^{-} \rightarrow$&$ 0^{+}+\pi$&forbidden&-\\
  &$ 1^{+}+\pi$&$0$&$\frac{1}{\sqrt{3}} g_{\mu_{1} \nu_{1}}^{\perp}$\\
  &&$2$&$\sqrt{\frac{3}{2}}(p_{\mu_{1}}^{\perp} p_{\nu_{1}}^{\perp}-\frac{1}{3}
         p_{\perp}^{2}g_{\mu_{1} \nu_{1}}^{\perp})$\\
  &$ 0^{-}+\pi$&$1$&$ p_{\mu_{1}}^{\perp}$\\
  &$ 1^{-} +\pi$&$1$&$\frac{1}{\sqrt{2}} \varepsilon (\mu_{1} \nu_{1} pv)$\\
\hline
$2^{-} \rightarrow $&$ 0^{+}+\pi$&$2$&$\sqrt{\frac{3}{2}}p_{\mu_{1}}^{\perp}
                       p_{\mu_{2}}^{\perp}$\\
  &$ 1^{+}+\pi$&$2$&$p_{\mu_{2}}^{\perp} \varepsilon (\mu_{1} \nu_{1} p v )$\\
 &$ 0^{-}+\pi$&forbidden&-\\
 &$1^{-}+\pi$ &$1$&$\sqrt{\frac{3}{5}}g_{\mu_{1}\nu_{1}}^{\perp} p_{\mu_{2}}^
   {\perp}$\\
 &&$3$&$\sqrt{\frac{5}{2}} \{p_{\mu_{1}}^{\perp} p_{\mu_{2}}^{\perp} p_
      {\nu_{1}}^{\perp}-\frac{1}{5}(p_{\perp}^{2} g_{\mu_{1} \mu_{2}}^{\perp}
      p_{\nu_{3}}^{\perp}+$cycl.$(\mu_{1} \mu_{2} \nu_{1}))\}$\\
 &$ 2^{-}+\pi$ &$1$&$\sqrt{\frac{2}{5}}g_{\mu_{1}\nu_{1}}^{\perp}\varepsilon
      (\mu_{2} \nu_{2} p v )$\\
 &&$3$&$\sqrt{\frac{2}{5}}(p_{\mu_{1}}^{\perp}p_{\nu_{1}}^{\perp}-\frac{1}{5}
      g_{\mu_{1}\nu_{1}}p_{\perp}^{2})\varepsilon(\mu_{2} \nu_{2} p v)$\\
\hline \hline
\end{tabular}
\end{center}
\renewcommand{\baselinestretch}{1}
\small \normalsize
\end{table}

The one--pion transition amplitudes between heavy baryons can then be
written as
\begin{eqnarray}
M^{\pi}&=& \langle \pi (\vec{p}), B_{Q2}(v) \mid T \mid B_{Q1}(v) \rangle
\nonumber\\[2mm]
&=&{\bar \psi}_{2}^{\nu_{1} \dots \nu_{j_{2}}}(v) \psi_{1}^{\mu_{1} \dots
\mu_{j_{1}}}(v) (\sum_{l_{\pi}} f_{l_{\pi}} t^{l_{\pi}}_{\mu_{1} \dots
\mu_{j_{1}};\nu_{1} \dots \nu_{j_{2}}}) \label{pi-trans1}
\end{eqnarray}
where the heavy--side baryon wave functions $\psi^{\mu_{1} \dots \mu_{j}}$
have been given in Tables 4 and 5 and the relevant light--side
tensors $t^{l_{\pi}}_{\mu_{1} \dots \mu_{j_{1}};\nu_{1} \dots \nu_{j_{2}}}$
are listed in Table 6. They are tensors of rank $(j_{1}+j_{2})$ build
from the building blocks $g_{\perp \mu \nu} = g_{\mu \nu}-v_{\mu} v_{\nu},
\; p_{\perp \mu}=p_{\mu}-p \hspace{-0.7mm}\cdot \hspace{-0.7mm}v v_{\mu}$
and, depending on parity, from the Levi--Civita tensor.
They are put together such that they have the correct parity and project out
the correct partial wave amplitude with amplitude $f_{l_{\pi}}$. The
normalization of the amplitudes $f_{l_{\pi}}$ is such that a given partial
wave amplitude $f_{l_{\pi}}$ contributes as $\mid f_{l_{\pi}}\mid^2
\mid \vec{p}\mid^{2l_\pi}$ to the spin summed square of the diquark
transition amplitude.

As a first application we consider the ground state to ground state
transition $\{ \Sigma_{c} \} \rightarrow
\Lambda_{c}+\pi$. The one--pion transition amplitudes are easily written
down using Eq.(\ref{pi-trans1}), the relevant heavy--side spin wave
functions from Tables 4 and 5, and the $1^{+} \rightarrow 0^{+} + \pi$
covariant pion coupling in Table 6. One has
\beq\label{sila-trans}
M^{\pi}={\bar u}_{2}(v) \left \{
   \begin{array}{c}
       \frac{1}{\sqrt{3}} \gamma_{\perp}^{\mu} \gamma_{5} u_{1}(v)\\
       u^{\mu}_1(v)
   \end{array}
\right \} f_{p} p_{\mu}^{\perp}
\eeq
The decay rates can be calculated using the general rate formula
\beq\label{rate}
\Gamma = \frac{1}{2J_{1}+1} \quad \frac{ \mid \vec{p} \mid}{8 \pi M_{1}^{2}}
         \sum_{spins} \mid M^{\pi} \mid^{2}.
\eeq
One then obtains
\beq\label{equal rates}
\Gamma_{\Sigma_{c}^{*} \rightarrow \Lambda_{c} + \pi}=\Gamma_{\Sigma_{c}
\rightarrow \Lambda_{c} + \pi}=\frac{1}{6\pi} \frac{M_{2}}{M_{1}} \mid
f_{p} \mid^{2} \mid \vec{p} \mid^{3}
\eeq
where $\mid \vec{p} \mid$ is the CM momentum of the pion. The equality of
the rates for $\Sigma_{c}^{*} \rightarrow \Lambda_{c} + \pi$ and $\Sigma_{c}
\rightarrow \Lambda_{c} + \pi$ true in the heavy quark mass limit (when
$M_{\Sigma_{c}^{*}}=M_{\Sigma_{c}}$) is a general result for transitions
into a Heavy Quark Symmetry spin singlet state. This general result is most
easily derived in the 6--j
symbol formalism discussed at the end of this section.

Differences in the phase space factors $ \mid \vec{p} \mid^{3}$ in the two
decays constitute
${\cal O}(1/m_{c})$ effects which may be important when one wants
to model $1/m_{c}$--effects in phenomenological applications. Similarly the
final form of the rate (\ref{equal rates}) depends on at what stage of the
rate calculation one has dropped the zero recoil approximation inherent to
the Heavy Quark Symmetry approach. This explains the ${\cal O}(1/m_{c})$
differences between the results of
\cite{yan92} and Eq.(\ref{equal rates}). We have
retained the zero recoil approximation in the calculation of the squared
matrix element $\mid M^{\pi} \mid^{2}$.

An estimate of the coupling strength $f_{p}$ can be obtained in the
constituent quark model approximation \cite{yan92}. The one--pion transition
between the dipion states is resolved into one--pion transitions of the
constituent quarks as drawn in Fig.9. The coupling of the pion to the
constituent quarks can be obtained from PCAC and is given by $g_{A} f_{\pi}
^{-1} \pslash_{\perp} \gamma_{5} \; (f_{\pi}=93$~MeV)  where $g_{A}$ is
a phenomenological factor ($g_{A}=0.75)$ which is introduced to get the
$g_{A}/g_{V}$ ratio in neutron
$\beta$--decay right. The coupling strength $f_{p}$ can then be computed by
using the constituent spin wave functions ${\hat \chi}^{0}$ and ${\hat \chi}
^{1,\mu}$ introduced in Sec.4.1. For the transition $1^{+}
\rightarrow 0^{+}+\pi$ one needs the trace
\beq\label{trace1}
Tr \{ {\bar {\hat \chi}}^{0} \pslash_{\perp} \gamma_{5} {\hat \chi}^{1}_{\mu}
\}=p_{\mu}^{\perp}.
\eeq
and for the $1^{+}\rightarrow 1^{+} + \pi$ transition one needs the trace
\beq\label{trace2}
Tr \{ {\bar {\hat \chi}}^{1}_{\nu} \pslash_{\perp} \gamma_{5} {\hat \chi}
^{1}_{\mu}\}=i \varepsilon ( \mu \nu p v).
\eeq
Returning to Eq.(\ref{sila-trans}) one then obtains
$f_{p}=g_{A} f_{\pi}^{-1}$ from a comparison with Eq.(\ref{trace1}).
This results in a width value of e.g.
\beq\label{lasi-width}
\Gamma_{\Sigma_{c}^{0}\rightarrow \Lambda_{c} \pi^{-}}=2.45 \, \mbox{MeV}
\eeq
where we quote the numerical result of the calculation of \cite{yan92}.
A QCD sum rule calculation results in a width value which includes the
constituent quark model value within its large error bound \cite{gy}.
A slightly different value is obtained upon using the rate formula
(\ref{equal rates}), due to a difference in the treatment of recoil
corrections, as remarked on earlier. From the width estimate
Eq.(\ref{lasi-width}) one concludes that the $\Sigma_{c}$ is very likely so
narrow that
it will not be an easy task to experimentally determine its absolute width.

As a further application consider the transition $ \{\Lambda_{CK1}\}
\rightarrow \{\Sigma \}+\pi$ with $J^{P}$ quantum numbers $ \{ \frac{1}{2}
^{-},\frac{3}{2}^{-}\}\rightarrow\{\frac{1}{2}^{+},\frac{3}{2}^{+}\}+0^{-}$.
The transition matrix element can easily be written down using Eq.(\ref
{pi-trans1}) and reads
\beq\label{la-si trans}
M^{\pi}=
      \left \{ \begin{array}{c}
        -\frac{1}{\sqrt{3}}{\bar u}_{2} \gamma_{5} \gamma_{\perp}^{\nu}\\
        {\bar u}_{2}^{\nu}
      \end{array} \right \} \left\{
      \begin{array}{c}
        \frac{1}{\sqrt{3}}\gamma_{\perp}^{\mu} \gamma_{5} u_{1}\\
        u_{1}^{\mu}
      \end{array} \right\}
      ({\textstyle \frac{1}{\sqrt3}}
       f_{s}g^{\perp}_{\mu \nu} +
       {\textstyle \sqrt{\frac32}}f_{d}(p_{\mu}^{\perp}
       p_{\nu}^{\perp}
      -{\textstyle \frac{1}{3}}g_{\mu \nu}^{\perp} p_{\perp}^{2} ))
\eeq
As mentioned above, the decays $ \frac{1}{2}^{-} \rightarrow \frac{1}{2}^{+}
+\pi$ and $ \frac{3}{2}^{-} \rightarrow \frac{1}{2}^{+} +\pi$ are
kinematically allowed and are thus interesting from the experimental point of
view. Their matrix elements can be read off from (\ref{la-si trans}) and are
\beq
\begin{array}{lll}
  M^{\pi}(\frac{1}{2}^{-} \rightarrow \frac{1}{2}^{+} +\pi)&=& -
  {\textstyle \frac{1}{\sqrt3}}f_{s}{\bar u}
  _{2} u_{1}\\
  M^{\pi}(\frac{3}{2}^{-} \rightarrow \frac{1}{2}^{+} + \pi)&=&\frac{1}
  {\sqrt{2}}f_{d}{\bar u}_{2} \gamma_{5} p_{\mu}^{\perp} \pslash^{\perp}
  u_{1}^{\mu}
\end{array}
\eeq
As expected the covariant couplings project out the correct orbital angular
momenta $l_{\pi} =L_{\pi}$.
Using the rate formula (\ref{rate}) one obtains
\begin{eqnarray}
\Gamma\left({\textstyle \frac{1}{2}^{-}}
\rightarrow {\textstyle \frac{1}{2}^{+}} +\pi\right) &=&f_{s}^{2}
      \frac{\mid \vec{p} \mid}{6\pi} \frac{M_{2}}{M_{1}}\label{rate2}\\
\Gamma\left({\textstyle \frac{3}{2}^{-}}
\rightarrow {\textstyle \frac{1}{2}^{+}} +\pi\right) &=&f_{d}^{2}
      \frac{\mid \vec{p} \mid^{5}}{8\pi}\frac{M_{2}}{M_{1}}.\label{rate3}
\end{eqnarray}
One notes that the rates (\ref{rate2}) and (\ref{rate3}) exhibit the correct
threshold behaviour $\mid \vec{p} \mid^{2L_{\pi}+1}$ where $\vec{p}$ is the
CM momentum of the pion.

Using $M_{\Sigma_{c}}=2.453$ GeV one finds $\mid \vec{p} \mid=1.62 \times
10^{-2}$ GeV and $\mid \vec{p} \mid^{5}=1.06 \times 10^{-5}$ GeV for the two
respective threshold factors. If the scale of the coupling constants were
1~GeV one would in fact have a $10^{-3}$ suppression of $\Lambda_{c}(2627)
\rightarrow \Sigma_{c} \pi$ relative to $ \Lambda_{c}(2593) \rightarrow
\Sigma_{c} \pi$ in agreement with the observation
\cite{CLEOLAMBDA93,E687LAMBDA93}. However, for the
soft pion emission in these decay processes $f_{\pi} \cong m_{\pi}$ is
frequently a more appropriate scale. One would then have $\mid \vec{p} \mid
/m_{\pi}=0.117 $ and $ \mid \vec{p} \mid^{5} / m_{\pi}^{5}=0.204$. In such a
case the d--wave decay in $\Lambda_{c}(2593) \rightarrow \Sigma_{c} \pi$
would not be suppressed as seems to be the case in the ARGUS result
\cite{ARGUSLAMBDA93}. Hopefully
future experiments can clarify the situation about the $ \Lambda_{c}(2593)
\rightarrow \Sigma_{c} \pi$ branching fraction.
\begin{figure}
\vspace{5.5cm}
\caption[dummy14]{ Light--side one--pion transition resolved into
constituent quark transitions. Equal velocities of diquark and quarks are
implied.}
\end{figure}

Next consider the one--pion transitions from the p--wave multiplet $\{
\Lambda_{Ck2} \}$ down to the ground state multiplet $\{\Sigma\}$. In terms
of $J^{P}$ quantum numbers one has the transitions $\{ \frac{3}{2}^{-},
\frac{5}{2}^{-}\} \rightarrow \{ \frac{1}{2}^{+}, \frac{3}{2}^{+} \} + \pi$.
Here the pion is emitted in a d--wave. For the transition amplitude one now
has
\beq \label{pi-trans2}
M^{\pi}=
   \left \{ \begin{array}{c}
    - \frac{1}{\sqrt{3}}{\bar u}_{2} \gamma_{5} \gamma_{\perp}^{\nu_{1}}\\
     {\bar u}_{2}^{\nu_{1}}
   \end{array} \right \} \left \{
   \begin{array}{c}
     \frac{1}{\sqrt{10}} \gamma_{5}( \gamma_{\perp}^{\mu_{1}} u_{1}^{\mu_{2}}
        +\gamma_{\perp}^{\mu_{2}} u_{1}^{\mu_{1}})\\
     u_{1}^{\mu_{1} \mu_{2}}
   \end{array} \right \}
   f'_{d} p^{\perp}_{\mu_{1}} \varepsilon( \mu_{2} \nu_{1} p^{\perp} v)
\eeq
One can then use the amplitude (\ref{pi-trans2}) to calculate the ratio of
the four one--pion transitions described by (\ref{pi-trans2}). After a
little bit of  algebra \cite{IKKL94} one obtains
\beq\label{ratio of rates1}
\Gamma_{\frac{3}{2}^{-}\rightarrow \frac{1}{2}^{+}+\pi}:
\Gamma_{\frac{3}{2}^{-}\rightarrow \frac{3}{2}^{+}+\pi}:
\Gamma_{\frac{5}{2}^{-}\rightarrow \frac{1}{2}^{+}+\pi}:
\Gamma_{\frac{5}{2}^{-}\rightarrow \frac{3}{2}^{+}+\pi}
=9:9:4:14
\eeq
and
\beq\label{ratio of rates2}
\Gamma_{\frac{3}{2}^{-}\rightarrow \frac{1}{2}^{+}+\pi}+
\Gamma_{\frac{3}{2}^{-}\rightarrow \frac{3}{2}^{+}+\pi}=
\Gamma_{\frac{5}{2}^{-}\rightarrow \frac{1}{2}^{+}+\pi}+
\Gamma_{\frac{5}{2}^{-}\rightarrow \frac{3}{2}^{+}+\pi}.
\eeq
This result agrees with the conventional approach using Clebsch--Gordan
coefficients \cite{IW91} or, in a more recent and compact guise, using 6--j
symbols \cite{Z93,fp93}. The sum rule (\ref{ratio of rates2}) is a general
result for Heavy Quark Symmetry doublet to doublet transitions and is easily
derived in the 6--j coupling approach.

Let us briefly review the reasoning of Ref.\citer{IW91,fp93} that leads to
the introduction of Wigner's 6--j symbols. Having Fig.8 in mind one first
compounds the spins in the initial and final state $j_{1}+S_Q
\rightarrow J_{1}$
and $j_{2}+S_Q\rightarrow J_{2}$, where $S_Q=1/2$ is the heavy quark spin,
and then
combines these with the transition $j_{1} \rightarrow j_{2}+l_{\pi} \quad
(l_{\pi}=L_{\pi})$ using the appropriate Clebsch--Gordan coefficients. One
then obtains
\begin{eqnarray}
M^{\pi}(J_{1}J_{1}^{z}\rightarrow J_{2}J_{2}^{z}+L_{\pi}m)
  &=&M_{L_{\pi}}\sum_{s^{z},j_{1}^{z},j_{2}^{z}} \langle J_{2}
     J_{2}^{z} \mid j_{2} j_{2}^{z}S_Q S^{z}_Q \rangle \langle L_{\pi}
     m j_{2} j_{2}^{z}\mid j_{1} j_{1}^{z} \rangle\nonumber\\[1mm]
  & &\hspace{0.5cm}\langle j_{1}j_{1}^{z} S_QS^{z}_Q \mid J_{1} J_{1}^{z}
     \rangle\nonumber\\[2mm]
  &=&M_{L_{\pi}} (-1)^{L_{\pi}+j_{2}+S_Q+J}(2j_{1}+1)^{1/2}(2J_{2}+1)^{1/2}
     \nonumber\\[1mm]
  & &\hspace{0.5cm}\left \{ \begin{array}{ccc}
       j_{2}&j_{1}&L_{\pi}\\
       J_1  &J_{2}&S_Q
     \end{array} \right \}
     \langle L_\pi mJ_{2} J_{2}^{z}\mid J_{1} J_{1}^{z} \rangle.
\label{pi-trans3}
\end{eqnarray}
The reduced matrix elements $M_{L_{\pi}}$ correspond to the coupling factors
$f_{l_{\pi}}$ used in Eq.(\ref{pi-trans1}). In the second step of Eq.(\ref
{pi-trans3}) one has rewritten the first part of Eq.(\ref{pi-trans3}) in
terms of the Wigner 6--j symbol.

After spin--averaging over the initial spin and summing over final spins one
obtains the rate
\beq
\frac{1}{2J_{1}+1} \sum_{spins} \mid M^\pi (L_\pi) \mid^{2}=
(2j_{1}+1) (2J_{2}+1)
\left| \left \{  \begin{array}{ccc}
   L_{\pi}&j_{2}&j_{1}\\
   S_Q&J_{1}&J_{2}
\end{array} \right  \} \right|^{2}
\mid M_{L_\pi} \mid^{2}
\eeq
for a transition involving a given orbital angular momentum $L_{\pi}$ of
the pion.
Using the standard orthogonality relation for 6--j symbols
\beq\label{orth6j}
\sum_{J_{2}} (2j_{1}+1)(2J_{2}+1)
  {\left| \left \{   \begin{array}{ccc}
      L_\pi & j_2 & j_1\\
      S_Q & J_1 & J_2
  \end{array} \right\} \right|}^{2}
=1
\eeq
one can immediately appreciate the significance of the result Eq.(\ref
{ratio of rates2}). The total rate of pionic decays from any of the two
doublet states $J_{1}=j_{1} \pm S_{Q}$ is independent of $J_{1}$. Also one
immediately concludes that the transitions into a heavy quark symmetry
singlet state from any of the two  heavy quark symmetry doublet states are
identical to one another.
A general proof of the equivalence of the covariant coupling scheme and the
6--j coupling scheme (\ref{pi-trans3}) is presently being worked out
\cite{IKL94}.

It is quite apparent that the 6--j approach to one--pion transitions is much
easier to handle from a calculational point of view and is structurally
more transparent than the covariant approach. In the covariant approach, on
the other hand, one may more readily include ${\cal O}(1/m_{c})$ recoil and
phase space corrections according to one's own intuition and experience (or
prejudice).
Also the covariant approach lends itself more easily to a transcription
into the usual field theoretic formulation of the pion's coupling in terms
of effective Lagrangians.
In fact one can easily turn the covariant couplings written down in this
section into chirally invariant pion couplings. This can be done
by enacting the following substitutions in Table 6
\beq\label{chiral}
\begin{array}{lcl}
p_{\mu_{1}} \dots p_{\mu_{k}} &\Rightarrow& \frac{1}{F} \partial_{\mu_{1}}
\dots \partial_{\mu_{k}} \Phi_{\pi} + \dots\\
g_{\mu_{1}\mu_{2}} &\Rightarrow& g_{\mu_{1} \mu_{2}}
\end{array}
\eeq
and, for k=0, constant $\Rightarrow \frac{1}{F} v \hspace{-0.7mm}\cdot
\hspace{-0.7mm}\partial \Phi_{\pi}$. The ellipses in Eq.(\ref{chiral}) stand
for higher order contribution in the chiral expansion (see \cite{IKKL94}).

\subsection{Photon Transitions Between Heavy Baryons}
In addition to the pion transitions between heavy baryons treated in
Sec.\ref{pi-trans} photon transitions between heavy baryon states are
also of interest. In fact, because of phase space limitations there  are many
more levels that can be reached via photon transitions than via pion
transitions for a given higher lying heavy baryon initial state. In some
cases where the pion mode is not available the total rate of the heavy baryon
state is entirely in terms of the photon decay mode. Examples are the
$\Xi'_{c}$ and $\Omega_{c}^{*}$ charm baryon states which are expected to
decay electromagnetically because pion emission is kinematically forbidden
according to present mass estimates (see Fig.10). Furthermore hyperfine
splitting effects can be expected to have become so small in the bottom
sector that transitions between the two partners in a heavy quark spin
multiplet can be mediated by photons alone.

In the following we set up the formalism necessary to describe photon
transitions between heavy baryon states in the heavy quark limit. Our
discussion will be limited to the treatment of leading effects in the
$1/m_{Q}$ expansion, although a consideration  of nonleading effects
certainly warrants future attention \cite{KPS93}. In the heavy mass limit
the (real!) photon couples only to the light diquark side since the photon
coupling to the heavy quark involves a spin--flip transition down by
$1/m_{Q}$. Although the
formalism in this section applies both to the heavy charm and bottom sectors
we shall stay in the charm sector when we work out a few definite one photon
transition examples. The reason is clearly experimental. While we are just at
the threshold of being able to observe photon transitions in the charm baryon
sector the corresponding physics in the bottom sector lies a few years ahead
of us.

The photon transition amplitude between heavy baryon states can be written
down in complete analogy to the corresponding one--pion transition
amplitude Eq.(\ref{pi-trans1}) in Sec.\ref{pi-trans}. The physics underlying
the heavy quark description of one--photon transitions is depicted in Fig.
8. The photon is emitted from the light diquark side while the heavy
quark remains unaffected. The transition amplitude can thus again be written
in a factorized form
\beq\label{pho-trans}
M^{\gamma}=\langle \gamma (k), B_{Q_{2}}(v) \mid T \mid B_{Q_{1}}(v)\rangle={
\bar\Psi}_{2}^{\nu_{1} \dots \nu_{j_{2}}} (v) \Psi_{1}^{\mu_{1} \dots
\mu_{j_{1}}}(v) (\sum_{J_{\gamma}}f^{J_{\gamma}}
t^{J_{\gamma}}_{\mu_{1}\dots\mu_{j_{1}};\nu_{1}\dots\nu_{j_{2}}})
\eeq
where the appropriate spin coupling factors are determined by performing the
tensor contractions in Eq.(\ref{pho-trans}). The index $J_{\gamma}$ in
(\ref{pho-trans}) denotes the total angular momentum of the photon (spin of
the photon plus its orbital angular momentum). We choose to work in terms of
multipole amplitudes $f^{J_{\gamma}}$
with multipolarities
$2^{J_{\gamma}}$ with $J_\gamma=J_{\gamma \, min},\;(J_{\gamma\, min}+1),
\dots  J_{\gamma\, max}$. The tensors $t^{J_{\gamma}}_{\mu_{1} \dots \mu_{j
_{1}}; \nu_{1} \dots \nu_{j_{2}}}$ project onto the multipole amplitudes. The
reason for working in terms of multipole amplitudes is simply that the
multipole amplitudes $f^{J_{\gamma}}$ contribute to the decay rate with
definite powers of the photon momentum $\mid\vec{k}\mid^{2J_{\gamma}+1}$.
This allows one to classify the transitions in terms of decreasing
importance. The reasoning is similar to the one--pion transition case
treated in Sec.\ref{pi-trans} where we used an orbital momentum
classification.
\begin{table}
\caption[dumy16]{\label{tab4}Tensor structure of photon couplings to
diquark states. Photon is in definite multipole state EJ (electric)
MJ (magnetic). Sign of
the product of naturalities determines whether coupling is to field
strength tensor $F_{\alpha \beta} \; (n_{1} \! \cdot \! n_{2}=+1)$ or to its
dual ${\tilde F}_{\alpha \beta} \; (n_{1} \! \cdot \! n_{2}=-1)$. Tensor
structure of transitions with $(j_{1}^{P_{1}}, j_{2}^{P_{2}}) \rightarrow
(j_{1}^{-P_{1}}, j_{2}^{-P_{2}}) \rightarrow (j_{2}^{P_{2}} j_{1}^{P_{1}})
\rightarrow (j_{2}^{-P_{2}} j_{1}^{-P_{1}})$ are identical and are not always
listed separately. }
\begin{center}
\vspace{5mm}
\renewcommand{\baselinestretch}{1.2}
\small \normalsize
\begin{tabular}{rlcccl}
\hline \hline
\multicolumn{2}{c}{\begin{tabular}{c}
                      diquark transition\\
                      $j_{1}^{P_{1}} \rightarrow j_{2}^{P_{2}}+\gamma$
                   \end{tabular}}
  &multipoles&$n_{1}n_{2}$&\hspace{0.5cm}&
                   \begin{tabular}{c}
                      covariant coupling\\
                      $t^{i}_{\mu_{1}\dots \mu_{j_{1}};\nu_{1}
                      \dots\nu_{j_{2}}}$
                   \end{tabular}\\
 \hline \hline
 $0^{+} \rightarrow$&$0^{+}+\gamma$&forbidden&$+1$&&\\
 \hline
 $1^{+} \rightarrow$&$0^{+}+\gamma$&M1&$-1$&$$&$\frac{1}{\sqrt{2}} {\tilde F}_
        {\alpha \beta}g_{\mu_{1}}^{\alpha}v^{\beta}$\\
     &$1^{+}+\gamma$&M1&$+1$&$$&$ \frac{1}{2}F_{\alpha \beta}g_{\mu_{1}}
        ^{\alpha}g_{\nu_{1}}^{\beta}$\\
     &&E2&$+1$&$$&$ \frac{1}{2}F_{\alpha
\beta}(2k_{\mu_{1}}g_{\nu_{1}}^{\alpha}
        v^{\beta}+k \hspace{-0.7mm}\cdot\hspace{-0.7mm} vg_{\mu_{1}}^{\alpha}
        g_{\nu_{1}}^{\beta})$\\
 \hline
 $0^{-} \rightarrow$&$0^{+}+\gamma$&forbidden&$-1$&$$&\\
 \hline
 $1^{-} \rightarrow $&$ 0^{+} + \gamma$&E1&$+1$&$$&$\frac{1}{\sqrt{2}}F_{\alpha
       \beta} g_{\mu_{1}}^{\alpha} v^{\beta}$\\
   &$1^{+} +\gamma$&E1&$-1$&$$&$\frac{1}{2}{\tilde F}_{\alpha \beta}
       g_{\mu_{1}}^{\alpha} g_{\nu_{1}}^{\beta}$\\
   &&M2&$-1$&$$&$\frac{1}{2}{\tilde F}_{\alpha \beta}(2k_{\mu_{1}}
       g_{\nu_{1}}^{\alpha}v^{\beta}+k \hspace{-0.7mm}\cdot\hspace{-0.7mm}v
       g_{\mu_{1}}^{\alpha} g_{\nu_{1}}^{\beta})$\\
 \hline
 $2^{-} \rightarrow$&$0^{+} +\gamma$&M2&$-1$&$$&${\tilde F}
       _{\alpha \beta} k_{\mu_{1}} g_{\mu_{2}}^{\alpha} v^{\beta}$\\
   &$1^{+} + \gamma$&E1&$+1$&$$&$ \sqrt{\frac{3}{10}}F_{\alpha \beta}
          g_{\mu_{1}}^{\alpha} g_{\mu_{2}\nu_{1}}v^{\beta}$\\
   &&M2&$+1$&$$&$\sqrt{\frac{1}{6}}F_{\alpha \beta} ( v \! \cdot \! k
          g_{\mu_{2} \nu_{1}} g_{\mu_{1}}^{\alpha}v^{\beta} + 2k_{\mu_{2}}
          g_{\mu_{1}}^{\alpha}g_{\nu_{1}}^{\beta} )$\\
   &&E3&$+1$&$$&$
        \sqrt{\frac{1}{30}}F_{\alpha \beta} ( (v \! \cdot \! k)^2  g_{\mu_{2}
          \nu_{1}}g_{\mu_{1}}^{\alpha} v^{\beta} + \frac{5}{4} v \! \cdot \! k
          k_{\mu_{2}} g_{\mu_{1}}^{\alpha} g_{\nu_{1}}^{\beta} $\\
   && &&&$\qquad + \frac{15}{4} v^{\beta} k_{\mu_{2}} ( k_{\nu_{1}}
          g_{\mu_{1}}^{\alpha} + k_{\mu_{1}} g_{\nu_{1}}^{\alpha} ))$\\
 \hline
 $2^{-} \rightarrow$&$0^{-}+\gamma$&E2&$+1$&$$&$F_{\alpha \beta}k_{\mu_{1}}
          g_{\mu_{2}}^{\alpha}v^{\beta}$\\
   &$1^{-}+\gamma$&M1&$-1$&$$&$\sqrt{\frac{3}{10}}{\tilde F}_{\alpha \beta}
          g_{\mu_{1}}^{\alpha}g_{\mu_{2} \nu_{1}} v^{\beta}$\\
   &&E2&$-1$&$$&$\sqrt{\frac{1}{6}}{\tilde F}_{\alpha \beta} ( v \! \cdot \! k
          g_{\mu_{2} \nu_{1}} g_{\mu_{1}}^{\alpha}v^{\beta} + 2 k_{\mu_{2}}
          g_{\mu_{1}}^{\alpha}g_{\nu_{1}}^{\beta} )$\\
   &&M3&$-1$&$$&$\sqrt{\frac{1}{30}}{\tilde F}_{\alpha \beta} ((v\!\cdot\!k)^2
          g_{\mu_{2} \nu_{1}} g_{\mu_{1}}^{\alpha} v^{\beta} + \frac{5}{4} v \!
          \cdot \! k k_{\mu_{2}} g_{\mu_{1}}^{\alpha} g_{\nu_{1}}^{\beta} $\\
   && &&&$\qquad + \frac{15}{4} v^{\beta} k_{\mu_{2}} ( k_{\nu_{1}}
          g_{\mu_{1}}^{\alpha} + k_{\mu_{1}} g_{\nu_{1}}^{\alpha} ))$\\
   &$2^{-}+\gamma$&M1&$+1$&$$&$\sqrt{\frac{1}{5}} F_{\alpha \beta} g_{\mu_{1}
          \nu_{1}}g_{\mu_{2}}^{\alpha}g_{\nu_{2}}^{\beta}$\\
   &&E2&$+1$&$$&$\sqrt{\frac{3}{7}}F_{\alpha \beta} g_{\mu_{1} \nu_{1}}
          (2 k_{\mu_{2}} v^{\beta}g_{\nu_{2}}^{\alpha} + v \! \cdot \! k
          g_{\mu_{2}}^{\alpha} g_{\nu_{2}}^{\beta})$\\
   &&M3&$+1$&$$&$\sqrt{\frac{3}{10}}F_{\alpha \beta} g_{\mu_{2}}^{\alpha}
          g_{\nu_{2}}^{\beta} ( (v \! \cdot \! k)^2  g_{\mu_{1} \nu_{1}} +
          \frac{5}{2} k_{\mu_{1}} k_{\nu_{1}})$\\
   &&E4&$+1$&$$&$\sqrt{\frac{1}{14}}F_{\alpha \beta} (2 k_{\mu_{2}} v^{\beta}
          g_{\nu_{2}}^{\alpha} + v \! \cdot \! k g_{\mu_{2}}^{\alpha}
          g_{\nu_{2}}^{\beta})$ \\
   &&&&$$&$\hphantom{\sqrt{\frac{1}{14}}F_{\alpha \beta}}
           ( (v \! \cdot \! k)^2  g_{\mu_{1} \nu_{1}} +
           \frac{7}{2} k_{\mu_{1}} k_{\nu_{1}} )$\\
 \hline \hline
\end{tabular}
\end{center}
\renewcommand{\baselinestretch}{1}
\small \normalsize
\end{table}

In Table 7 we list explicit forms of the covariant multipole
tensors for radiative diquark transitions $j_{1}^{P_{1}} \rightarrow
j_{2}^{P_{2}} + \gamma$ for the cases of interest. The covariant tensors
project onto magnetic $(MJ_{\gamma})$ or electric $(EJ_{\gamma})$ multipole
transitions. The nature of the multipole transition is determined by the
parity of the photon in a given multipole state which is
$P(EJ_{\gamma})=(-1)^{J_{\gamma}}$ and $P(MJ_{\gamma})=(-1)^{J_{\gamma}+1}$.
For example, for $P_{1} \! \cdot \!P_{2}=+1$ parity conservation implies an
even and an odd $J_{\gamma}$ for electric and magnetic multipole transitions,
respectively. The covariants
are written in terms of the momentum representation of the field strength
tensor $F_{\alpha\beta}=k_{\alpha} \varepsilon_{\beta}-k_{\beta} \varepsilon
_{\alpha}$ or its dual
$\frac12{\tilde F}_{\alpha\beta}=\varepsilon_{\alpha \beta
\gamma \delta}F^{\gamma \delta}$ where $\varepsilon_{\alpha}$ is the
polarization  vector of the photon. The use of the field strength tensor
guarantees the appropriate coupling to (three--) transverse photons as can be
easily appreciated by rewriting the antisymmetric field strength tensor as an
one--index object $F_{\alpha \beta} \rightarrow \varepsilon_{\mu \alpha
\beta \gamma} k^{\alpha} \varepsilon^{\beta} v^{\gamma}$, in analogy to
Eq.(\ref{one ind}). Whether the coupling is to $F_{\alpha \beta} \quad
(n_{1}n_{2}=+)$ or to its dual ${\tilde F}_{\alpha \beta} \quad
(n_{1}n_{2}=-)$ is determined
by the product of normalities $n_{1}\cdot n_{2}$ also listed in Table
4.
The counting of covariants and thereby the counting of the number of
multipole
amplitudes can be done by helicity or multipole amplitude counting and is
given by
\beq
\begin{array}{lll}
i) &j_{1}=j_{2}:&N=2j\\
ii)&j_{1}\neq j_{2}:& N=2j_{min}+1
\end{array}
\eeq
The normalization is such that a given multipole amplitude $f^{J_{\gamma}}$
contributes as $\mid \!f^{J_{\gamma}}\!\mid^{2}\mid\vec{k}
\mid^{2J_{\gamma}+1}$
to the spin summed
square of the diquark transition amplitude \cite{IKLA}.

It is then an easy matter to derive the heavy quark symmetry structure of
photon transitions between heavy baryon states using Eq.(\ref{pho-trans}) and
Tables 4,5 and 7. As a first application we
write down the amplitude for the ground state transition $\{ \Sigma_{Q}\}
\rightarrow \Lambda_{Q} + \gamma$. One has
\beq\label{gr-trans}
M^{\gamma}={\bar u}_{2}
    \left \{ \begin{array}{c}
      \frac{1}{\sqrt{3}} \gamma_{\perp}^{\mu_{1}}\gamma_{5} u_{1}\\
      u_{1}^{\mu_{1}}
    \end{array} \right \}
{\textstyle \frac{1}{\sqrt{2}}}
f^{M1} {\tilde F}_{\alpha \beta} g_{\mu_{1}}^{\alpha} v^{\beta}
\eeq
Using standard $\varepsilon_{\alpha \beta \gamma \delta}$--tensor identities
one obtains
\begin{eqnarray}
  \Sigma_{c} \rightarrow \Lambda_{c}+\gamma &:&\quad M^{\gamma}=i
  {\textstyle \frac{1}{\sqrt{6}}}f^{M1}
  {\bar u}_{2} \kslash \eslash^{*} u_{1}\label{gr-trans1}\\
  \Sigma_{c}^{*} \rightarrow \Lambda_{c}+\gamma &:& \quad M^{\gamma}=
 {\textstyle \frac{1}{\sqrt{2}}}f^{M1}
  {\bar u}_{2} \varepsilon(\mu_{1} v k\varepsilon^{*}) u_{1}^{\mu_{1}}
  \label{gr-trans2}
\end{eqnarray}
The transition (\ref{gr-trans2}) can be checked to have the correct M1
coupling structure. The rate follows from the two body decay rate formula
Eq.(\ref{rate}). In the degeneracy limit $M_{\Sigma_{c}^{*}}=M_{\Sigma_{c}}$
one finds
\beq
\Gamma_{\Sigma_{c} \rightarrow \Lambda_{c}+\gamma}=\Gamma_{\Sigma_{c} ^{*}
\rightarrow \Lambda_{c}+\gamma}=\frac{1}{6\pi} \mid \! f^{M1} \! \mid^{2}
\frac{M_{2}}{M_{1}} \mid \vec{k} \mid^{3}
\eeq
where $\mid \vec{k} \mid =(M_{1}^{2}-M_{2}^{2})/2M_{1}$. The equality of the
decay rates of heavy quark symmetry partners into the ground state $\Lambda
_{c}$ is again a general result that can easily be derived in the 6--j
formalism as applied to photon transitions.

The photonic coupling $(\Lambda_{c};\Lambda_{c} \gamma)$ vanishes in the
heavy quark limit as Table 7 shows. The reason is simply that a (real!)
photon cannot couple to the $0^{+}$ diquark state.  For the ground state
transition $\{ \Sigma_{Q} \} \rightarrow \{ \Sigma_{Q} \} + \gamma$ we
calculate the M1
contribution to the kinematically accessible transition $\Sigma_{c}^{*}
\rightarrow \Sigma_{c} + \gamma$. One obtains
\beq\label{gr-trans3}
\Sigma_{c}^{*}\rightarrow \Sigma_{c} + \gamma: \quad M^{\gamma}=
{\textstyle \frac{1}{2\sqrt{3}}}
f'^{M1}{\bar u}_{2}\gamma_{5} \gamma^{\alpha}u_{1}^{\mu_{1}}
(k_{\mu_{1}}\varepsilon^{*}_{\alpha}-k_{\alpha}\varepsilon^{*}_{\mu_{1}})
\eeq
where we denote the M1 diquark amplitude by $f'^{M1}$ to set it aside from
the amplitude $f^{M1}$ used in Eq.(\ref{gr-trans}). It can again be checked
that Eq.(\ref{gr-trans3}) has the correct M1 coupling structure even though
the proportionality of the M1--covariants in Eq.(\ref{gr-trans2}) and
(\ref{gr-trans3}) is not apparent. The rate of the transition
(\ref{gr-trans3}) can be computed to be
\beq\label{gr-trans4}
\Gamma_{\Sigma_{c}^{*}\rightarrow \Sigma_{c} +\gamma}=\frac{1}{36\pi}\mid \!
f'^{M1} \! \mid^{2} \frac{M_{2}}{M_{1}}\mid\vec{k}\mid^{3}
\eeq
\begin{figure}
\vspace{5.5cm}
\caption[dummy17]{ Light--side one--photon transition resolved into
constituent quark
transitions. Equal velocities of light diquark and heavy quark are implied.}
\end{figure}
It is quite clear that heavy quark symmetry can tell us nothing about the
strength of the "reduced matrix elements" $f^{M1}$ and $f'^{M1}$. In order
to obtain a rough estimate for the magnitude of the couplings $f^{M1}$ and
$f'^{M1}$ we resort to the constituent quark model as has been done in
\cite{yan92}. In the constituent quark model the coupling of the photon to
the diquark state is resolved into the sum of the couplings of the photon to
the constituent quarks as shown in Fig.11. The photon couples to the
constituent quarks
with a M1 coupling structure and a known coupling strength
$\frac12\mu_{q} \sigma_{\alpha \beta} F^{\alpha \beta}$ where $\mu_{q}$
is the magnetic moment of the quark q given by $\mu_{q}=e_{q}e/2m_{q}$.
The diquark coupling strengths $f^{M1}$ and $f'^{M1}$ can then
be obtained by evaluating the photonic quark coupling sandwiched between the
spin 0 and spin 1 constituent diquark states ${\hat \chi}^{0}$ and
${\hat\chi}^{1}_{\mu_{1}}$ introduced in Sec \ref{ground state wf}. One
obtains
\begin{eqnarray}
Tr \{ {\bar {\hat \chi}}^{0} {\textstyle \frac{1}{2}} \mu_{q}
\sigma_{\alpha \beta}F^{\alpha
\beta} \chi_{\mu}^{1}\}&=&\mu_{q}{\tilde F}^{\alpha \beta}
g_{\mu\alpha}v_{\beta}\label{di-trans1}\\
Tr \{ {\bar {\hat \chi}}^{1}_{\nu} \frac{1}{2} \mu_{q} \sigma_{\alpha
\beta}F^{\alpha\beta} \chi_{\mu}^{1}\}&=&i\mu_{q} F^{\alpha \beta}
g_{\nu\alpha}g_{\mu\beta}\label{di-trans2}
\end{eqnarray}
for the photon coupling to one of the quark lines.

 From comparing (\ref{di-trans1}) and (\ref{di-trans2}) with the covariant
structure in Table 7 one finds
\begin{eqnarray}
f^{M1}&=&\sqrt{2}e\left( \frac{e_{q_{1}}}{2m_{q_{1}}}
- \frac{e_{q_{2}}}{2m_{q_{2}}}\right)\label{photon form1}\\
f'^{M1}&=&2ie\left( \frac{e_{q_{1}}}{2m_{q_{1}}}
+ \frac{e_{q_{2}}}{2m_{q_{2}}}\right)
\label{photon form2}
\end{eqnarray}
after adding in flavour factors and the contribution of the second
constituent quark. Note
that the constituent quark model approach predicts a vanishing E2 amplitude
for the $\{ \Sigma_{Q} \} \rightarrow \{\Sigma_{Q}\}$ transitions.

Using standard values for the constituent quark masses the authors of
\cite{yan93} have calculated the photonic width of $\Sigma_{c}^{+} \to
\Lambda_{c}+\gamma$. They obtain
\beq\label{rate1}
\Gamma_{\Sigma_{c}^{+} \rightarrow \Lambda_{c}+\gamma}=93\quad
\mbox{keV}.
\eeq
Comparing with the pionic width $\Gamma_{\Sigma_{c}^{+} \rightarrow \Lambda
_{c}+ \pi^{0}}=2.43$~MeV calculated in the same constituent approximation one
finds a photonic branching ratio of $\cong 4\%$ for the $\Sigma_{c}^{+}$
which compares favourably with the $\cong 0.5\%$ branching ratio found for
$\Delta \rightarrow N + \gamma$. From the minus sign in Eq.(\ref{photon
form1}) one predicts a severe rate suppression for ${\Xi'}^{0}_{c}
\rightarrow \Xi_{c}^{0} + \gamma$ relative to ${\Xi'}^{+}_{c} \rightarrow
\Xi_{c}^{+} + \gamma$ due to an almost complete cancellation of the
contributions of the d and s quarks. From \cite{yan93} the two $\Xi'
_{c}\rightarrow \Xi_{c} + \gamma$ rates are in the ratio 0.3/16.

\begin{figure}
\vspace{10.5cm}
\caption[dummy18]{ One--photon transitions between s--wave
charm baryon states including the lowest lying $\Lambda$--type p--wave state.
Multipolarities drawn in plot are multipolarities predicted by Heavy Quark
Symmetry. In some cases there are no restrictions on number of multipole
transitions from Heavy Quark Symmetry.}
\end{figure}

In Fig.10 we have drawn all possible one--photon
transitions involving the ground state charm baryons. We have included the
two
newly discovered $\Lambda$--type p--wave states in the plot. The multipole
structure of the photonic transitions indicated in Fig.10
 refers to the multipole structure predicted by Heavy Quark
Symmetry. For example, the decay $\Lambda_{cK1} \rightarrow \Lambda_{c}
+\gamma$
can in general be a E1 and M2 transition but Heavy Quark Symmetry tells us
that
the transition is purely E1. This will not be an easy task to check
experimentally. The remaining  E2 and M2 quadrupole transitions indicated in
Fig.10 are not forbidden by Heavy Quark Symmetry but
are expected to be small in the constituent quark model approximation.

As a last application consider the one--photon transitions $\{ \Lambda_{cK1}
\} \rightarrow \Lambda_{c}+\gamma$. Using again Table 4 and
7 we find
\begin{eqnarray}
 \Lambda_{cK1} \rightarrow \Lambda_{c}+\gamma &:& \quad M^{\gamma}=-
{\textstyle \frac{1}{\sqrt{6}}}
f^{E1} {\bar u}_{2} \eslash^{*} \gamma_{5}u_{1} \mid \vec{k}\mid
 \label{la-trans1}\\
 \Lambda_{cK1}^{*} \rightarrow \Lambda_{c}+\gamma &:& \quad M^{\gamma}=-
{\textstyle \frac{1}{\sqrt{2}}}
 f^{E1} {\bar u}_{2} \varepsilon_{\mu_{1}}^* u_{1}^{\mu_{1}} \mid \vec{k}\mid
 \label{la-trans2}
\end{eqnarray}
One can check that  Eqs.(\ref{la-trans1}) and (\ref{la-trans2}) have the
correct E1 coupling structure. In the mass degeneracy limit the two rates
are equal as remarked on before. One finds
\beq
\Gamma_{\Lambda_{cK1} \rightarrow \Lambda_{c}+\gamma}=\Gamma_{\Lambda_{cK1}
^{*} \rightarrow \Lambda_{c}+\gamma}=\mid f^{E1} \mid^{2} \frac{1}{6\pi}
\frac{M_{2}}{M_{1}} \mid \vec{k} \mid^{3}
\eeq
When phase space effects are taken into account the ratio of rates of
$\Lambda_{cK1} \rightarrow \Lambda_{c}+\gamma$ and $\Lambda_{cK1}^{*}
\rightarrow \Lambda_{c}+\gamma$ gets lowered by $\cong25\%$. A very rough
estimate of the rate for  $\Lambda_{cK1}^{+*} \rightarrow \Lambda_{c}+\gamma$
can be obtained by comparison with the rate estimate for $\Sigma_{c}^{+}
\rightarrow \Lambda_{c}+\gamma$ given in Eq.(\ref{rate1}). Phase space
enhances the former rate by a factor of $\cong7$.
Setting $f^{E1} \cong f^{M1}$ one can
thus expect a rate of ${\cal O}(700~\mbox{keV})$ for the decay
$\Lambda_{cK1}^{+*} \rightarrow \Lambda_{c}+\gamma$.
If the hadronic width of the $\Lambda_{cK1}^{*}$ is suppressed as much as is
argued for in \cite{CHO94} one can indeed expect a substantial one--photon
branching fraction of the  $\Lambda_{cK1}^{*}$. At any rate, one can hope to
extract a great deal of interesting physics from the analysis of one--photon
transitions between charm baryon states in the future.

We close this subsection by expressing the photonic transition amplitudes in
terms of 6--j symbols, in complete analogy to the pionic case discussed in
Sec.\ref{pi-trans} . In fact one just needs to replace $L_{\pi}$ in
Eqs.(\ref{pi-trans3}) and (\ref{orth6j}) by the total angular momentum of the
photon $J_{\gamma}$. Skipping the first step in the derivation of
Eq.(\ref{pi-trans3}) one now has
\begin{eqnarray}
M^{\gamma}(J_{1} J_{1}^{z} \rightarrow J_{2} J_{2}^{z}+J_{\gamma}m)
&=&M_{J_{\gamma}}(-1)^{J_{\gamma}+j_{2}+S_Q+J_{1}} (2j_{1}+1)^{\frac{1}{2}}
   (2J_{2}+1)^{\frac{1}{2}}\nonumber\\[1mm]
& &\left \{ \begin{array}{ccc}
             J_{\gamma}&j_{2}&j_{1}\\
             S_Q&J_{1}&J_{2}
           \end{array} \right \}
   \langle J_{\gamma}mJ_{2}J_{2}^{z}\mid J_{1}J_{1}^{z}\rangle\label{pho6j}
\end{eqnarray}
The symbols appearing in (\ref{pho6j}) are explained at the end of
Sec.\ref{pi-trans}. The reduced matrix elements $M_{J_{\gamma}}$ correspond
to the multipole amplitudes $f^{J_{\gamma}}$ in Eq.(\ref{pho-trans}). As
discussed in the beginning of this subsection parity determines whether the
transition $M^{\gamma}(J_{\gamma})$ is a magnetic or electric multipole
transition.

\newpage
\vspace*{1cm}\hspace*{3cm}
\section{Semileptonic Decays}
\vspace*{2cm}
\newcommand{\xht}{\mbox{$\vert H_{-\frac{1}{2}t}\vert^2$}}
\newcommand{\xhti}{\mbox{$\vert H_{\frac{1}{2}t}\vert^2$}}
\newcommand{\hpli}{\mbox{$\vert H_{\frac{3}{2}\frac{1}{2}}\vert^2$}}
\newcommand{\hmn}{\mbox{$\vert H_{-\frac{1}{2}-1}\vert^2$}}
\newcommand{\hmni}{\mbox{$\vert H_{-\frac{3}{2}-1}\vert^2$}}
\newcommand{\ho}{\mbox{$\vert H_{-\frac{1}{2}0}\vert^2$}}
\newcommand{\hpl}{\mbox{$\vert H_{\frac{1}{2}1}\vert^2$}}
\newcommand{\hoi}{\mbox{$\vert H_{\frac{1}{2}0}\vert^2$}}
\newcommand{\mbmax}{\mbox{{\scriptsize max}}}
\newcommand{\mbmin}{\mbox{{\scriptsize min}}}
\newcommand{\mbos}{\mbox{{\scriptsize off-shell}}}
\newcommand{\mlab}{\mbox{{\scriptsize lab}}}
\newcommand{\mmbmax}{\mbox{{\tiny max}}}
\newcommand{\zmin}{z_{\mbox{{\tiny min}}}}
\newcommand{\lc}{\Lambda_{c}^{+}}
\newcommand{\ksicp}{\Xi_{c}^{+}}
\newcommand{\ksicn}{\Xi_{c}^{0}}
\newcommand{\omec}{\Omega_{c}^{0}}
\newcommand{\ra}{\rightarrow}
\newcommand{\lam}{\Lambda_{c}^{+}}
\newcommand{\xip}{\Xi_{c}^{+}}
\newcommand{\xio}{\Xi_{c}^{0}}
\newcommand{\sip}{\Sigma^{+}}
\newcommand{\sio}{\Sigma^{0}}
\newcommand{\omo}{\Omega_{c}^{0}}
\newcommand{\kbar}{\overline{K^0}}
\newcommand{\xw}{x_W}
\newcommand{\vb}{v_b}
\newcommand{\ab}{a_b}
\newcommand{\bup}{b^\uparrow}
\newcommand{\bdo}{b^\downarrow}
\newcommand{\qup}{\bar{q}^\uparrow}
\newcommand{\qdo}{\bar{q}^\downarrow}

\subsection{Inclusive Semileptonic Rates}

  The main motivation for studying the inclusive semileptonic decays of
bottom hadrons is to learn more about the two fundamental constants of
the Standard Model $V_{cb}$ and $V_{ub}$. To extract these parameters, it
is important to have a precise calculation of the electron spectrum in the
inclusive decays $b\to\{c,u\}e\bar\nu$. The simplest description of
these processes assumes that the lepton spectrum is that of the free heavy
quark decay (FQD). The light quark is assumed to play no role at all and
is regarded as being just a ``spectator''.
Refinements of this approach include taking into account the one--loop
radiative corrections and of the internal motion (Fermi motion) of the heavy
quark inside of the hadron in the framework of a nonrelativistic bound--state
model. It is, however, only recently that a systematic calculation of
the nonperturbative corrections to this picture has become possible,
in terms of an expansion in powers of the inverse heavy quark mass
$m_b^{-1}$.
The aim of this chapter is to give a brief account of these developments.

  We begin by presenting the predictions of the FQD picture for the decay
of a polarized $b$ quark. This provides a lowest--order description to be
improved upon and is useful for understanding the gross features of the
lepton spectrum. The weak interaction Lagrangian responsible for the
decays we are considering is
\beq\label{a1}
{\cal L}_W=V_{jb} 2\sqrt{2}G_F J_\mu
   [\bar\ell\gamma^\mu{\textstyle \frac{1}{2}}(1-\gamma_5)\nu_\ell]
\eeq
where $J_\mu=\bar j\gamma_\mu\frac{1}{2}(1-\gamma_5)b$ is the charged
current and $j=u,c$, $\,\ell=e,\mu,\tau$. $V_{jb}$ is the corresponding
Kobayashi--Maskawa matrix element. This yields for the decay rate of
the process $b(m_bv)\to j(p_j)e(p_e)\bar\nu(p_\nu)$ the following
expression (under the assumption that the lepton mass vanishes $m_\ell=
0$)
\bea\label{a2}
\mbox{d}\Gamma=32G_F^2|V_{jb}|^2(p_e\cdot p_j)[(p_\nu\cdot v)+
(p_\nu\cdot s)] \mbox{d}Lips\,.
\eea
The heavy quark spin $s$ satisfies $v\cdot s=0$. The phase space element of
the final state particles is
\bea\label{a3}
\mbox{d}Lips &=& (2\pi)^4\delta^{(4)}(p_B-p_e-p_\nu-p_j)
 \frac{\mbox{d}^3p_e}{(2\pi)^32E_e}
 \frac{\mbox{d}^3p_\nu}{(2\pi)^32E_\nu}
 \frac{\mbox{d}^3p_j}{(2\pi)^32E_j}\nonumber\\
& &\to\frac{E_eE_\nu\mbox{d}E_e\mbox{d}\Omega_e\mbox{d}\Omega_\nu}
{8(2\pi)^5[m_b+E_e(\cos\theta_{e\nu}-1)]}\,.
\eea
We have chosen to parameterize the final state by giving the electron
energy $E_e$ and the flight directions of the electron and neutrino in the
$b$--quark rest frame with $\mbox{d}\Omega=\mbox{d}\cos\theta\mbox{d}\phi$.

  In some applications it might be more convenient to choose the neutrino
energy $E_\nu$ as an independent variable. This is done most easily by
choosing $p_e$ as the $z$--axis and specifying the neutrino direction by
giving $\theta_{e\nu}$ and a polar angle $\phi_\nu$ whose origin might be
chosen corresponding to the configuration where the
three vectors $p_e,p_\nu,s$ are coplanar. The direction of $p_e$ is given
with respect to $\vec s$ by $(\theta,\phi)$ and will be considered for the
moment being as fixed. Now $\theta_{e\nu}$ can be replaced by $E_\nu$ by
making use of
\bea\label{a4}
E_\nu=\frac{m_b^2-m_j^2-2m_bE_e}{2[m_b+E_e(\cos\theta_{e\nu}-1)]}\,,\qquad
\mbox{d}E_\nu=\frac{E_eE_\nu\mbox{d}\cos\theta_{e\nu}}
{m_b+E_e(\cos\theta_{e\nu}-1)}
\eea
and (\ref{a2}) can be rewritten as
\bea\label{a5}
\mbox{d}\Gamma=2G_F^2|V_{jb}|^2(m_b^2-m_j^2-2m_bE_\nu)(1+\cos\theta_{s\nu})
\frac{E_\nu\mbox{d}E_e\mbox{d}E_\nu\mbox{d}\Omega_e\mbox{d}\phi_\nu}{(2\pi)
^5}
\eea
with
\beq\label{a6}
\cos\theta_{s\nu}=\cos\theta_{e\nu}\cos\theta+\sin\theta_{e\nu}\sin\theta
\cos\phi_\nu
\eeq
It is a simple matter now to integrate over $\phi_\nu$. The
$\phi$--integration is also trivially performed. The result can be most
conveniently represented in terms of the scaled variables
\bea\label{a7}
y_e=\frac{2E_e}{m_b}\,,\qquad y_\nu=\frac{2E_\nu}{m_b}\,,\qquad
\rho=\frac{m_j^2}{m_b^2}
\eea
and is
\bea\label{a8}
\frac{\mbox{d}\Gamma}{\mbox{d}y_e\mbox{d}y_\nu\mbox{d}\cos\theta}=
\frac{G_F^2m_b^5|V_{jb}|^2}{4(2\pi)^3}y_\nu(1-\rho-y_\nu)
\left[1+\frac{2\cos\theta}{y_ey_\nu}\left(1-\rho-y_e-y_\nu
+\frac{1}{2}y_ey_\nu\right)\right]\,.
\eea
An integration over the neutrino energy within the limits $y_{\nu_{min}}=
1-\rho-y_e$ and $y_{\nu_{max}}=1-\rho/(1-y_e)$ gives
\bea\label{a9}
\frac{\mbox{d}\Gamma}{\mbox{d}y_e\mbox{d}\cos\theta}&=&
\frac{G_F^2m_b^5|V_{jb}|^2}{24(2\pi)^3}
\frac{y_e^2(1-\rho-y_e)^2}{(1-y_e)^3}
\left([(1-y_e)(3-2y_e)+\rho(3-y_e)]\right.\nonumber\\
& &\left.+\cos\theta[(1-y_e)(1-2y_e)-\rho(1+y_e)]
\right)\,.
\eea
The total decay rate is obtained after integrating this distribution over
$\theta$ and over $y_e$ from 0 to $1-\rho$ with the result
\beq\label{a10}
\Gamma_b=\frac{G_F^2m_b^5|V_{jb}|^2}{24(2\pi)^3}(1-8\rho+8\rho^3-\rho^4-12
\rho^2\log\rho)\,.
\eeq

   The corrections to these predictions are of three types: i) radiative
corrections of order $\alpha_s(m_b)$ due to hard gluon exchange between
the initial and final state quarks; ii) corrections of order $\Lambda/m_b$
due to the interaction of the heavy quark with the light quarks in the
hadron and their gluon field; iii) corrections proportional to the lepton
mass of order ${\cal O}(m_\ell^2/m_b^2)$.

   The radiative corrections have been computed in \cite{ACM,JK} and their
effect is to change the electron spectrum (\ref{a9}) into
\bea\label{a11}
\frac{\mbox{d}\Gamma}{\mbox{d}y_e}=\frac{\mbox{d}\Gamma^{(0)}}{\mbox{d}y_e}
\left(1-\frac{2\alpha_s}{3\pi}G(y_e,\rho)\right)\,.
\eea
Here $\mbox{d}\Gamma^{(0)}/\mbox{d}y_e$ is the FQD spectrum of an
unpolarized $b$ quark and $G(y_e,\rho)$ is in general a complicated function
of its arguments. For $\rho\simeq 0.08$, corresponding to a $b\to c$
transition, it is almost constant ($\simeq 2.5$) for $y_e<0.7$ and diverges
logarithmically for $y_e\to 1-\rho$. However, this divergence is not
very problematic in this case, because in this region the uncorrected
spectrum vanishes and furthermore, the bound--state effects average the
effect away.
The situation is more serious for the $b\to u$ case, where the divergence
is doubly--logarithmic
\beq\label{a12}
G(y_e,0)=\ln^2(1-y_e)+
{\textstyle \frac{7}{2}}\ln(1-y_e)+\mbox{terms regular as $y_e\to
1$}\,,
\eeq
and the FQD spectrum (\ref{a9}) does not vanish for $y_e=1$ (actually at this
point it has a very steep fall, which in the limit $m_u\to 0$ is described
by a step--function $\theta(1-y_e)$). The authors of \cite{ACCMM} have
therefore proposed to change the correction (\ref{a11}) into
\bea\label{a13}
\frac{\mbox{d}\Gamma}{\mbox{d}y_e}=\frac{\mbox{d}\Gamma^{(0)}}{\mbox{d}y_e}
\left(1-\frac{2\alpha_s}{3\pi}\tilde G(y_e,\rho)\right)\exp\left(-
\frac{2\alpha_s}{3\pi}\ln^2(1-y_e)\right)\,.
\eea
Here the exponential sums the double--logarithms of the form $\alpha_s^{n}
\ln^{2n}(1-y_e)$ (Sudakov logarithms) to all orders in $\alpha_s$ and
$\tilde G(y_e,0)$ is the remainder of $G(y_e,0)$ (\ref{a12}) after the first
term has been subtracted away. After this modification, the QCD corrected
spectrum vanishes at the end--point $y_e=1-\rho$ in a smoother way than a
step--function $\theta(1-y_e)$.
We mention that the ${\cal O}(\alpha_s)$ radiative corrections to the
electron spectrum (\ref{a9}) in semileptonic decays of polarized $b$
quarks have recently been calculated \cite{CJKK93}.

\begin{figure}
\vspace{7.5cm}
\caption[dummy19]{
Inclusive semileptonic $H_b\to X_c$ decays in QCD. The squared
transition amplitude is expressed as the absorptive part of the handbag
diagram where the $b$ quark interacts with the light degrees of freedom
in $H_b$ and the $c$ quark propagates in the background field of the
light constituents.}
\end{figure}

   The corrections proportional to $\Lambda/m_b$ are partly due to the
internal motion of the $b$ quark inside of the heavy hadron, which Doppler
shifts the FQD predictions (\ref{a9},\ref{a11}) and partly to the
interactions of the $b$ quark with the background gluonic field created by
the light constituents of the hadron (see Fig.12). Recent theoretical
advances have made it
possible to study these effects in an essentially model--independent
way \cite{CGG,6}. We present in the following a brief account of the method,
as applied to the inclusive semileptonic decays of polarized $\Lambda_b$
baryons.
 The decay rates and the lepton angular
distributions can be expressed in terms of the hadronic tensor
\bea\label{a14}
\lefteqn{W^{\mu\nu}=(2\pi)^3\sum_X\delta^{(4)}(p_{\Lambda_b}-q-p_X)
\langle\Lambda_b(v,s)|J^{\mu\dagger}|X\rangle
\langle X|J^\nu|\Lambda_b(v,s)\rangle=}\\
& &-g^{\mu\nu}W_1+v^\mu v^\nu W_2-i\epsilon^{\mu\nu\alpha\beta}v_\alpha
q_\beta W_3+q^\mu q^\nu W_4+(q^\mu v^\nu+q^\nu v^\mu)W_5\nonumber\\
& &-q\cdot s[-g^{\mu\nu}G_1+v^\mu v^\nu G_2-i\epsilon^{\mu\nu\alpha\beta}
v_\alpha q_\beta G_3+q^\mu q^\nu G_4+(q^\mu v^\nu+q^\nu v^\mu) G_5]
\nonumber\\
& &(s^\mu v^\nu+s^\nu v^\mu)G_6+(s^\mu q^\nu+s^\nu q^\mu)G_7+
i\epsilon^{\mu\nu\alpha\beta}v_\alpha s_\beta G_8+
i\epsilon^{\mu\nu\alpha\beta}q_\alpha s_\beta G_9\,,\nonumber
\eea
with $q$ the total momentum of the lepton pair, $s$ the spin vector of
the $\Lambda_b$ and $W_{1-5}(q^2, q\cdot v)$, $G_{1-9}(q^2, q\cdot v)$
invariant form--factors.

  The hadronic tensor $W^{\mu\nu}$ (\ref{a14}) can be expressed as the
discontinuity across the physical cut of the transition amplitude
\bea\label{a15}
T^{\mu\nu}=-i\int\mbox{d}^4xe^{-iq\cdot x}\langle\Lambda_b(v,s)|
\mbox{T}J^{\mu\dagger}(x)J^\nu|\Lambda_b(v,s)\rangle
\eea
which can be decomposed into covariants $T_{1-5}$, $S_{1-9}$ similarly to
(\ref{a14}) and are related to $G_i$ by
\beq\label{a16}
W_i=-\frac{1}{\pi}\mbox{Im}\,T_i\,
     \qquad G_i=-\frac{1}{\pi}\mbox{Im}\,S_i\,.
\eeq
The physical cut
for the inclusive process under consideration extends from $q\cdot v=
\sqrt{q^2}$ up to $q\cdot v=(m_{\Lambda_b}^2+q^2-m_{j_{min}}^2)/
(2m_{\Lambda_b})$ with $m_{j_{min}}$ the mass of the lightest hadron
containing a $j$--type quark.

  The time ordered product in (\ref{a15}) can be expanded into an operator
product
expansion (OPE) \cite{CGG}. In momentum space, the operators of higher
dimension which appear are suppressed by increasing (inverse) powers of
$m_b$ and $\Delta=(m_bv-q)^2-m_j^2$,
so the OPE can be meaningfully applied only in regions of the phase space
where the second parameter is large enough, in comparison with the QCD scale
$\Lambda$. In particular, this is not true near the boundaries of the
allowed phase space, where special methods have been devised to deal with
this problem \cite{SF}.

   The leading order term in the OPE consists of the operators $\bar b
\gamma_\mu b$ and $\bar b\gamma_\mu\gamma_5 b$, whose matrix elements
in a polarized $\Lambda_b$ state are given by $v_\mu$ and $s_\mu$,
respectively, to leading order in $1/m_b$. By taking the imaginary part of
the transition matrix element (\ref{a15}) according to (\ref{a16}), the old
results of the FQD picture (\ref{a2}-\ref{a10}) are recovered. This procedure
is shown in graphical form in Fig.12.

  The operators contributing to the next order in the $1/m_b$ expansion
have dimension five. Actually, their contribution can be shown to vanish,
which gives the remarkable prediction that the FQD results obtain
corrections only at order $\Lambda^2/m_b^2$ \cite{CGG}. This is a constraint
set by QCD on any model--dependent description of the inclusive semileptonic
 of heavy
hadrons. It has been recently shown \cite{true} that the ACCMM model
\cite{ACCMM} can accomodate this feature.

  Up to now, the OPE method has been pushed up to order $1/m_b^2$.
The corrections which appear to this order arise from two sources:
i) from the matrix element of the kinetic energy of the $b$ quark
$\hbbar (iD)^2\hb$; ii) the corrections to the matrix element of the
axial current in a polarized $\Lambda_b$ state. The first one has been
calculated with the help of constituent quark models or from QCD sum rules,
whereas the second one has only been
estimated from a constituent quark model in \cite{FN93}.
For the case of the inclusive bottom meson decays, there
is one more source of corrections, due to the chromomagnetic interaction
operator $\hbbar g\sigma\cdot F\hb$, whose matrix elements can be
determined in a model--independent way from the hyperfine splitting
$m_{B^*}-m_B$. The effect of these corrections on the lepton spectrum is
shown on Fig.13 for the two cases $b\to c$ and $b\to u$.

\begin{figure}
\vspace{8cm}
\caption[dummy20]{ Electron spectrum in inclusive semileptonic decays of
bottom mesons for $b\to c$ and $b\to u$ transitions. The spectra are
normalized to $\Gamma_b(\rho=0)$. Full line: the free quark decay
result (\protect\ref{a9}). Dashed line: ${\cal O}(1/m_b^2)$ nonperturbative
corrections included.}
\end{figure}

  Finally, the last type of corrections to the FQD predictions which
we discuss are due to a nonvanishing lepton mass. As said before, they
are of order ${\cal O}(m_\ell^2/m_b^2)$. Therefore they are most important
for the case $\ell=\tau$ and, to a lesser extent, for $\ell=\mu$.
The combined corrections of order  ${\cal O}(\Lambda^2/m_b^2,m_\ell^2/m_b^2)$
have been calculated, to all orders in $m_\ell^2/m_b^2$, in \cite{tau}.
The combined effect of these corrections is to lower the total FQD decay rate
for $B\to X_c\tau\bar\nu_\tau$ by 6.5\%, whereas the neglect of the $\tau$
mass would have given a decrease of only 3.7\%.

  On the experimental side, the totally inclusive semileptonic branching
ratio for $\Lambda_{c}\to e^{+} + X $ has been measured by the
 Mark II collaboration \cite{VELLA} as $(4.5 \pm 1.7 )\%$.
 This then leads to the inclusive rate
\begin{eqnarray}
 \Gamma_{\Lambda_{c}}^{sl}(\mbox{inclusive})&=&
 (22.5 \pm 8.7)\times 10^{10}s^{-1} \mbox{ . }
\end{eqnarray}
 The corresponding inclusive semileptonic charm meson rates
 are \cite{pdg}
 $ \Gamma_{D^{\pm}}^{sl}(\mbox{inclusive})=
 (16.1 \pm 1.54)\times 10^{10}s^{-1} $ and
 $ \Gamma_{D^{0}}^{sl}(\mbox{inclusive})=
 (18.5 \pm 2.9)\times 10^{10}s^{-1} $.

    We mention that the Mark II collaboration \cite{VELLA}
 has also given results on semileptonic semi--inclusive
 branching ratios. They find $BR_{\Lambda _{c}\to p e^{+} +X}=
 (1.8\pm 0.9)\%$ including protons from $\Lambda $ decay,
 and $BR_{\Lambda _{c}\to \Lambda e^{+} +X}=
 (1.2 \pm 0.4)\%$.

\subsection{Exclusive Semileptonic Decays}
\subsubsection{Amplitudes, Rates and Angular Decay Distributions}

 Let us first stake out the spin complexity of the problem that one is
concerned
with. This entails an enumeration of the number of independent amplitudes
in exclusive semileptonic decays
and a discussion of how to measure them. To set the scope we remind the
reader that the separate determination of the three respective form
factors in semileptonic B$\to \mbox{D}^{\ast }$ \citer{alb,cleo}
and D$\to \mbox{K}^{\ast }$ \cite{cv}
transitions through angular measurements has been eminently
important for the development of a plausible theory for heavy meson
transition form factors \citer{cj,cp}.

  Let us begin our discussion by defining a standard set of invariant form
factors for the weak current--induced baryonic $1/2^{+} \to 1/2^{+}$
and $1/2^{+} \to 3/2^{+}$ transitions. One has
\begin{eqnarray}\label{5-2}
 \langle \Lambda _{s}(P_{2})\vert J_{\mu}^{V+A} \vert
 \Lambda _{c}(P_{1})\rangle
 &=& \bar u(P_{2})[ \gamma _{\mu}(F_{1}^{V}+F_{1}^{A} \gamma_{5})
        +i\sigma _{\mu \nu }q^{\nu }(F_{2}^{V}+F_{2}^{A}\gamma_{5})
\nonumber\\
 & & \mbox{\hspace*{1.5cm}}+q_{\mu }(F_{3}^{V}+F_{3}^{A}\gamma _{5})]u(P_{1})
 \\[2mm]\label{5-3}
 \langle \Omega ^{-}(P_{2})\vert J_{\mu}^{V+A} \vert
 \Omega _{c}(P_{1})\rangle
 &=& \bar u^{\alpha }(P_{2})[ g_{\alpha \mu}(G_{1}^{V}+G_{1}^{A} \gamma _{5})
    +P_{1\alpha }\gamma _{\mu }(G_{2}^{V}+G_{2}^{A} \gamma _{5})\nonumber\\
& & + P_{1\alpha }  P_{2\mu }(G_{3}^{V}+G_{3}^{A} \gamma _{5})
 +P_{1\alpha}q_\mu(G_{4}^V+G_4^A \gamma_5)]\gamma_5 u(P_{1})
\end{eqnarray}
where $J^{V}_{\mu }$ and $J^{A}_{\mu }$ are vector and axial vector
currents and $q_{\mu }=(P_{1}-P_{2})_{\mu }$ is the 4--momentum
transfer. We have found it convenient to use particle
labels from baryonic $c\rightarrow s$ decays instead of generic names. The
form factors $F_{i}^{V,A}$ and $G_{i}^{V,A}$ are functions of $q^{2}$.

   The invariants $F_{3}^{V},F_{3}^{A},G_{4}^{V}$ and $G_{4}^{A}$
multiplying $q_\mu$ contribute to semileptonic decays at ${\cal O}(m^2_\ell
/q^2)$ and are thus difficult to measure. Muon effects have been investigated
in the corresponding semileptonic D$\to $K($\mbox{K}^{\ast}$) decays and
have been
found to be $\leq 5\%$ in the total rate \cite{ck}. Muon effects in charm
baryon decays are of similar size. The biggest effect occurs
for the longitudinal rate, where the muon mass effect amounts to ${\cal O}
(10\%)$, and is largest at small $q^{2}$. This is different in semileptonic
$b\to c$ decays where lepton mass effects can be conveniently probed
in the $\tau$--channel \cite{cq,cr}.
The branching ratio into the $\tau$--modes is typically $\simeq$ 20-30\%
of the $e$--mode and now there can be large contributions from the
$q_\mu$--form factor terms. Nevertheless, the invariants $F_{3}^{V,A}$ and
$G_{4}^{V,A}$ multiplying $q_\mu$ are important in
nonleptonic decays where they contribute through the so--called factorizing
contributions in the nonleptonic ${\mbox{B}}_{Q_{1}}\to {\mbox{B}}
_{Q_{2}}+\mbox{M}(0^{-})$ decays.

Rates and angular decay distributions are given
 in terms of bilinear forms of the form factors. We
 first consider the (cascade) decay of an unpolarized
 charm baryon $\mbox{B}_{c}\to \mbox{B}(\to \mbox{B}'\mbox{M})
 +l^{+}+\nu _{l}$ where the cascade decay
 $\mbox{B}(\to \mbox{B}'\mbox{M})$
 is used as an analyzer of the polarization
 of the daughter baryon B. For semileptonic
 $1/2^{+} \to 1/2^{+}$ transitions the full
 four--fold decay distribution
 differential in the momentum transfer squared $q^{2}$
 and the angles $\Theta , \chi $ and $ \Theta _{B}$
 shown in Fig.14 reads \cite{cs,polen}
\begin{eqnarray}\label{this}
 &&\frac{\mbox{d}\Gamma (\Lambda _{c}^{+}\to \Lambda (\to p\pi ^{-})+l^{+}
  +\nu _{l})}{\mbox{d}q^{2} \mbox{d}\cos\Theta \mbox{d}\chi
  \mbox{d}\cos\Theta _{\Lambda }}
 =B_{\Lambda \to p\pi ^{-}} \frac{1}{2\pi }
 \nonumber \\
 &&\left[ \frac{3}{8}(1+\cos^{2}\Theta )
  \frac{\mbox{d}\Gamma _{U}}{\mbox{d}q^{2}}
  (1+\alpha _{c}^{U}\alpha _{\Lambda} \cos\Theta _{\Lambda})
  +\frac{3}{4} \sin^{2} \Theta \frac{\mbox{d}\Gamma _{L}}{\mbox{d}q^{2}}
  (1+\alpha _{c}^{L}\alpha _{\Lambda} \cos\Theta _{\Lambda}) \right.
 \nonumber \\
 &&-\frac{3}{2\sqrt{2}} \sin(2\Theta )\cos\chi \sin\Theta _{\Lambda})
 \alpha _{\Lambda} \frac{\mbox{d}\Gamma _{I}}{\mbox{d}q^{2}}
 -\frac{3}{4} \cos\Theta \frac{\mbox{d}\Gamma _{P}}{\mbox{d}q^{2}}
  (1+\frac{\mbox{d}\Gamma _{U}}{\mbox{d}\Gamma _{P}}
  \alpha_\Lambda\cos\Theta_\Lambda)
 \nonumber \\
 && \left.
 \frac{3}{\sqrt{2}}\sin\Theta\cos\chi\sin\Theta_\Lambda\alpha_\Lambda
 \frac{\mbox{d}\Gamma _{A}}{\mbox{d}q^{2}} \right]
\end{eqnarray}
 where
\begin{eqnarray}\label{5-4}
 \frac{\mbox{d}\Gamma _{i}}{\mbox{d}q^{2}}=\frac{1}{2}
 \frac{G^{2}}{(2\pi )^{3}}
 \vert V_{cs} \vert ^{2} \frac{pq^{2}}{12M_{1}^{2}}H_{i}
\end{eqnarray}
 and where $G$ is the Fermi coupling constant
 ($G=1.026\cdot 10^{-5}m_{\mbox{p}}^{-2}$) and $p$ is the momentum of the
daughter baryon in the B$_c$ rest frame, $p=\sqrt{Q_{+}Q_{-}}/2M_{1}$ with
 $Q_{\pm }=2(P_{1}P_{2} \pm M_{1}M_{2}) =(M_1\pm M_2)^2-q^2 $.
 Again we have used particle labels in Equation (\ref{this}) instead of
 a generic notation.

\begin{figure}
\vspace{7.5cm}\caption[dummy21]{ Definition of polar angles $\theta_\Lambda$
and $\theta$ and azimuthal angle $\chi$ in the decay $\Lambda_c^+\to\Lambda
(\to p\pi)+W^+(\to\ell^+\nu_\ell)$. For $b\to c$ decays change the lepton
side to $W^-(\to\ell^-\bar\nu_\ell)$.}
\end{figure}

The helicity rates (or structure functions) $H_{i}$ in Equation (\ref{this})
 are defined as follows
\begin{eqnarray}
 H_{U}&=&|H_{\frac{1}{2},1}|^2+|H_{-\frac{1}{2},-1}|^2\nonumber \\
 H_{L}&=&|H_{\frac{1}{2},0}|^2+|H_{-\frac{1}{2},0}|^2\nonumber \\
 H_{I}&=&\frac{1}{2}\mbox{Re}\,\left( H_{-\frac{1}{2},0}H_{\frac{1}{2},1}^*-
         H_{\frac{1}{2},0}H_{-\frac{1}{2},-1}^*\right)\nonumber\\
 H_{P}&=&|H_{\frac{1}{2},1}|^2-|H_{-\frac{1}{2},-1}|^2\nonumber \\
 H_{A}&=&\frac{1}{2}\mbox{Re}\,\left( H_{-\frac{1}{2},0}
        H_{\frac{1}{2},1}^*+H_{\frac{1}{2},0}H_{-\frac{1}{2},-1}^*
        \right)\label{5-5}
\end{eqnarray}
 where the $H_{\lambda_2,\lambda_W}=H_{\lambda_2,\lambda_W}^V+
 H_{\lambda_2,\lambda_W}^A$ are the
 helicity amplitudes of the current induced transition,
 $\lambda _{W}$ is the helicity of the current
 ($\lambda _{W}=0$ longitudinal, $\lambda _{W}=\pm 1$ transverse )
 or, equivalently, of the off--shell W--boson and $\lambda_2$ is the helicity
of the daughter baryon. The relation between the set of helicity
 and invariant form factors is given by
\bea\label{5-6}
\sqrt{q^2}H_{\frac{1}{2},0}^V & = & \sqrt{Q_-}\left((M_1+M_2)F_1^V-
    q^2F_2^V\right)\\\label{5-7}
H_{\frac{1}{2},1}^V & = & \sqrt{2Q_-}\left(-F_1^V+
    (M_1+M_2)F_2^V\right)\\\label{5-8}
\sqrt{q^2}H_{\frac{1}{2},0}^A & = & \sqrt{Q_+}\left((M_1-M_2)F_1^A+
    q^2F_2^A\right)\\\label{5-9}
H_{\frac{1}{2},1}^A & = & \sqrt{2Q_+}\left(-F_1^A-
    (M_1-M_2)F_2^A\right)
\eea
 The remaining
helicity amplitudes can be obtained with the help of the parity relations
\beq\label{par}
H_{-\lambda_2,-\lambda_W}^{V(A)} = +(-)H_{\lambda_2,\lambda_W}^{V(A)}\,.
\eeq
The labeling of the helicity rates $H_i$ describes the polarization of the
off--shell $W$ boson in the decay: U (unpolarized transverse), L
(longitudinal), I (transverse--longitudinal interference), P (parity odd),
A (parity asymmetric).
For semileptonic decays involving the leptons $\ell^-+\bar\nu_\ell$ (as
in $\bar c\to\bar s$ and $b\to c$ transitions) the sign in the last two
terms of Eq.(\ref{this}) have to be reversed.

We have also introduced $\alpha_c^U$ and $\alpha_c^L$, the $q^{2}$--dependent
transverse and longitudinal asymmetry parameters
\bea\label{5-11}
\alpha_{c}^{U}&=&\frac{|H_{\frac{1}{2},1}|^2-|H_{-\frac{1}{2},-1}|^2}
                       {|H_{\frac{1}{2},1}|^2+|H_{-\frac{1}{2},-1}|^2}
\\\label{5-12}
\alpha_c^L&=&\frac{|H_{\frac{1}{2},0}|^2-|H_{-\frac{1}{2},0}|^2}
                  {|H_{\frac{1}{2},0}|^2+|H_{-\frac{1}{2},0}|^2}\,.
\eea
 $\alpha_\Lambda$ is the asymmetry parameter in the parity--violating
 nonleptonic decay $\Lambda \rightarrow p +\pi^{-}$ defined in analogy to
 Eq.(\ref{5-12}). Its experimental value is $\alpha_{\Lambda}=0.64$
\cite{pdg}.
 Triple, double and single decay distributions as well
 as the rate may be obtained from Eq.(\ref{this}) by the appropriate
 integrations. For example, integrating over the azimuthal angle $\chi$ and
 the lepton--side polar angle $\theta$ one obtains
\beq\label{polar}
\frac{\mbox{d}\Gamma}{\mbox{d}q^2\mbox{d}\cos\Theta_\Lambda}\propto
1+\alpha\alpha_\Lambda\cos\Theta_\Lambda
\eeq
where the asymmetry parameter $\alpha$ is defined by
\beq\label{5-14}
\alpha=\frac{|H_{\frac{1}{2},1}|^2-|H_{-\frac{1}{2},-1}|^2+
              |H_{\frac{1}{2},0}|^2-|H_{-\frac{1}{2},0}|^2}
            {|H_{\frac{1}{2},1}|^2+|H_{-\frac{1}{2},-1}|^2+
              |H_{\frac{1}{2},0}|^2+|H_{-\frac{1}{2},0}|^2}\,.
\eeq

   The asymmetry parameter $\alpha$ is nothing but the longitudinal
(``alignment'') polarization of the daughter baryon $\Lambda$ which is
being analyzed by its subsequent decay according to the above polar angle
distribution (\ref{polar}). For example the ARGUS \cite{ARGUS94}
 and CLEO \cite{CLEO94} Collaborations have
recently measured the mean of the asymmetry parameter $\alpha$ in the
semileptonic $\Lambda_c\to\Lambda+\ell^++\nu_\ell$ decay using the above
polar decay distribution. They found a large negative value of the asymmetry
parameter $\alpha$ in agreement with earlier theoretical predictions. We
shall return to this point in Sec.5.2.2.

 In Eq.(\ref{this}) we have assumed that the form factors and
 helicity amplitudes are real since the physical threshold is at
 $q^{2}=(M_{1}+M_{2})^{2} > q^{2}_{max}=(M_{1}-M_{2})^{2}$.
 We have thus omitted so--called T--odd contributions in the
 decay distribution which are proportional
 to $\sin\Theta \sin\chi \sin\Theta_{\Lambda }$ and
 $\sin(2\Theta ) \sin\chi \sin\Theta_{\Lambda }$ \cite{cs}.
 We note in passing that the presence of such contributions
 could signal possible CP--violations in the decay process \cite{polen}.

   The structure of the decay distribution Eq.(\ref{this}) is quite
similar to the corresponding four--fold decay distribution for the cascade
 decay D$\to \mbox{K}^{\ast }$($\to $K$\pi$)$+l^{+}+\nu _{l}$
\cite{ck},\citer{cn,cp} which has been proven so useful in
disentangling the form factor structure in the semileptonic D$\to
\mbox{K}^{\ast }$
 decays \cite{ct,cv}. The angular distribution Eq.(\ref{this})
 defines a set of eight observables which are bilinears in the
 four independent $q^{2}$--dependent real form factors.
 A measurement of these eight observables would
 considerably overdetermine the form factors.
 Note though that the complexity of the problem is reduced
 close to the phase space boundaries. At zero recoil
 $q^{2}\approx q^{2}_{max}$ only the s--wave contribution
 remains, and at $q^{2}\approx 0$ only the longitudinal contribution
survives. The relevant dynamical information may be extracted by either
one of the following methods:
 (i) moment analysis
 (ii) analysis of suitably defined asymmetry ratios as in \cite{cu} or
 (iii) angular fits to the data as in \cite{cv} depending
 on the quantity and quality of the data. All of this carries over to the
$b\rightarrow c$ sector with the requisite sign changes as mentioned after
Eq.(\ref{par}).

An additional set of polarization observables
 can be defined for the decay of polarized charm and bottom baryons.
 For example, bottom and charm quarks produced on the $Z^0$ resonance
are 94\% and 67\% negatively polarized. It is quite likely
that some of this polarization is retained when the bottom and charm quarks
fragment into bottom and charm baryons. This will be discussed in more
detail in Sec.5.3.. Also, hadronically produced $\Lambda$'s have been
observed to be polarized where the polarization necessarily has to be
transverse to the production plane because of parity invariance in the
production process. It may well be that hadronically produced $\Lambda_c^+$'s
show a similar polarization effect \cite{cw,cx}. Also, charm baryons from
weak decays of bottom baryons are expected to be polarized.

For polarized $\Lambda _{c}$--decays one orients the decay products
 and the subsequent decay of the daughter baryon
 relative to the $\Lambda _{c}$ polarization as drawn in Fig.15
  where the orientation angles
 $\Theta _{p}, \Theta _{\Lambda }$ and $\chi $ are defined.
 For the corresponding four--fold angular decay distribution one finds
\begin{eqnarray}\label{5-15}
 &&\frac{\mbox{d}\Gamma (\Lambda _{c}^{\uparrow }\to \Lambda
  (\to p\pi ^{-})+l^{+}
  +\nu _{l})}{\mbox{d}q^{2}
  \mbox{d}\cos\Theta _{p} \mbox{d}\chi \mbox{d}\cos\Theta _{\Lambda}}
 =\frac{1}{8\pi } B_{\Lambda \to p\pi ^{-}} \nonumber \\
 && \left[ \frac{\mbox{d}\Gamma _{U+L}}{\mbox{d}q^{2}}
  +\alpha_\Lambda\cos\Theta_\Lambda
  \left( \alpha _{c}^{U} \frac{\mbox{d}\Gamma _{U}}{\mbox{d}q^{2}}
   +\alpha _{c}^{L} \frac{\mbox{d}\Gamma _{L}}{\mbox{d}q^{2}}
   \right)\right.\nonumber \\
 &&-P_{c}\cos\Theta_p\left(\alpha _{c}^{U}
   \frac{\mbox{d}\Gamma _{U}}{\mbox{d}q^{2}}
   -\alpha _{c}^{L} \frac{\mbox{d}\Gamma _{L}}{\mbox{d}q^{2}} \right)
   +P_{c} \alpha _{\Lambda}\cos\Theta_p\cos\Theta_\Lambda
  \left( -\frac{\mbox{d}\Gamma _{U}}{\mbox{d}q^{2}}
  +\frac{\mbox{d}\Gamma _{L}}{\mbox{d}q^{2}} \right)
 \nonumber \\
 && \left. -P_c\alpha_\Lambda\sin\Theta_{p}\sin\Theta _{\Lambda}
  \cos\chi \frac{\mbox{d}\Gamma _{LI}}{\mbox{d}q^{2}} \right]
\end{eqnarray}
 where $P_{c}$ is the degree of polarization  of the $\Lambda _{c}^{+}$.
 The longitudinal interference rate is
 $\mbox{d}\Gamma _{LI}/\mbox{d}q^{2}=2\mbox{Re}\,(H_{\frac{1}{2},0}
 H_{-\frac{1}{2},0}^*)$.
 A corresponding four--fold angular decay distribution formula can be
 derived for the lepton--side \cite{polen}.

\begin{figure}
\vspace{8cm}\caption[dummy22]{Definition of polar angles $\theta_\Lambda$
and $\theta_p$ and azimuthal angle $\chi$ in the decay of a polarized
$\Lambda_c^+\to\Lambda(\to \pi^-)+X$. The left plane is determined by
polarization vector $\vec P_{\Lambda_c}$ of the $\Lambda_c$. The unobserved
state X stands for $W_{off-shell}$ in semileptonic decays and for a meson
in nonleptonic decays.}
\end{figure}

 Measurements of the angular decay distribution relative to
 the initial spin polarization vector would allow
 one to either measure the degree and sign of the polarization
 of the $\Lambda _{c}$ if its decay structure is known,
 or, vice versa, if $P_{c}$ were known, further constrain the
 decay amplitudes of the $\Lambda _{c}$.
 The information contained in the "decay"
 $W^{+} \to \ell^{+}+ \nu _{\ell}$ has
 not been used in Equation (\ref{5-15}),
 i.e. the angular dependence on the orientation angles $(\Theta ,\chi ')$ of
the $W^+ \to \ell^+ +\nu_\ell$ decay have been integrated out.
As the lepton--side angular distribution goes unanalyzed the distribution
(\ref{5-15}) holds for both $\Lambda_{c} \rightarrow \Lambda$ and
$\Lambda_{b}\rightarrow \Lambda_{c}$ transitions without any sign change.
If this angular dependence is kept one would have a six--fold differential
distribution. Corresponding decay distributions for semileptonic $1/2^{+}
\to 3/2^{+}$
transitions (polarized and unpolarized) can be found in \cite{polen}.
 Let us mention that the decay distributions
 Equations (\ref{this}) and (\ref{5-15}) can also be derived using a
 frame independent representation in terms of spin and
 momenta correlations \cite{cy}.

\subsubsection{Model Results $c\to s$}

   The accessible $q^2$--range in semileptonic charm baryon decays is not
small ($m_\ell^{2}\leq q^{2}\leq (M_{1}-M_{2})^{2}$). One thus has a large
experimental leverage to study the $q^2$--dependence of the form factors.
Also, the full spin dependent form factor structure can be investigated as
the C.M. momentum $p=|\vec p|$ becomes large enough when $q^2$ moves away
from the zero recoil point (or pseudo--threshold point) $q^2=(M_1-M_2)^2$
to populate all partial waves in the decay $B_1\to B_2 + W_{off-shell}
(q^2)$.

   In fact the momentum dependence of the semileptonic transitions close to
the zero recoil edge of phase space can be conveniently classified by doing
a partial wave analysis of the two--body decay of $1/2^+\to 1/2^++W$ as
drawn in Fig.16. Since the $W$ is off--shell it has four degrees of freedom
for each of its vector and axial components. The $J^P$ content of the
off--shell $W$ are $J^P=(1_V^-,1_A^+)$ for the spin 1 pieces and $J^P=(0_V^+
,0_A^-)$ for the spin 0 piece. We remind the reader that the spin 0 pieces
are not active in the decay in the limit of zero lepton mass.

\begin{figure}
\vspace{7cm}\caption[dummy23]{ Partial wave analysis of quasi--two body decay
$1/2^+\to 1/2^++W_{off-shell}$.}
\end{figure}

One can then determine the partial wave content of the decay $1/2^+\to 1/2^+
+W$. For the spin 1 part one has (in brackets the final state  spin sum
$S=S_1+S_2$ )
\bea\label{nuclear1}
{\textstyle \frac{1}{2}}^+\to{\textstyle \frac{1}{2}}^+
+ 1_V^-\,:& &\mbox{p--wave } (S=1/2,3/2)
 \quad \mbox{``first forbidden Fermi transition''}\nonumber\\
{\textstyle \frac{1}{2}}^+\to{\textstyle \frac{1}{2}}^+
+ 1_A^+\,:& &\mbox{s--wave } (S=1/2)
   \quad\hspace{9mm}\mbox{``allowed Gamow--Teller transition''}\\
& &\mbox{d--wave } (S=3/2)\quad\hspace{9mm}\mbox{``second forbidden
Gamow--Teller transition''}\nonumber
\eea
whereas for the spin 0 pieces one has
\bea\label{nuclear2}
{\textstyle \frac{1}{2}}^+\to{\textstyle \frac{1}{2}}^+
+ 0_V^+\,:& &\mbox{s--wave } (S=1/2)
  \quad\hspace{7.5mm}\mbox{``allowed Fermi transition''}\\
{\textstyle \frac{1}{2}}^+\to{\textstyle \frac{1}{2}}^+
+ 0_A^-\,:& &\mbox{p--wave } (S=1/2)
  \quad\hspace{7.5mm}\mbox{``first forbidden Gamow--Teller
 transition''}\nonumber
\eea

  Let us limit our discussion to the case $m_\ell=0$ which is a very good
approximation for $e$ and a good approximation for $\mu$ for $c\to s$
decays. There are then two vector and axial amplitudes
each for $1/2^+\to 1/2^++W(1_V^-)$ and $1/2^+\to 1/2^++W(1_A^-)$ in agreement
with the number of  covariant form factor in Eq.(\ref{5-2}) when the $q_\mu$
form factors are disregarded. As $p\to 0$ and the phase space
closes\footnote{Readers old enough will remember that the zero recoil point
$q_{max}^2=(M_1-M_2)^2$ used to be also called the pseudo--threshold.
``Pseudo'' because phase space closes when the zero recoil point is
approached in contrast to the normal threshold where phase space opens.}
only the axial vector s--wave contribution in $1/2^+\to 1/2^++W(1_A^+)$
survives. We want to call to mind that in nuclear physics parlance the vector
and axial vector transitions are referred to as Fermi and Gamow--Teller
transitions, respectively. They are further classified according to their
partial wave threshold behaviour by ``allowed'', ``first forbidden'', etc.
The nuclear physics classification has been included in Eq.(\ref{nuclear1},
\ref{nuclear2}). In contrast to nuclear physics transitions (where the
$Q$--values of the transitions are comparable to the electron mass and the
spin 0 components do contribute) the $Q$--values of the $c\to s$ transitions
are so big that only the spin 1 pieces are active, to a good approximation.
We shall return to the partial wave classification when discussing the
measurement of the KM matrix element $V_{bc}$ in the exclusive semileptonic
decay $\Lambda_b\to\Lambda_c+\ell^-+\bar\nu_\ell$ in Sec.5.2.3. Returning to
the $c\to s$ decays one can certainly state that for large enough momenta $p$
all partial waves in the decay $B_1\to B_2+W_{off-shell}(q^2)$ come into
play. This is different in ordinary hyperon decays, where the accessible
$q^2$--range is small and only the low partial waves contribute to any
significant degree. How to actually extract the various form factors through
polarization type measurements has been dealt with before in Section 5.2.1.

   To begin with we discuss rates. There have been a number of theoretical
attempts to model the form factors in the semileptonic $1/2 \to 1/2^+$ and
 $1/2^{+}\to 3/2^{+}$ transitions employing flavour symmetry and/or quark
models. In Table 8 we have listed the rate predictions of various models for
the semileptonic decay $\Lambda_{c}^{+}\to\Lambda +\ell^{+}+\nu_\ell,
\hspace{3mm}\Xi_{c}^{+}\to \Xi^{0}+\ell^{+}+\nu_\ell$ and the  $1/2^{+}\to
3/2^{+}$
decay $\Omega_c\to\Omega^-+\ell^++\nu_\ell\mbox{\hspace*{0.4cm}}(m_\ell=0)$.
   The first column contains early predictions which exploited SU(4) flavour
symmetry at $q^{2}=0$ to relate $\Delta C=1$ to the known $\Delta C=0$
amplitudes \cite{cz}. The results were then continued to $q^{2}=0$
by using suitable form factors. The predictions of \cite{da} are similar.
The rates come out too large due to the use of SU(4) at $q^{2}=0$. \cite{db}
and \cite{dc} have shown that there are large mass breaking corrections
to the SU(4) limit
at $q^{2}=0$ which brings the rates down as column 2 in Table 8
shows. Flavour symmetry should rather be applied at $q^{2}_{max}$.
Nonrelativistic quark model results calculated close to $q^{2}_{max}$
and then continued to $q^{2}\not=q^{2}_{max}$ via form factors
 were presented in \cite{dd,de}. Explicit quark model calculations as the
ones in \cite{dd,de} tends to show approximate unit overlap at zero recoil,
regardless of the masses of the quarks involved in the transition. Thus
they tend to mimic the HQET zero recoil normalization, even for $c
\rightarrow s$ decays. The model of \cite{df} uses such a HQET zero recoil
normalization and dipole form factors to continue to $q^2 \neq q^2_{max}$.
In order to be able to compare predictions we have taken the liberty to
rescale the results of \cite{dd} by taking away their assumed large QCD
correction which must be considered to be unrealistically large. The rate
values
for $\Lambda_c^+\to\Lambda$ then scatter around $20\times 10^{10} s^{-1}$
except for the rate prediction of Singleton \cite{de}. These rates would
imply a saturation of the total semileptonic inclusive rate by the exclusive
mode. The semi--inclusive rates quoted in Sec.5.1, however, preclude such a
possibility.
A more precise measurement of the experimental semileptonic $\Lambda_c^+\to
\Lambda$ (exclusive or inclusive) rate would help to pin down this issue.
\begin{table}
\begin{center}
\caption[dummy24]{\label{tablec12}
Exclusive semileptonic decay rates in units of
10$^{10}\,s^{-1}$.}
\renewcommand{\baselinestretch}{1.33}
\small \normalsize
\vspace{5mm}
\begin{tabular}{lrrrrrr}
\hline\hline
 & Buras     & Gavela  & PHGA     & PHGA & Singleton  & HK \\
 & \cite{cz} &\cite{db}& NRQM \cite{dd} &
  MBM \cite{dd} & \cite{de} & \cite{df} \\
\hline
 $\Lambda_c^+\to\Lambda$ & 60(20) & 15(5) & 17(5.6) & 13(4.3) & 10 &
22(7.3) \\
 $\Xi_c^+\to\Xi^0$       & 235(118)& 28(14) & 28(14) & 22(11) & 17 &
33(16.5) \\
 $\Omega_c^0\to\Omega^-$ &280 & 49 &   -- & -- & -- & 48 \\
 \hline\hline
\end{tabular}
\renewcommand{\baselinestretch}{1}
\small \normalsize
\end{center}
\end{table}

The calculation of Singleton \cite{de} (see also \cite{KQS89})
differs from the other calculations in one important respect
in that he has taken a spin--flavour suppression factor in the
$\Lambda_c\to\Lambda$ and $\Xi_c \to\Xi$ transitions into account
which other authors have not included. In order to understand the
spin--flavour suppression factor present in the calculation of \cite{de}
consider the amplitude $c_s$ of the spin 0 light diquark configuration
in the various charm and strange baryons. Whereas the spin 0 light
diquark configuration has the amplitude $c_s=1$ in $\Lambda_c$ and
$\Xi_c$ it is only one of the many possible diquark configurations
in the light baryons $\Lambda$ and $\Xi$. In order to determine the
amplitude of the spin 0 diquark configuration in the $\Lambda$ and
$\Xi$ state one appeals to SU(6) for guidance. Consider the spin--flavour
wave functions of the $\Lambda_c,\Xi_{c},\Lambda$ and $\Xi$. They are
given by (see e.g. \cite{CLOSE})
\bea
\Lambda_{c}^{\uparrow}&=&c^{\uparrow}[ud]_0\,,\qquad
\Xi_{c}^{+ \uparrow}=c^\uparrow [us]_0\\
\Lambda^\uparrow &=&
 {\textstyle \frac{1}{\sqrt3}}s^\uparrow [ud]_0
+{\textstyle \frac{1}{2\sqrt3}}d^\uparrow [us]_0
+{\textstyle \frac{1}{2\sqrt3}}u^\uparrow [ds]_0
-{\textstyle \frac{1}{2\sqrt3}}d^\uparrow \{us\}_0 \nonumber\\
                 &+&
 {\textstyle \frac{1}{2\sqrt3}}u^\uparrow  \{ds\}_0
+{\textstyle \frac{1}{\sqrt6}}d^\downarrow \{us\}_{+1}
-{\textstyle \frac{1}{\sqrt6}}u^\downarrow \{ds\}_{+1}\\
\Xi^{0 \uparrow} &=&
 {\textstyle \frac{1}{\sqrt2}}s^\uparrow [us]_0
-{\textstyle \frac{1}{\sqrt{18}}}s^\uparrow\{us\}_0
+{\textstyle \frac{1}{\sqrt9}}s^\downarrow \{us\}_{+1}
+{\textstyle \frac{1}{\sqrt9}}u^\uparrow\{ss\}_0
-{\textstyle \sqrt{\frac{2}{9}}}u^\downarrow \{ss\}_{+1}
\eea
where the suffix label denotes the $m$--quantum number of the diquark
states. The $[ud]_0$ is the totally symmetric spin 0--isospin 0 combination
$[ud]_0=\frac{1}{2}(u^\uparrow d^\downarrow -u^\downarrow d^\uparrow +
d^\downarrow u^\uparrow -d^\uparrow u^\downarrow )$ and the $\{us\}_{0,+1}$
are the corresponding totally symmetric spin 1--flavour symmetric
combinations with $m=0,+1$. The spin--flavour wave functions can be checked
to be normalized to 1. The transitions $\Lambda_c\to\Lambda$ and
$\Xi_c\to\Xi$ can be seen to acquire factors of $\sqrt{\frac{1}{3}}$ and
$\sqrt{\frac{1}{2}}$ in amplitude, respectively, from the $[ud]_0$ and
$[us]_0$ components in the $\Lambda$ and $\Xi^0$ wave functions. The
authors of \cite{db,dc,dd,df} instead used SU(8)--type wave functions
for the charm baryons $\Lambda_c$ and $\Xi_c$ treating all quarks
democratically,  i.e. they used wave
functions for $\Lambda_c$ and $\Xi_c$ that are identical to the above
$\Lambda$ and $\Xi$ wave functions except for replacing $s$ by $c$.
In such an approach there are no a priori spin--flavour suppression factors.
In order to be able to compare the rates we have accordingly multiplied the
high rates by factors of 1/3 and 1/2 and have included the lowered rates
in brackets in Table 8. One can only hope for more accurate measurements
on exclusive semileptonic charm baryon rates in the near future to be
able to resolve this issue. No such adjustment
is required for the transition $\Omega_c^0\to\Omega^-$. The semileptonic
rates for $\Omega_c^0\to\Omega^-+\ell^++\nu_\ell$ are predicted to be
quite large \cite{df}. This is basically because there are several
possibilities for the initial $c$ quark to make a transition to the final
$s$ quark, regardless of the model.

   Next we turn to the polarization type observables measurable through the
joint angular decay distributions Eqs.(\ref{this}) and (\ref{5-15}). First
observe that the angular decay distributions become quite uninteresting
close to zero recoil point $q^2\to q^2_{max}=(M_1-M_2)^2$ where the axial
vector s--wave contribution dominates ($H_{\lambda_2,\lambda_W}^V\to 0$,
$H_{\frac{1}{2},1}^A\to -\sqrt{2}H_{\frac{1}{2},0}^A$). For example, the
asymmetry Eq.(\ref{polar}) (or the polarization of the $\Lambda$) vanishes at
the zero recoil point and the corresponding polar angle distribution
becomes flat.
At the other end of phase space as $q^2\to 0$ (or more exactly $q^2\to
m_\ell^2$) the longitudinal helicity amplitudes dominate as
Eqs.(\ref{5-6}--\ref{5-9})
show. There exists a very interesting Heavy Quark Symmetry prediction for
the structure of heavy to light $\Lambda_c\to\Lambda_s$ transitions at $q^2=
0$. Take the relevant heavy to light form factor structure from
Eq.(\ref{4-122}) and substitute it into Eq.(\ref{5-14}). At $q^2=0$ one
finds $H_{\frac{1}{2}
,0}=0$, i.e. the daughter baryon $\Lambda$ is predicted to emerge 100\%
(negatively) polarized from the decay at $q^2=0$. For the mean value of
the polarization averaged over $q^2$ Ref.\cite{cs} quote a theoretical
range $-0.52$ to $-0.94$ (depending on the assumed form factor ratio),
$\langle\alpha\rangle=-0.82$ being a preferred value.

There have been two recent measurements of the mean polarization of the
$\Lambda$ in semileptonic $\Lambda_c\to\Lambda$ decays by ARGUS
\cite{ARGUS94} and CLEO \cite{CLEO94} who quote
\bea
\langle\alpha\rangle
&  = & \left\{\!\begin{array}{lc} -0.91 \pm 0.49 &
\mbox{ARGUS}~\cite{ARGUS94}\\[2mm]
 -0.89 {+ 0.17 + 0.09 \atop - 0.11 - 0.05}& \mbox{CLEO}~\cite{CLEO94}
\end{array}\right.
\eea
where we refer to the original papers for a discussion of the phase space
region in which the mean is taken. Both collaborations conclude that their
results imply that $\alpha$ is close to $-1$ at $q^2=0$ in agreement with
the Heavy Quark Symmetry prediction.

   One can check that the 100\% polarization prediction of Heavy Quark
Symmetry remains intact even if one includes $1/m_c$ and naively applied
$1/m_s$ corrections according to Eq.(\ref{50}--\ref{55}). This may be
taken as an indication of the stability of the HQET result. A
semiphenomenological analysis that includes terms of order $1/m_c$ in the
HQET expansion but still treats the $\Lambda$ as light leads to a very small
departure from $-1$ at $q^2=0$ \cite{MANNEL93}.

   It will not be an easy task experimentally to completely disentangle
the form factor structure of semileptonic baryonic $c\to s$ transitions
through the angular correlation measurements in Eq.(\ref{this}) and
(\ref{5-15}). However, one can hope for more and better data in the future.
The two recent measurements \cite{ARGUS94} and \cite{CLEO94} on the
polarization of the $\Lambda$ in the $\Lambda_c\to\Lambda+\ell^++\nu_\ell$
and their interpretation in terms of HQET foreshadow things to come.

\subsubsection{Model Results $b\to c$}

   For the heavy--to--heavy $b\to c$ transitions it is more convenient to
work entirely in terms of velocity variables. Correspondingly we define
a new set of form--factors in terms of velocity covariants, which we can
choose to be the six form--factors introduced in (\ref{48}--\ref{49}). Note
that now
all six ``velocity'' form--factors contribute to the transition in the zero
lepton mass case, i.e. one has not separated out the two scalar form--factors
(multiplying $q_\mu$) as in the representation (\ref{5-2},\ref{5-3}) used
before.
Let us first state the linear relation between the ``velocity''
form--factors defined in (\ref{48}--\ref{49}) and the helicity amplitudes
that enter into the formulas for physical observables. One has
\bea\label{5-30}
\sqrt{q^2}H_{\frac{1}{2},0}^{V,A} &=& \sqrt{2M_1M_2(\omega\mp 1)}
   \left( (M_1\pm M_2)f_1^{V,A} \pm M_2(\omega\pm 1)f_2^{V,A}
   \pm M_1(\omega\pm 1)f_3^{V,A}\right)\nonumber\\
H_{\frac{1}{2},1}^{V,A} &=& -2\sqrt{M_1M_2(\omega\mp 1)}f_1^{V,A}
\eea
where $H_{\lambda_2,\lambda_W}^{V,A}$ are the helicity amplitudes for the
vector (V) and axial vector (A) current induced $1/2^+\to 1/2^+ +
W_{off-shell}^-$ transitions. The upper and lower signs in (\ref{5-30})
stand for the vector (V) current and axial vector (A) current contributions,
respectively, where the total helicity amplitude is given by
\bea\label{5-31}
H_{\lambda_2,\lambda_W} = H_{\lambda_2,\lambda_W}^V +
H_{\lambda_2,\lambda_W}^A
\eea
 The remaining helicity amplitudes are related to the above two helicity
amplitudes by parity as given in Eq.(\ref{par}).
For the differential decay rate one then obtains
\bea\label{5-33}
\frac{\mbox{d}\Gamma}{\mbox{d}\omega} = \frac{G_F^2}{(2\pi)^3}
|V_{bc}|^2\frac{q^2pM_2}{12M_1}\left( |H_{\frac{1}{2},1}|^2+
|H_{-\frac{1}{2},-1}|^2 + |H_{\frac{1}{2},0}|^2 + |H_{-\frac{1}{2},0}|^2
\right)
\eea
where $p$ is the CM momentum of the daughter baryon $\Lambda_c$
 ($p=M_2\sqrt{(\omega+1)
(\omega-1)}$).

   The structure of the rate formula becomes very simple at the zero recoil
point $\omega=1$. Using again $H_{\lambda_2,\lambda_W}^V\to 0$ and
$H_{\frac{1}{2},1}^A\to -\sqrt{2}H_{\frac{1}{2},0}^A$ one finds that
\bea\label{5-34}
\frac{\mbox{d}\Gamma}{\mbox{d}\omega} = \frac{G_F^2}{(2\pi)^3}
2|V_{bc}|^2 M_2^3(M_1-M_2)^2\sqrt{\omega^2-1} |f_1^A(1)|^2 + \cdots
\eea
at $\omega=1$. At zero recoil only the "allowed Gamow--Teller transition"
$f_{1}^{A}$ survives (see Eq.(\ref{nuclear1})). Since HQET predicts that
$f_1^A(1)=1$ up to order ${\cal O}(1/m_Q)$ (see Eq.(\ref{57})), this relation
would allow one to extract
the value of $|V_{bc}|$ up to order ${\cal O}(1/m_Q)$ accuracy from e.g.
semileptonic $\Lambda_b\to \Lambda_c$ transitions if the data can be reliably
continued to the zero recoil point. In doing so one has to also account for
the small change in normalization resulting from vertex renormalization as
discussed in Sec.3. This and the corrections of order ${\cal O}(1/m_Q^2)$
to (\ref{5-34}) have been carefully discussed in \cite{FN93}.

  It is interesting to keep in the rate formula (\ref{5-33}) not only
the terms corresponding to $\omega=1$ (as has been done in deriving
(\ref{5-34})), but also the contributions linear in $(\omega-1)$ including
also p--wave contributions (see Eq.(\ref{nuclear1})). Using the predictions
of HQET (\ref{50}--\ref{55}) for the ${\cal O}(1/m_Q)$ structure of the
$\Lambda_b \to \Lambda_c$ transition one obtains
\bea\label{5-35}
\frac{\mbox{d}\Gamma}{\mbox{d}\omega} & = & \frac{G_F^2}{(2\pi)^3}
2|V_{bc}|^2 M_2^3(M_1-M_2)^2\sqrt{\omega^2-1} \nonumber \\
&&   \left( 1+(\omega-1)
\left[-2\rho^2+1-\frac{2}{3}\frac{M_1M_2}{(M_1-M_2)^2}+\bar\Lambda
\frac{M_1+M_2}{2M_1M_2}\right]\right) + \cdots
\eea
where the ellipsis stand for ${\cal O}((\omega-1)^2)$ contribution.
Here $\rho$ is the so--called ``charge radius'' of the $f_1^A$
form--factor defined by
\bea\label{5-36}
f_1^A(\omega)=f_1^A(1)-\rho^2(\omega-1)+{\cal O}((\omega-1)^2)\,.
\eea
Eq.(\ref{5-35}) is useful when one wants to extrapolate experimental
$\Lambda_b \to \Lambda_c$ data into the zero recoil point for a given
model value of the charge radius $\rho^2$.

As concerns the joint angular decay distribution for unpolarized
$\Lambda_b\to\Lambda_c$ transitions it can be taken from Eq.(\ref{this})
with the
requisite sign changes when going from a final state $\ell^++\nu_\ell$
to $\ell^-+\bar\nu_\ell$, as remarked on already there. The angular decay
distribution for polarized $\Lambda_b$ decay is identical to Eq.(\ref{5-15}).
Corresponding formulas including lepton mass effects relevant for
semileptonic decays involving the $\tau$--lepton can be found in
\cite{polen,kkk94}. Rate formulas and decay distributions for
$1/2^+\to 3/2^+$ transitions relevant for the decay $\Omega_b\to
\Omega_c^*+\ell^-+\bar\nu_\ell$ have been worked out in \cite{ct,polen}.

   There have been a number of attempts to calculate the decay properties
of the semileptonic decays $\Lambda_b\to \Lambda_c+\ell^-+\bar\nu_\ell,
\hspace{3mm}\Xi_b\to \Xi_c+\ell^-+\bar\nu_\ell$ and $\Omega_b\to\Omega_c+\ell
^-+\bar\nu_\ell$ using a variety of model assumptions and elements of HQET as
a guiding principle. We shall not attempt to provide an exhaustive
discussion of all the model calculations, in particular since there is no
data to compare with yet. Instead, we list the results of a few
representative model calculations for $\Lambda_b\to \Lambda_c+\ell^-+\bar
\nu_\ell$ (in the zero lepton mass case) in Table 9.

   The model of \cite{KKKK} uses infinite momentum frame wave functions
and determines the transition form factors at $q^2=0$ by making use of the
${\cal O}(1)$ and ${\cal O}(1/m_Q)$ structure of HQET. The form factors
are then continued to the whole $q^2$--region by using the HQET scaling
properties of the form factors. The model of \cite{GK93} is similar.
 For comparison we list the results of a very simple dipole form factor
ansatz described after Eq.(\ref{5-39}). The model of
Singleton uses a constituent quark model approach with a harmonic
oscillator potential \cite{de}. The transition form factors are evaluated
in the
$\Lambda_b$ rest frame. For comparison we also give the results of free
quark decay where we have taken $m_b=m_{\Lambda_b}=5.64$ GeV and
$m_c=m_{\Lambda_c}=2.285$ GeV in order to get the kinematics right. The
free quark decay result corresponds to taking structureless form factors
in the dipole model or in the ${\cal O}(1)$ HQET calculation and shows the
influence of the form factor effect on the rate and the polarization.
If one takes $m_b=4.73$ GeV and $m_c=1.55$ GeV one obtains a rate of
7.52$\times 10^{10}$ s$^{-1}$ instead. Judging from the numbers in Table 9
the exclusive semileptonic decay rate $\Lambda_b\to\Lambda_c+\ell^-+\bar
\nu_\ell$ would be predicted to amount to about 37\%--73\% of the total
inclusive semileptonic rate if one compares to the above two parton model
rates. The difference in rate between the form factor models and the
``structureless'' rate $\Gamma_{tot}\simeq (7.52-11.73)\times 10^{10}$
s$^{-1}$ would have to be filled in by the contribution of higher
$\Lambda_c$ resonances and continuum states.

 ${\cal O}(1/m_Q)$ effects in the IMF model \cite{KKKK} are small and tend to
increase the rates and the value of the polarization. The value of the
polarization of the
daughter baryon $\Lambda_c$ does not appear to be very model dependent
and is predicted to lie in the range $-0.70$ to $-0.80$, close to the
polarization of the charm quark in the free quark decay.
\begin{table}
\begin{center}
\caption[dummy25]{Rate for semileptonic decay
     $\Lambda_b\to\Lambda_c+e^-+\bar\nu_e$ and polarization $\langle\alpha
     \rangle$ for daughter baryon $\Lambda_c$ in the models
     \protect{\cite{KKKK,de}}, dipole form factor model and in the free
quark decay model with
     $m_b=m_{\Lambda_b}=5.64$ GeV and $m_c=m_{\Lambda_c}=2.285$ GeV. We have
     used $|V_{bc}|=0.044$.}
\renewcommand{\baselinestretch}{1.33}
\small \normalsize
\vspace{5mm}
\begin{tabular}{lrr}
\hline\hline
 Decay: $\Lambda_b\to\Lambda_c+e^-+\bar\nu_e$ &
 $\Gamma$ [10$^{10}$ s$^{-1}$] & $\langle\alpha\rangle$ \\
\hline
 Infinite momentum frame \cite{KKKK}\\
 ${\cal O}(1)$ & 3.70 & --0.71 \\
 ${\cal O}(1)+ {\cal O}(1/m_Q)$ & 4.57 & --0.77 \\
 Dipole form factor  & 5.14 & --0.72 \\
 Quark model \cite{de} & 3.48 & --0.71 \\
 Free quark decay & 11.73 & --0.81 \\
 \hline\hline
\end{tabular}
\renewcommand{\baselinestretch}{1}
\small \normalsize
\end{center}
\end{table}
We shall return to the subject of the $\Lambda_c$ polarization in
semileptonic $\Lambda_b$ decays in Sec.5.3 where we discuss attempts
to determine the chirality of $b\to c$ transitions using polarized
$\Lambda_b$ decays.

  To conclude this section we briefly discuss the $\Sigma$--type $b\to c$
transitions $\Omega_b^-\to \Omega_c^0$ and $\Omega_b^-\to \Omega_c^{*0}$.
The leading order HQET structure of these decays can easily be worked out
from the spinor expression Eq.(\ref{curs3}). One obtains
\bea\label{5-37}
\Omega_b^-(1/2^+)\to\Omega_c^0(1/2^+):&& \nonumber\\
\qquad\langle\Omega_c(v_2)|J_\lambda^{V+A}
   |\Omega_b(v_1)\rangle &=&-{\textstyle \frac{1}{3}}\bar u_2
    \left[F_L\gamma_\lambda(1-\gamma_5)
   -\frac{2}{\omega+1}(F_L+F_T)(v_{1_\lambda}+v_{2_\lambda})\right.
\nonumber\\
 &&+ \left.
\frac{2}{\omega-1}(F_L-F_T)(v_{1_\lambda}-v_{2_\lambda})\gamma_5\right]u_1
\\[2mm]
\Omega_b^-(1/2^+)\to\Omega_c^{*0}(3/2^+):&&\nonumber\\
\langle\Omega_c^{*0}(v_2)|J_\lambda^{V+A}|\Omega_b(v_1)\rangle &=&
{\textstyle \frac{1}{\sqrt{3}}}\bar u_2^\nu
\left[2F_Tg_{\nu\lambda}(1+\gamma_5)+\frac{1}{\omega+1}
  (F_L+F_T)v_{1_\nu}\gamma_\lambda\gamma_5\right.\nonumber\\
&&\left.-\frac{1}{\omega-1}(F_L-F_T)v_{1_\nu}\gamma_\lambda+
   \frac{2}{\omega^2-1}(F_L-\omega F_T)v_{1_\nu}v_{2_\lambda}(1+\gamma_5)
\right] u_1\,,\nonumber\\\label{55-38}\eea
where $F_L(\omega)=g_1^{(0)}(\omega)$ and $F_T(\omega)=\omega
 g_1^{(0)}(\omega)
-(\omega^2-1)g_2^{(0)}(\omega)$ describe the longitudinal and
transverse spin 1 diquark transitions. As remarked on earlier $F_L$ and
$F_T$ diagonalize the transition rates. We have dropped any reference to the
$\omega$--dependence in the form--factors in Eq.(\ref{5-37}). In the heavy
quark limit there are thus two universal form factors $F_L$ and $F_T$
compared to the many independent factors in the general covariant expansion
(\ref{5-2},\ref{5-3}) to which they can be related. The two form factors
are normalized to one at zero recoil $q^2=q^2_{max}=(M_1-M_2)^2$ or
$\omega=1$, that is $F_L(\omega=1)=F_T(\omega=1)=1$. The $1/m_Q$
corrections to the limiting structure described by Eqs.(\ref{5-37}) and
(\ref{55-38}) have
been worked out in \cite{BB1}. At ${\cal O}(1/m_Q)$ there are altogether
seven universal form factors compared to the two leading order form factors
$F_L$ and $F_T$. According to Luke's theorem \cite{LUKE,KT} one retains
the zero recoil normalization condition at ${\cal O}(1/m_Q)$.

   In order to perform a quick appraisal of the structure of the
$\Sigma$--type transitions and their rates we turn to the constituent
quark model approximation discussed in Sec.4.3. The longitudinal and
transverse form factors $F_L$ and $F_T$ can be seen to be related to
the residual form factor $f(\omega)$ by
\bea\label{5-39}
F_L(\omega)=F_T(\omega)=\frac{\omega+1}{2}f(\omega)\,.
\eea
For the residual form factor we make a dipole ansatz, i.e. we write
\bea\label{5-40}
f(\omega)=\left( 1+\frac{2M_1M_2(\omega-1)}{m_{FF}^2-(M_1-M_2)^2}
\right)^{-2}
\eea
where $f(\omega)$ is evidently normalized at $\omega=1$. By writing this
formula in terms of the momentum transfer variable $q^2$ one recovers the
familiar dipole representation $F^{dipole}(q^2)=N(q^2)(1-q^2/m_{FF}^2)^{-
2}$ where $N(q^2)$ normalizes the dipole form factor to one at the zero
recoil point $q^2=(M_1-M_2)^2$. As pole masses we take $m_{FF}=6.34$ GeV
and 6.73 GeV for the vector and axial vector form factors, respectively.
These pole masses correspond to the expected masses of the $(b\bar c)$
vector and axial vector mesons. Using $|V_{bc}|=0.044$ one obtains the
following rate values
\beq\label{5-41}
\begin{array}{lll}
\Lambda_b^0\to \Lambda_c^-\, &:&\qquad (5.14 ; 1.54)\times 10^{10}
                                \mbox{s}^{-1}\\
\Xi_b^0\to \Xi_c^+\,         &:&\qquad (5.21 ; 1.55)\times 10^{10}
                                \mbox{s}^{-1}\\
\Omega_b^-\to \Omega_c^0\,   &:&\qquad (1.52; 0.52)\times 10^{10}
                                \mbox{s}^{-1}\\
\Omega_b^-\to \Omega_c^{*0}\,&:&\qquad (3.41; 0.99)\times 10^{10}
                                \mbox{s}^{-1}
\end{array}
\eeq
In Eq.(\ref{5-41}) we also list predictions for the $\Lambda$--type
transitions in the same constituent approximation. We have also included
the corresponding rate predictions for the $\tau$--mode which can be seen
to be down by a factor of approximately three. Similar results have been
obtained in \cite{de}. The rate predictions of the QCM model are higher
\cite{EIKL92} due to the fact that the QCM form factors are much harder
than our
dipole form factors. As argued in Sec.4.3 the high rate predictions of
\cite{EIKL92} for the $\Sigma$--type transitions cannot be trusted since
the QCD form factors violate one of the bounds due to  Bjorken.

\subsection{Polarization effects}

   According to the Standard Model, $b$ quarks from the $Z^0$ resonance
$e^+e^- \ra Z^0 \ra b\bar{b}$ have almost complete longitudinal
polarization, given by
\begin{equation}\label{plb}
P_{L}^{b} = \frac{(8\xw^2-4\xw+1)2\vb\ab\beta(1+\cos^2\Theta)
                  +2(4\xw-1)(\vb^2+\beta\ab^2)\cos\Theta}
                 {(8\xw^2-4\xw+1)((\vb^2+\beta^2\ab^2)(1+\cos^2\Theta)
                  +\vb^2(1-\beta^2)\sin^2\Theta)+4(4\xw-1)\vb\ab\beta
                  \cos\Theta}
\end{equation}
where $\vb=-1+\frac{4}{3}\xw$, $\ab=1$, $\xw=\sin^2\Theta_W$,
$\beta^2=1-4m_{b}^{2}/m_{Z}^{2}$ and $\Theta$ is the $CM$
scattering angle relative to $e^-$. For $\xw=0.23$ this
gives a mean value $< P_{L}^{b} > = -0.94$ with
very little angular dependence. A small transverse polarization
of order $0.02$ is predicted in the scattering plane; there is
no polarization normal to the plane if we neglect $\gamma-Z$
interference and loop corrections. Noteworthy is the absence
of a longitudinal component proportional to $\sin^2\Theta$ in
the numerator of (\ref{plb}) which comes about because there is
no axial vector induced amplitude into the positive helicity $b$
(or $\bar{b}$) to lowest order in QCD. The lowest order result
for $< P_{L}^{b} >$ is not changed by soft or hard
gluon emission, or by loop effects, to any order in $\alpha_s$
if (and only if!) the $b$ quark can be treated as massless.
However, while doing the ${\cal O}(\alpha_s)$ radiative corrections
to $< P_{L}^{b} >$ \cite{kpt93}, it was noticed that
one cannot naively set the quark mass to zero ab initio when
doing radiative corrections. There is a spin flip contribution
proportional to $m_{b}^{2}$ which survives the limit
$m_b \ra 0$ since it gets promoted to a constant term
by a would--be collinear singularity $\propto m_{b}^{-2}$.
As a result $< P_{L}^{b} >$ is corrected at
${\cal O}(\alpha_s)$ \cite{kpt93} even for $m_b/m_{Z^0} \ra 0$.
We mention that the orientation dependent $P_{L}^{b}(\cos\Theta)$
will be changed by hard gluon emission regardless of the above
spin flip contribution. For charm the Standard Model predicts a
somewhat smaller value $< P_{L}^{c} > = -0.67$.

The question is whether this big polarization can be exploited and
whether $b$ quark polarization can survive hadronization to give
bottom hadron polarization \cite{ckps92,fp93}. Hadronization to
{\em mesons} is hopeless for our purposes. It is possible that
high--spin mesons formed from $b$ quarks may retain some of the
initial $b$ polarization, but they will decay by parity--conserving
processes (that cannot give polarization--dependent asymmetries)
down to a spin--0 $B$ meson that retains no spin information.
It is instructive however to consider how the $b$ spin information
is lost. Suppose at $t=0$ a spin--up $b$ combines with a spin--down
$\bar{q}$ forming the state $\bup\qdo$. This is not an eigenstate
of total spin $S$ but can be decomposed into a sum of $S=0$ and $S=1$
eigenstates:
\begin{equation}\label{bupqdo}
\bup\qdo = \frac{1}{\sqrt{2}}
           \left (\frac{\bup\qdo - \bdo\qup}{\sqrt{2}}\right )
         + \frac{1}{\sqrt{2}}
           \left (\frac{\bup\qdo + \bdo\qup}{\sqrt{2}}\right )
\end{equation}
If we add a common space wave function, then these two terms
represent a spin--0 meson and a spin--1 meson coherently superposed.
If indeed the physical meson states were precisely degenerate
(as in the $M_Q \ra \infty$ limit of HQET), the time evolution
of the $S=0$ and $S=1$ terms would be identical and the coherence
would be preserved; the spin wave function would then remain
$\bup\qdo$, the coherent superposition of meson states would
preserve the $b$ spin and total spin $S$ would be irrelevant.
In reality however $M_b \not= \infty$; the pseudoscalar $B$
and vector $B^*$ mesons have different masses and therefore
different time evolutions; at times $t\gg (M_{B^*} - M_B)^{-1}$,
the $S=0$ and $S=1$ amplitudes become effectively incoherent
and the $b$ quark is depolarized over a period of time by
spin--spin forces within the mesons (indeed the same forces
that generate the $B-B^*$ mass splitting).

One can make this more precise by defining a decoherence
time scale $t_{decoherence} = (M_{B^*} - M_{B})^{-1}$ and a
decay time scale $t_{decay} = \Gamma^{-1}$. For the case
in question one has
\begin{eqnarray}
t_{decoherence} & = & \frac{1}{M_{B^*}-M_B}
                \cong 2\cdot 10^{-2}~\mbox{MeV}^{-1}\nonumber \\
t_{decay} & = & \frac{1}{\Gamma_{B^* \ra B\gamma}}
                \cong (10^{3}-10^{4})~\mbox{MeV}^{-1}
\end{eqnarray}
where the width of the $B^*$ is determined by its one--photon decay.
The $B^*$ width can be estimated in the constituent model as
described in Sec.4.6. The authors of \cite{yan93} quote
$\Gamma(B_{u}^{*+} \ra B_{u}^{+} + \gamma)=0.84$~keV and
$\Gamma(B_{d}^{*0} \ra B_{d}^{0} + \gamma)=0.28$~keV similar to the
estimates in \cite{a92,cg92}. Thus one has $t_{decay} \gg
t_{decoherence}$ for the $B$ mesons, i.e. the $t=0$ coherent
superposition in Eq.(\ref{bupqdo}) will have become completely
decoherent by the time $B^*$ decays (the $B$ decays weakly and is
much longer lived).

Hadronization to bottom {\em baryons} is more promising.
The Pauli principle implies that if a $b$ quark combines with a
spin--0 combination of one $u$ plus one $d$, a $\Lambda_b$ is
formed; if the light--quark pair has spin 1 then $\Sigma_b$ or
$\Sigma_{b}^{*}$ result with total spin 1/2 or 3/2. The crucial
feature of this system is that in the heavy--quark limit the
$\Sigma_Q$, $\Sigma_{Q}^{*}$ become degenerate and some
200~MeV more massive than the $\Lambda_Q$ with the
result that they decay to $\Lambda_Q$ by the strong interaction,
preserving the $b$ polarization.

Suppose first that the polarized $b$ quark picks up a spin--0
$ud$ pair to form a 'prompt' $\Lambda_b$. Due to its $ud$ pair
having spin 0, all of the $\Lambda_b$ spin resides on the
valence $b$ quark and we expect $b$ polarization to become
$\Lambda_b$--polarization (in the heavy--quark limit where $b$
spin--flip is suppressed during hadronization). Suppose instead
that the polarized $b$ quark had combined with a spin--1 $ud$
pair. In this case we would have to decompose the $bqq$ wave
functions into superpositions of eigenstates of different
total spin $S= 1/2, 3/2 (\Sigma,\Sigma^*)$. Take e.g. the state
$\bup\{uu\}_0$ which, at $t=0$, is a coherent superposition
of $S=1/2$ and $S=3/2$ states according to
\begin{eqnarray}
\bup\{uu\}_0 &=& \sqrt{{\textstyle \frac{1}{3}}}\left
                (\sqrt{{\textstyle \frac{1}{3}}}\bup\{uu\}_0 -
                 \sqrt{{\textstyle \frac{2}{3}}}\bdo\{uu\}_{+1}\right
                                                )_{S=1/2}\nonumber\\
             & & +\sqrt{{\textstyle \frac{2}{3}}}\left
                 (\sqrt{{\textstyle \frac{2}{3}}}\bup\{uu\}_0 +
                  \sqrt{{\textstyle \frac{1}{3}}}\bdo\{uu\}_{+1}\right
                                                )_{S=3/2}
\end{eqnarray}
At $t>(M_{\Sigma^{*}_{b}}-M_{\Sigma_b})^{-1}$ the $b$ quark
would become depolarized in the $\Sigma_b$, $\Sigma_{b}^{*}$
systems for reasons analogous to those outlined for the $B,B^*$
bottom mesons above (in the present case the depolarization is
only partial). Subsequent decays to $\Lambda_b$ would produce
partially depolarized $\Lambda_b$.

Quantitatively one has
\begin{eqnarray}
t_{decoherence} &=& (M_{\Sigma_{b}^{*}}-M_{\Sigma_b})^{-1} \cong
                    5 \cdot 10^{-2}~\mbox{MeV}^{-1} \nonumber\\
t_{decay}       &=& \Gamma^{-1}_{(\Sigma_{b}^{*},\Sigma_b) \ra
                    \Lambda_b + \pi} \cong
                    5 \cdot 10^{-2}~\mbox{MeV}^{-1}
\end{eqnarray}
taking the constituent one--pion width estimate from
\cite{yan92} as discussed in Sec.4.5. That the two time--scales
come out approximately equal is an accident since the width is
independent of the heavy quark mass while
$\Delta M(\Sigma_{b}^{*},\Sigma_b)$ is proportional to $1/m_b$.
The outcome of the above estimate is that the $b$ quark will
have become partially depolarized when it finally ends up in
the $\Lambda_b$, because of the $t_{decoherence}\cong t_{decay}$.
The situation is more favourable for fragmentation into the
higher lying excited $\Sigma^{**}_{b}$--states because of phase
space, but may be less favourable for fragmentation into excited
$\Lambda^{**}_{b}$ since these can decay to $\Lambda_b$ only via
two--pion emission. According to a rough estimate presented in
\cite{fp93} the polarization transfer from $b$ to $\Lambda_b$ can
be expected to be all in all about 70\%.

The reason that there has been such a wide interest in the
polarization of the $b$ or the $\Lambda_b$ from $Z^0$--decays
is that one can hope to turn this polarization information into
an effective tool to analyze bottom or $\Lambda_b$--decays. One of
the issues that has been discussed in this context is the hope
to be able to determine the $b \ra c$ chirality using polarized
$b$--decay. In order to set the stage let us gather together pieces
of information and arguments concerning $b \ra c$ chirality (see
also the review of \cite{GRONAU94}).

The prediction of the Standard Model that the $b \ra c$
transition is left chiral has recently been confirmed by a
determination of the sign of the lepton's forward--backward (FB)
asymmetry in the ($l^-\overline{\nu}_l$) rest system in the
semileptonic decay $\overline{B} \ra D^* + l^- +\overline{\nu}_l$
\cite{argus,cleo}. In this analysis one uses the Standard Model
left--handedness of the lepton current as input. However, if one
leaves the realms of the Standard Model, the same FB asymmetry
would arise if both quark and lepton currents were taken to be
right--chiral, i.e. if one would switch from a
   $H_{\mu\nu}(V - A) L^{\mu\nu}(V - A)$ coupling to a
   $H_{\mu\nu}(V + A) L^{\mu\nu}(V + A)$ coupling.\footnote{A viable
model involving a right-handed $W_R$ that is consistent with all
present data has recently been proposed \cite{gw}.}

The FB asymmetry measure alluded to above constitutes a
momentum--momentum correlation measure $< \vec{l}\cdot\vec{p}>$
which clearly is not a truly parity--violating measure.\footnote{
For example, it is well--known that in
$e^+e^-$--annihilation the two photon exchange contribution
also gives rise to nonvanishing FB asymmetries despite
of the fact that QED is parity conserving.} What is needed to
distinguish between the two above options is to define truly
parity--violating spin--momentum correlation measures of the type
$<\vec{\sigma}\cdot \vec{p}>$.

Some such possible parity--violating measures that have
been discussed recently exploit the polarized bottom quarks
produced on the $Z_0$ resonance. One then defines spin--momentum
correlations w.r.t. the longitudinal spin direction of the
decaying $b$ or $\Lambda_b$ using the momenta of the decay
products of the $b$ or $\Lambda_b$. For the semileptonic decays
$\Lambda_b \ra \Lambda_c + l^- + \bar\nu_l$
or $b \ra c + l^- + \bar\nu_l$ this has been done using the
lepton momentum \cite{ckps92,jap} and the $c$ or $\Lambda_c$ momentum
\cite{jap,cs}. The sign of these correlations or the sign of the
correspondingly defined FB asymmetries allow one to differentiate
the above two options which remain after the analysis of the
mesonic experiments, \cite{argus,cleo}, i.e. the
   $H_{\mu\nu}(V - A) L^{\mu\nu}(V - A)$ or the
   $H_{\mu\nu}(V + A) L^{\mu\nu}(V + A)$ option.
A problem with the suggested analysis is that they require the
reconstruction of the $\Lambda_b$ rest frame which will be a
difficult experimental task.

Alternatively one can consider the shape of the lepton spectrum
directly in the lab system \cite{sample}. The spin--lepton--momentum
correlation effects referred to above have the effect that the
emitted leptons in the semileptonic decay $\Lambda_b \ra \Lambda_c
+ l^- + \bar\nu_l$ (or $b \ra c + l^- + \bar\nu_l$) tend to
counteralign and align with the polarization of the $b$ for
   $H_{\mu\nu}(V - A) L^{\mu\nu}(V - A)$ and
   $H_{\mu\nu}(V + A) L^{\mu\nu}(V + A)$
interactions, respectively, leading to harder and softer lepton
spectra in the lab system relative to unpolarized decay allowing
one
to distinguish between the two options in principle. However, as has
been emphasized in \cite{ckps92}, a lack of knowledge of the precise
form of the $b \ra \Lambda_b$ fragmentation function precludes a
decision whether the lepton spectrum is harder or softer than that of
unpolarized decay, in particular since there is no unpolarized decay
sample to compare with.

Another possibility to distinguish between the
$H_{\mu\nu}(V - A) L^{\mu\nu}(V - A)$ and
$H_{\mu\nu}(V + A) L^{\mu\nu}(V + A)$
options via a parity--violating measure is to determine the
polarization of the lepton in the semileptonic decays
$B \ra D(D^*) + l^- + \overline{\nu}_l$ \cite{wakai92} or
$\Lambda_b \ra \Lambda_c + l^- + \bar\nu_l$ \cite{kkk94}.
This will be a difficult experiment but may be feasible in the
not too distant future for semileptonic decays involving the
$\tau$--lepton.

In \cite{kkk93} two of us together with B.~K{\"o}nig proposed yet a
fourth variant of a truly parity--violating spin--momentum correlation
measure in $b \ra c$ decays. They proposed to look at the decay
cascade $\Lambda_b \ra \Lambda_c (\ra a_1 + a_2 + \cdots)
+ \l^- +\overline{\nu}_l$ to determine the chirality of $b \ra c$
decays where $\Lambda_c \ra a_1 + a_2 + \cdots$ are nonleptonic
decays of the $\Lambda_c$. The weak nonleptonic decays of the
$\Lambda_c$ serve to analyze the polarization of the $\Lambda_c$
through the correlation of their momenta with the polarization of the
decaying $\Lambda_c$. Ideal in this regard are the nonleptonic decays
$\Lambda_c \ra \Lambda \pi$ and $\Lambda_c \ra \Sigma \pi$ the
analyzing power of which has recently been determined
\cite{cleo,argus,proc}. As a further analyzing channel
they discussed the decay modes $\lc \ra p \bar K^{*0}$ and $\lc \ra
\Delta^{++} K^-$ which could make up a large fraction of the dominant
decay mode $\Lambda_c \ra p  K^- \pi^+$. The analyzing power of these
channels has not yet been determined experimentally but can be
estimated using the theoretical quark model ansatz of \cite{kk92b}.

Consider the semileptonic decay of an unpolarized $\Lambda_b$.
Possible polarization effects due to polarized $\Lambda_b$--decays
average out if one integrates over all possible momentum directions
of the $\Lambda_c$ in the decay $\Lambda_b \ra \Lambda_c + l^- +
\overline{\nu}_l$. Possible $\Lambda_b$ polarization effects due to
incomplete averaging because of experimental cut biases have been found
to be very small.
 From simple helicity arguments the longitudinal polarization $P_L$
(also called $\alpha$ in Sec.5.2.1) of the daughter baryon $\Lambda_c$
is expected to be negative (positive) in most of the phase space
region for left--chiral ($\xi = 1$) (right--chiral ($\xi = -1$))
$b\ra c$ transitions, respectively. For the mean value of $P_L$ one
finds
\begin{equation}\label{plmean}
< P_L >
 = \xi \bigg\{ \!\begin{array}{ll} -0.77 & \mbox{IMF}\; \cite{KKKK}
        \\ -0.81 & \mbox{FQD} \end{array} \quad .
\end{equation}
The two polarization values refer to the Heavy Quark Effective Theory
(HQET) improved infinite momentum frame (IMF) model of
                                                 Ref.\cite{KKKK}
and free quark decay (FQD) where we use $m_b = M_{\Lambda_b} = 5.64$
GeV and $m_c = M_{\Lambda_c} = 2.285$ GeV in order to get the phase
space right.

The longitudinal polarization of the $\Lambda_c$ can be probed by
looking at the angular distribution of its subsequent nonleptonic
decays as written down in Eq.(\ref{polar}). Ideal in this regard are the
nonleptonic modes $\Lambda_c \ra \Lambda\pi$ and $\Lambda_c \ra
\Sigma\pi$ since the analyzing power of these decays has recently
been determined. For $\Lambda_c \ra \Lambda\pi$ one has
\begin{equation}\label{alp1}
\alpha_{\Lambda_c\ra\Lambda\pi}
 = \bigg\{ \!\begin{array}{lc} -1.0 {+ 0.4 \atop - 0.0}& \cite{cleo}
        \\ -0.96 \pm 0.42 & \cite{argus} \end{array} \quad .
\end{equation}
For $\Lambda_c \ra \Sigma\pi$ we quote the preliminary value \cite{proc}
\begin{equation}\label{alp}
\hphantom{deca }
\alpha_{\Lambda_c\ra\Sigma\pi} = - 0.43 \pm 0.23 \pm 0.20 \quad .
\end{equation}
In correspondence to the decay distribution Eq.(\ref{polar}) one can define a
forward--backward (FB) asymmetry by averaging over the daughter baryons
in the respective forward (F) $(0^\circ \le\Theta < 90^\circ)$ and
backward (B) $(90^\circ \le \Theta < 180^\circ)$ hemispheres to obtain
\begin{equation}\label{fb}
A_{FB} = {\textstyle \frac{1}{2}}    P_L \alpha_{\Lambda_c} \quad .
\end{equation}
Judging from the large numerical values of the mean of $P_L$
Eq.(\ref{plmean}) and of the asymmetry parameters $\alpha_{\Lambda_c}$
Eqs.(\ref{alp1},\ref{alp}) a measurement of the sign of $A_{FB}$
within reasonable errors should allow one to conclude for the sign of
$\xi$ and therefore for the chirality of the $b \ra c$ transition with
a good certainty.

What has been said up to now requires the reconstruction of the
$\Lambda_b$ rest system. This will not be an easy task for the
energetic $\Lambda_b$ bottom baryons produced on the $Z_0$ where the
analysis suggested in this paper is most likely to be done first.
There is some hope, though, that such a reconstruction can be done
with the newly installed vertex detectors in the CERN detectors.
At present it is more realistic to consider the LEP
environment with energetic longitudinally polarized $\Lambda_b$'s
whose rest frames cannot be reconstructed. The polarization of
the $\Lambda_c$'s in the semileptonic decays takes a more
complicated form in the laboratory frame than in the $\Lambda_b$
rest frame as given by Eq.(\ref{5-14}). In particular
negatively polarized $\Lambda_c$'s emerging backward in the
$\Lambda_b$ rest frame will turn into positively polarized
$\Lambda_c$'s in the lab frame because of the momentum reversal
due to the requisite Lorentz boost. Also, because of experimental
cuts and/or biases the $\Lambda_c$'s polarization dependence on the
polarization of the $\Lambda_b$ may no longer average out, i.e. one
has to address the question of polarization transfer under realistic
experimental conditions.

In order to study all these issues a Monte Carlo program has been
written that generates semileptonic decay events of polarized
$\Lambda_b$ into polarized $\Lambda_c$. It is then a simple matter
to adapt the calculation to the experimental conditions present in
the LEP environment including longitudinal and transverse lepton
momentum cuts. What one finds is that approximately 50\% of the
polarization information is retained when going from the $\Lambda_b$
rest frame to the lab frame ($Z^0$ rest frame). One obtains
$< P_L >_{lab frame} = - (0.3-0.4)\xi$ with little cut and
$\Lambda_b$ polarization sensitivity.
$\xi$ is the chirality parameter as before. With sufficient
statistics it should not be too difficult to pin down the chirality
of the $b \ra c$ transitions through some such measurements. We
emphasize that the quality of this experiment is crucially
dependent on the quality of the charm decay data that one is using
to analyze the $\Lambda_c$
polarization.

Regardless of what has been said about the potential use of polarized
$b$ or $\Lambda_b$ decays to measure the chirality of the $b \ra c$
transitions there is a wealth of interesting physics to be
investigated using polarized $b$ quarks, and for that matter,
polarized heavy baryon decays.

\newpage
\vspace*{1cm}\hspace*{3cm}
\section{Lifetimes and Inclusive Nonleptonic Decays}
\vspace*{2cm}
\subsection{Experimental Lifetimes}

   Let us quote the charm baryon lifetime values from the 1992 Particle
Data Group \cite{pdg}. The $\Lambda_c$ has a lifetime of $\tau(\Lambda_c)
=(1.91\pm{0.15\atop 0.12})\times 10^{-13}\,s$, the charged $\Xi_c^+$ has
a lifetime of $\tau(\Xi_c^+)=(3.0\pm{1.0\atop 0.6})\times 10^{-13}\,s$
and the neutral $\Xi_c^0$ lifetime is quoted at $\tau(\Xi_c^0)
=(0.82\pm{0.59\atop 0.30})\times 10^{-13}\,s$. There exist no lifetime
measurements for the $\Omega_c^0$ yet. For comparison, the lifetimes of
the charmed mesons are $\tau(D^0)=(4.20\pm 0.08)\times 10^{-13}\,s$,
$\tau(D^\pm)=(10.66\pm 0.23)\times 10^{-13}\,s$ and $\tau(D_s^\pm)=(4.50
\pm{0.30\atop -0.26})\times 10^{-13}\,s$.

   In the bottom baryon sector there exist lifetime measurements only
for the $\Lambda_b$. R.Forty \cite{FORTY93} quotes a LEP average
of $\tau(\Lambda_b^0)=(1.07\pm 0.15)\times 10^{-12}
\,s$. Ironically one now has a lifetime measurement of the $\Lambda_b$
although the $\Lambda_b$ has not yet been seen with certainty. Contrast
the $\Lambda_b$ lifetime with the bottom meson averages quoted by the
Particle Data Group \cite{pdg}:
$\tau(B^\pm)=(1.62\pm 0.13)\times 10^{-12}\,s$, $\tau(B^0)=(1.43\pm 0.12)
\times 10^{-12}\,s$ and $\tau(B_s^0)=(1.41\pm 0.22)\times 10^{-12}\,s$.

\subsection{Theoretical Lifetime Estimates}

 In the large mass limit, one expects all heavy hadrons of the
 same flavour to have identical lifetimes. The spread in the
 experimental lifetime values of the $\Lambda _{c}^{+}, \Xi _{c}^{0,+}$
 and $\Omega _{c}$  charm baryons signals that $1/m_c$ effects are still
 important in the weak inclusive decays of charm baryons (as they are
 for charm mesons). The preasymptotic effects
 enter in the form of W--exchange
 contributions \cite{cb}, and additional contributions come
 from the interference of decay quarks and spectator
 quarks. These are sensitive to the probability that the
 the charm  and light quarks
 in the baryon wave function will come together
 at one point: to the square of the wave function at the
 origin $\vert \Psi (0) \vert ^{2}$ with mass dimension $[m^{3}]$.

 From dimensional arguments, one then finds
\begin{eqnarray}
 \Gamma _{FQD} &\approx & G_{F}^{2} m_{c}^{5}
 \nonumber \\\label{mac}
 \Gamma _{exch, int} &\approx & G_{F}^{2} m_{c}^{2}
  \vert \Psi (0) \vert ^{2}
\end{eqnarray}
where $\Gamma _{FQD} $ denotes the "free quark decay"
 parton model decay rate, $\Gamma _{exch}$ and
 $\Gamma _{int}$ denote the W--exchange and interference rates, and $m_c$
 refers to the charm quark mass.

Explicit calculations \cite{cc,cd} show that
 $\Gamma _{FQD} \approx \Gamma _{exch, int} $ in the charm
 baryon sector and that $\Gamma _{FQD} \approx \Gamma _{int} $ in the charm
 meson sector \cite{cd}.
 Using the fact that the wave function at the origin of the heavy--light
 bound state becomes independent of the heavy quark mass as the heavy quark
 mass becomes large \cite{wipol88}, one can scale Equation (\ref{mac}) to the
 bottom quark sector. One then finds $\Gamma _{exch, int} /\Gamma _{FQD}
 \approx (m_{c}/m_{b})^{3} \approx {\cal O}(5\%)$,  which implies that the
 lifetime differences in the bottom sector are expected to be quite small.
 To some extend this is corroborated by the present bottom lifetime
 measurements.

The difficulty in obtaining reliable
 rate and life time estimates for the charm baryons is clearly
 evidenced by the fact that the preasymptotic effects,
 which are down by several powers of $1/m_{c}$,
 are so important. This makes an analysis
 in terms of a $1/m_{c}$ expansion difficult. Nevertheless
 one can attempt to obtain a qualitative
 picture of the life time differences of charm
 baryons in the form of a life time hierarchy \cite{cc,cd}.

The starting point in the analysis is the
 usual effective nonleptonic Hamiltonian
\begin{eqnarray}\label{hamiltonian}
 H_{eff}&=&\sqrt{2} G_{F} V_{cs} V_{ud}^{\ast }
  \left[c_{-}O_{-}+c_{+}O_{+}\right]
\end{eqnarray}
 where $O_{\pm }$ are local 4-quark operators
\begin{eqnarray}
  O_{\pm }&=&(\bar u_{L} \gamma _{\mu } d_{L})
    (\bar s_{L} \gamma ^{\mu } c_{L})
   \pm  (\bar s_{L} \gamma _{\mu } d_{L})
    (\bar u_{L} \gamma ^{\mu } c_{L})
\end{eqnarray}
 with $\bar q_{L} \gamma _{\mu } q_{L}
 =\frac{1}{2} \bar q \gamma _{\mu } (1-\gamma _{5})q$,
 and  $V_{\bar q_{\alpha }q_{\beta }}$ are
 elements  of the Kobayashi-Maskawa
 mixing matrix with
 $V_{cs} \simeq V_{ud}\simeq \cos\Theta _{c}$
 and $\Theta _{c}$ the Cabibbo angle. The coefficients
 $c_{\pm }$ describe the leading logarithmic evolution of the nonleptonic
 Hamiltonian from the W mass scale down to the charm
 mass scale $\mu \simeq {\cal O}(m_{c})$ \cite{cf}.
 We take $c_{+}=0.74$ and $c_{-}=1.80$ as in the work of Guberina et al.
 \cite{cd}.

\begin{figure}
\vspace{12.5cm}
\caption[dummy26]{
Free quark decay, W--exchange and interference contributions to
inclusive nonleptonic $\Lambda_c^+$ decays.}
\end{figure}

The effective nonleptonic Hamiltonian Eq.(\ref{hamiltonian}) induces
 the inclusive nonleptonic decay contributions drawn
 in Fig.17 for e.g. the inclusive
 $\Lambda _{c}^{+}$ decays.
 Simple expressions can be obtained for these
 rates when one neglects u, d, s quark masses
 and uses a nonrelativistic wave function for
 the charm baryons. For example, for the $\Lambda _{c}^{+}$ decay
 one then has a nonleptonic (n.l.) rate \cite{cd}
 \begin{eqnarray}
  \Gamma _{n.l.}^{\Lambda _{c}^{+}} &=&
  \Gamma _{FQD}^{\Lambda _{c}^{+}}
  +\Gamma _{exch}^{\Lambda_{c}^{+}}
  +\Gamma _{int}^{\Lambda_{c}^{+}}
 \nonumber \\
 &=&(2c_{+}^{2}+c_{-}^{2})\frac{G_{F}^{2}m_{c}^{5}}{192 \pi ^{3}}
  +c_{-}^{2} \frac{G_{F}^{2}m_{c}^{2} \vert \Psi (0) \vert ^{2}}
  {4 \pi }
   -c_{+}(2c_{-}-c_{+})
    \frac{G_{F}^{2}m_{c}^{2} \vert \Psi (0) \vert ^{2}}
  {4 \pi }
 \nonumber \\\label{idi}
 &=&(1.58 +3.01 R
 -0.99 R )\times 10^{-12} \mbox{s}^{-1}
\end{eqnarray}
 with $m_{c}=1.6$ GeV , $c_{\pm }$-values as above and
 $ R=\vert \Psi (0) \vert ^{2}/
 10^{-2}\mbox{GeV}^{3}$.

As is evident from Equation (\ref{idi}) the resulting nonleptonic rate
 is quite sensitive to the value of the
 wave function at the origin $\vert \Psi (0) \vert ^{2}$. From a
 fit to the hyperfine splitting,
 as discussed in Sec.2,
  one has $\vert \Psi (0) \vert ^{2}
 \simeq 10^{-2}\mbox{GeV}^{3}$.
 Adding a nominal semileptonic rate value $2\times 15\%$ of the
 nonleptonic FQD rate one finds
 $\tau _{\Lambda _{c}^{+}}=2.46\times 10^{-13}$s which is
 somewhat larger than the experimental value
 $\tau _{\Lambda _{c}^{+}}(exp.)=
  (1.91\pm {0.15\atop 0.12})\times 10^{-13}$s. Smaller values
 of the wave function at the origin are obtained
 in a bag model \cite{cd}
 ($\vert \Psi (0) \vert ^{2}
 \simeq 0.4 \times 10^{-2}\mbox{GeV}^{3}$)
 or if one equates the baryon's and meson's wave function
 at origin \cite{cc}
 $(\vert \Psi (0) \vert ^{2}
 \simeq 0.4 \times 10^{-2}\mbox{GeV}^{3}$
 with $f_{D}=165$ MeV). Values similar to the above
   $\vert \Psi (0) \vert ^{2}
 \simeq 10^{-2}\mbox{GeV}^{3}$ are also obtained
by using electromagnetic
 mass differences in the hyperfine formula \cite{cg}.
 It is clear that using the smaller values of
   $\vert \Psi (0) \vert ^{2} $ worsens
 the agreement with the experimental rate.

Applying the same calculation to
 the other weakly decaying charm baryons
 Guberina, R\"uckl and Trampetic \cite{cd} find a
 lifetime hierarchy
 $\tau (\Omega _{c}) \approx \tau (\Xi _{c}^{0})
  <\tau (\Lambda _{c}) < \tau (\Xi _{c}^{+})$
 whereas Voloshin and Shifman \cite{cc} obtained
 $\tau (\Omega _{c}) < \tau (\Xi _{c}^{0})
  <\tau (\Lambda _{c}) \approx  \tau (\Xi _{c}^{+})$.
 In a more recent analysis Blok and Shifman estimate the lifetime
 ratios at $\tau (\Xi_{c}^{0}) : \tau (\Lambda_{c}) : \tau (\Xi_{c}^{+})
 \approx 0.36 : 0.77 : 1$ and point out that the lifetime of the
 $\Omega_c$ can either be the most short--living and the most
 long--living among charmed baryons depending on the strength of the
 unknown spin--spin interaction in the $\Omega_c$ \cite{bs93}.
Present data favours the inequality chain of \cite{cd} and the new values
of Blok and Shifman \cite{bs93}.

The main effects leading to
 the lifetime extremes in the inequality chains are easily
 identified: the large $\Omega _{c}^{0}$--rate is due to
 a large positive interference effect among the
 s--quarks (the s--quark from the weak
 decay vertex can interfere with either of the
 spectator s--quarks) and the small $\Xi _{c}^{+}$--rate
 is due to the absence of a W--exchange contribution
 in this case.

One must stress that there are a number of theoretical
 uncertainties in the lifetime calculations of \cite{cc}
 and \cite{cd} related to the size of the preasymptotic
 effects which could not be discussed in detail here.
 Nevertheless, the authors of \cite{cc}
 are optimistic and claim that their inequalities can be replaced
 by equality relations with multiplicative factors
 of $1.5$ to $2$.

The absence of large preasymptotic
 effects in bottom baryon nonleptonic decay rates is gratifying.
Present evidence points to a somewhat larger difference in lifetimes
between bottom mesons and the $\Lambda_b$ bottom baryon than expected from
the naive dimensional analysis (\ref{mac}). Clearly one needs better
experimental data on bottom hadron lifetimes including the bottom baryons
$\Xi_b^{0,-}$ and $\Omega_b$ to be able to ascertain how big the spread
in lifetimes in the bottom sector is, and whether one can understand the
lifetime hierarchy from first theoretical principles.

\newpage
\vspace*{1cm}\hspace*{3cm}
\section{Exclusive Nonleptonic Decays}
\vspace*{2cm}
\subsection{Decay Rates and Decay Distributions}

Let us begin by counting the number of independent amplitudes
 in the four classes of two--body nonleptonic ground-\-state to
 ground-\-state transitions\\
                            $B_1(1/2^+)$ $\rightarrow$
 $ B_2(1/2^+,3/2^+) + M(0^-,1^-)$:
\begin{eqnarray}
 \mbox{(i)\hspace*{0.5cm}
}
 1/2^{+}\to 1/2^{+} + 0^{-} &:&
 \mbox{2 complex amplitudes}\nonumber \\
 \mbox{(ii)\hspace*{0.5cm}
}
 1/2^{+}\to 3/2^{+} + 0^{-} &:&
 \mbox{2 complex amplitudes}\nonumber \\
 \mbox{(iii)\hspace*{0.5cm}
}
 1/2^{+}\to 1/2^{+} + 1^{-} &:&
 \mbox{4 complex amplitudes}\nonumber \\
 \mbox{(iv)\hspace*{0.5cm}
}
 1/2^{+}\to 3/2^{+} + 1^{-} &:&
 \mbox{6 complex amplitudes}\nonumber
\end{eqnarray}

Using standard methods (e.g. \cite{jack}) one can then derive
 angular decay distributions which, upon integration,
 give the decay rates. Again we prefer an explicit frame--dependent
 representation of the decay distributions instead of the frame
 independent representation discussed in \cite{bjang}. We begin by
 considering the nonleptonic decay of unpolarized charm or bottom
 baryons. In the simplest case one has the decay
 $ 1/2^{+}\to 1/2^{+}(\to 1/2^{+} + 0^{-})+ 0^{-} $
 as for example in
 $\Lambda_{c}^{+}\to \Lambda (\to p \pi ^{-})+\pi^{+}$.
 Referring to Fig.14, one now sees  a cascade only on one side
 as the pion's decay goes unobserved. Consequently one has only a
 single polar angle distribution. One obtains
\begin{eqnarray}
 \frac{\mbox{d}\Gamma (\Lambda _{c}^{+}\to
                                    \Lambda (\to p\pi^{-})+\pi^{+})}
 {\mbox{d}\!\cos\Theta_{\Lambda }}
 &=&{\textstyle \frac{1}{2}}B_{\Lambda \to p\pi^{-}}
 \Gamma_{\Lambda_{c}^{+}\to \Lambda \pi^{+}}
  (1+\alpha_{c} \alpha_{\Lambda} \cos\Theta_{\Lambda })
\end{eqnarray}
 where $\alpha_{c}$ and $\alpha_{\Lambda }$ are the asymmetry
 parameters in the decays $\Lambda_{c}^{+} \to \Lambda \pi^{+}$
 and $\Lambda \to p\pi^{-}$, respectively, defined in analogy
 to Eqs.(\ref{5-11},\ref{5-12}). The definition of the polar angle
 $\Theta_{\Lambda }$ is given in Fig.~14 with the replacement
 $ W^{+} \to \pi^{+} $. The cascade decay $\Lambda \to p\pi^{-}$
 acts as an analyzer of the longitudinal polarization of the
 daughter baryon $\Lambda $ whose polarization is given by the
 asymmetry parameter $\alpha_{c}$. This single angular decay
 distribution was utilized experimentally to measure the asymmetry
 parameter $\alpha_{c}$ in the decay
 $\Lambda _{c}\to \Lambda (\to p\pi^{-})+\pi^{+}$ \cite{cleo,argus}.

Somewhat more complicated is the decay distribution in the double
 cascade
 $ 1/2^{+}\to 1/2^{+}(\to 1/2^{+} + 0^{-})+1^{-}(\to 0^{+}+0^{-})$
 as for example in
 $\Lambda_{c}^{+} \to \Lambda (\to p\pi^{-})+\rho ^{+}
 (\to \pi^{+}\pi^{0})$. One has \cite{kk92b,polen}
\begin{eqnarray}
 &&\frac{\mbox{d}\Gamma (\Lambda_{c}^{+}\to \Lambda (\to p\pi^{-})+
 \rho^{+}(\to \pi^{+}\pi^{0}))}
 {\mbox{d}\!\cos\Theta\mbox{d}\chi\mbox{d}\!\cos\Theta_{\Lambda
}}
 =\frac{1}{2 \pi}B_{\Lambda \to p\pi^{-}}B_{\rho ^{+}\to \pi^{+}\pi^{0} }
 \frac{p}{32\pi M_{1}^{2}}
 \nonumber \\
 && \cdot \left( \frac{3}{4}\sin^{2}\Theta H_{U}
 (1+\alpha_{c}^{U}\alpha _{\Lambda
} \cos\Theta_{B})
 +\frac{3}{2}\cos^{2}\Theta H_{L}
 (1+\alpha_{c}^{L}\alpha _{\Lambda
} \cos \Theta_{\Lambda
})\right.
 \nonumber \\
 &&\left.
 -\frac{3}{4\sqrt{2}}\sin(2\Theta )\cos\chi\cos\Theta_{\Lambda
} \alpha _{\Lambda
} H_{I}
 +\frac{3}{4\sqrt{2}}\sin(2\Theta )\sin\chi\cos\Theta _{\Lambda
}
 \alpha _{\Lambda
} H_{I'} \right)
\end{eqnarray}
 where the helicity rates $H_{U}$, $H_{L}$, and $H_{I}$
 and the asymmetry parameters $\alpha_{c}^{U}$
 and $\alpha_{c}^{L}$ are defined in analogy to
Eqs.(\ref{5-5},\ref{5-11},\ref{5-12}).
 Angles are defined as in Fig.14 with the replacement of
 $(W^+\rightarrow l^+\nu_l)$ by $(\rho^+\rightarrow \pi^+\pi^+)$.
 Clearly the six observables defined by the decay distribution
 do not suffice to determine the four complex decay amplitudes
 of the process.

The observable $H_{I'}$ is proportional to the {\em imaginary}
 part of the longitudinal--transverse interference term
 (see the third line of Equation (\ref{5-5})) and is thus a so--called
 T--odd observable. It obtains contributions from $CP$--violating
 interactions and/or from effects of final--state interaction.
 The Standard Model $CP$--violating contributions are expected
 to be quite small and thus $H_{I'}$ would be a good measure
 of the strength of final state interaction effects.
 Alternatively, one may extract possible $CP$--violating
 effects by comparing $\Lambda _{c}^{+}$ and
 $\bar \Lambda _{c}^{+}$ cascade decays.

We now briefly turn to the decays of {\em polarized} heavy
 baryons in which the cascade decay of the daughter baryon
 is used as an analyzer and the meson decay goes unanalyzed.
 The orientation angles are defined in Fig.15.
 The angular decay distribution for the
 $1/2^{+}\to 1/2^{+}(\to 1/2^{+}+0^{-})+0^{-}$ transition
 is well known from the analysis of the nonleptonic decays of
 the cascade hyperon $\Xi$ and reads \cite{polen,lednicky}
\begin{eqnarray}
 &&\frac{\mbox{d}\Gamma (\Lambda_{c}^{\uparrow }\to
                                              \Lambda (\to p\pi^{-})+
 \pi^{+})}
 {\mbox{d}\!\cos\Theta_{p}\mbox{d}\!\cos\Theta_{\Lambda
} \mbox{d}\!\sin\chi}
 =\frac{1}{8\pi}\Gamma _{\Lambda _{c}\to \Lambda \pi^{+}}
 B_{\Lambda \to p\pi^{-}}
 \nonumber \\
 &&
 \left[ 1+\alpha_{c}\alpha_{\Lambda }
 \cos\Theta_{\Lambda }+P_{c}(\alpha_{c}\cos\Theta_{p}
 +\alpha_{\Lambda }\cos\Theta_{p}\cos\Theta_{\Lambda }\right.
 \nonumber \\
 && \left.
  +\alpha_{\Lambda }\sin\Theta_{p}\sin\Theta_{\Lambda }
 (\gamma_{c}\cos\chi +\beta _{c}\sin\chi ))
 \right]
\end{eqnarray}
 where $\beta_{c}$ and $\gamma_{c}$ are the usual nonleptonic
 decay parameters (e.g. \cite{bjang}). In a noncascade charm
           baryon decay, for example $\Lambda _{c}\to p\bar K^{0}$,
 one would be left with a decay distribution
 $W(\Theta_{c})=(1+P_{c}\alpha_{c}\cos\Theta_{p})$. This would
 allow for a determination of the asymmetry parameter $\alpha _{c}$
 only if $P_{c}$ were known.

The remaining angular decay distributions involving the other nonleptonic
decay processes (polarized and unpolarized) can be found in \cite{polen}.

\subsection{Symmetry Considerations}

 In the nonleptonic Hamiltonian Eq.(\ref{hamiltonian})
 we included only the dominant contribution
 proportional to
 $\simeq \cos^{2}\Theta _{c}$. Once suppressed transitions
 proportional to $ \simeq cos\Theta _{c} \sin\Theta _{c} $,
 not written in Eq.(\ref{hamiltonian}),
 are the transition $c\to d u \bar d$ and $c\to s u \bar s$ ,
 and the doubly suppressed decay $c\to d u \bar s$
 proportional to $\simeq \sin^{2}\Theta _{c}$.\footnote{
 In the sum rule approach, one relates different nonleptonic decay
 amplitudes by using flavour symmetry relations based on the
 flavour symmetry group SU(4) and/or its SU(3) and SU(2) subgroups.}
 The $\Delta$C$=1$
 SU(3) content of the antisymmetric
 representation $20''$ is $3_a$
 and $6^{\ast }$, and
 that of the symmetric representation 84 is
 $3_s$ and 15. The dominant pieces are the $6^{\ast}$ and 15
 SU(3) representations.
 The I--, U--, and V--spin content of the dominant piece is
 $\Delta$I=1, $\Delta$U=1 and $\Delta$V=0, 1.
 Sum rules relating different charm changing
 nonleptonic amplitudes can be and have been
 written down using various techniques \citer{kokrawi,pakv};
 the simplest technique appears to be an analysis
 using the three SU$(2)_{\mbox{I,U,V}}$ subgroups \cite{kokrawi}.
The I--spin sum rules are expected to be quite accurate. For example one
predicts equal rates for $\Lambda_c\to\Sigma^+\pi^0$ and $\Lambda_c\to
\Sigma^0\pi^+$ from the $\Delta I=1$ rule \cite{kokrawi} which is borne
out by recent experiments \cite{proc}. The U--spin and V--spin sum rules are
not expected to be as good because of SU(3) breaking effects.
 Nonet symmetry for the mesons can be
 incorporated in the usual way by excluding
 disconnected flavour flow diagrams (see e.g. \cite{kokrawi}).

An interesting observation concerns the
 rates of the two members $\Lambda ^{+}_{c}$ and $\Xi ^{0}_{c}$
 of the same U--spin doublet. In the case of $20''$
 dominance of $H_{eff}$ it has been
 shown from U--spin arguments that for
 the dominating $c \to s u  \bar d$ transitions one can
 derive equality of total rates and partial
 rates into any particular spin channels \cite{altar}. Considering the
 present nonequality of $\Lambda ^{+}_{c}$ and $\Xi^{0}_{c}$ life times
 shows that $H_{eff}^{20''}$ dominance of the
 effective nonleptonic Hamiltonian may not be a good
 approximation.

Further sum rules may be obtained relating Cabibbo
 favoured, suppressed and doubly suppressed decay amplitudes
 when the Cabibbo suppression factors are removed.
 Similarly one may even attempt to
 relate charm changing $\Delta$C$=1$ processes to ordinary
 $\Delta$C$=0$ nonleptonic hyperon decays although the large mass difference
 between charm and ordinary baryons makes such an
 approach problematic.

Still another class of sum rules may be
 obtained by considering parity--conserving and parity--violating
 amplitudes separately and assuming
 the \underline{full} SU(4) symmetry of the transition
 in conjunction with the charge conjugation
 symmetry of $H_{eff}$ which is C=+1
 and C=$-1$ for the parity--conserving and parity--violating parts,
 due to CP--conservation \cite{kokrawi,iwa}.
One then e.g.
 obtains the result that the parity--violating amplitude $A$ in $\Lambda_c^+
\to\Lambda\pi^+$ vanishes. This is
 in direct conflict with the recent
 nonvanishing asymmetry measurement
 in this decay \cite{argus,cleo}. One concludes
 that SU(4) is not a useful flavour
 symmetry for charm changing weak decays due to the
 large mass breaking factor
 $(m_{c}-m_{s})/m_{c} \simeq 70\%$.

 While SU(4) is not a useful symmetry, SU(3) flavour symmetry
 may still be useful
 for charm changing decays \cite{kokrawi,sav}. But even
 then one encounters the problem of which mass
 dimension the SU(3) flavour symmetric amplitude
 should carry. Related to this is the extraction
 of supposed flavour symmetric
 amplitudes from rates where one does not
 know which mass scale $\tilde M$ is appropriate for
 the $(p/\tilde M)^{2l+1}$ phase space factor. In order
 to be able to answer this question reliably
 one is back to the dynamical problem.
 Hopefully one will learn more about the
 appropriate mass scaling factors for an amplitude in the
 future.
 Unfortunately one must conclude that the SU(3)
 flavour symmetry approach to nonleptonic and
 semileptonic decays involving $\Delta$C$=1$ transitions
 provides a rule of thumb at best.

\subsection{Quark Model and Current Algebra Results}

In the quark model the effective current$\times$current
 Hamiltonian (\ref{hamiltonian}) gives rise to the five types of
 flavour diagrams drawn in Fig.18. We have chosen to label
 the quark lines for the specific transition
 $\Lambda_{c}^{+} \rightarrow \Lambda \pi^{+}$
 for illustrative purposes. The wavy lines are included in
 order to indicate how the effective quark currents of the
 Hamiltonian (\ref{hamiltonian}) act. As a next step, one wants to
 interpret the diagrams as Feynman diagrams possibly with
 additional gluon exchanges added. The general dynamical
 problem in all its complexity is far from being solved, so
 one has to resort to some approximation. The quark lines in
 Fig.18 transmit spin information from one hadron to the other.
 This is realized in the spectator quark model, which postulates
 that there is no spin communication between quark lines.
 Quark pairs are created from the vacuum with $^{3}P_{0}$
 quantum numbers. Finally, these postulates can be cast into a
 covariant form if the quarks in a hadron are assumed to
 propagate with equal velocity which is also the hadron's velocity.

\begin{figure}
\vspace{7cm}
\caption[dummy27]{
Quark diagrams contributing to nonleptonic decay $\Lambda_c\to
\Lambda\pi^+$, including colour--flavour weight factors.}
\end{figure}

In terms of quark model spin wave functions, the decay
 amplitudes for the process $B_1 \rightarrow B_2 + M$ corresponding
 to Fig.18 can then be written as \cite{kokrawi}
\begin{eqnarray}\label{quarkflow}
T_{B_1 \rightarrow B_2+M}&=&H_1
                 \bar B_{2}^{ABC'}B_{1ABC}\bar M^{D'}_{D}(
O^{CD}_{C'D'}-\frac{1}{N_c}O^{CD}_{D'C'})\nonumber\\
& &+ \frac{1}{N_c}H_2
           \bar B_{2}^{AB'D}B_{1ABC}\bar M^{D'}_{D}O^{BC}_{B'D'}
\nonumber\\
& &+ \frac{1}{N_c}H'_2
              \bar B_{2}^{AB'C'}B_{1ABC}\bar M^{B}_{D}O^{CD}_{B'C'}
\nonumber\\
& &+ \frac{1}{N_c}H_3
             \bar B_{2}^{A'B'C'}B_{1ABC}\bar M^{C}_{C'}O^{AB}_{A'B'}
\end{eqnarray}
 where the first, second, third and fourth terms of
 (\ref{quarkflow}) correspond to the contributions of
 diagrams Ia,b, IIa, IIb and III in Fig.18 in that order.
 $B_{ABC}$ and $M_{A}^{B}$ are quark model wave functions for
 the baryons and mesons. Each index $A$ stands for a pair of
 indices $(\alpha,a)$, where $\alpha$ and $a$ denote the spin and
 flavour degrees of freedom. We have already summed over colour
 degrees of freedom which results in the typical factors $1/N_c$
 where $N_c=3$. We emphasize that the limit $N_c \ra \infty$ cannot
 be taken naively for the last three contributions in (\ref{quarkflow})
 (IIa,b and III in Fig.18). We shall return to this point later on.
 The matrix $O_{AB}^{CD}$ describes the spin--flavour structure of
 the effective current$\times$current Hamiltonian (\ref{hamiltonian}).
 $H_1, H_2, H'_2$ and $H_3$ are wave function overlap integrals
 corresponding to diagrams I, IIa,b and III which are expected to
 depend on the masses of a particular decay process.
 Eq.(\ref{quarkflow}) can be viewed as an algebraic realization
 of the diagrams shown in Fig.18: each line in Fig.18
 corresponds to a contraction of doubly occurring spin--flavour
 indices in (\ref{quarkflow}), where one sums over the spin--flavour
 indices.

The first term in (\ref{quarkflow}) corresponds to the
 so--called factorization contribution and can be calculated
 in terms of the current matrix elements of Sec.5.
 Bringing the contributions of the non--factorizing diagrams
 IIa,b and III into tenable forms with the above assumptions
 does not preclude the possibility that (\ref{quarkflow})
 can be derived from a more general point of view dropping some
 of the above assumptions. One should note that in the case of
 transitions between ground state baryons, the non--factorizing
 diagrams IIa,b and III obtain contributions only from $O^-$
 (transforming as $20''$ in SU(4)) because of the symmetric
 nature of the ground state baryons \cite{patiwoo}.
 Both operators $O^+$ and $O^-$ contribute to diagram Ia and Ib.
 The contributions of Ia and Ib add up such that the resulting
 contribution is proportional to $\chi_{\pm}=(c_+(1+1/N_C)\pm
 c_-(1-1/N_C))/2$ depending on whether the final state meson is
 charged (+) or neutral $(-)$.

The contribution of diagram Ib can be seen to be colour suppressed.
 Guided by the ana\-lysis of exclusive nonleptonic charm and bottom
 meson decays \cite{bauste,bugerru} it seems to be appropriate
 to take the $N_c \ra \infty$ limit and accordingly drop the
 contribution of diagram Ib in Fig.18. Superficially also the
 contributions of diagrams IIa,b and III appear to be colour suppressed.
 But considering the fact that baryons contain $N_c$ quarks as
 $N_c \ra \infty$ the denominator factor $N_c$ is balanced by
 combinatorial numerator expressions such that diagrams IIa,b and III
 occur at ${\cal O}(1)$ as $N_c \ra \infty$ and may not be dropped in
 this limit.

The results of calculating diagrams IIa, IIb, and III depend
 on the details of the quark model wave functions used as input.
 As a first approximation, one may use $\mbox{SU}(2)_W$ spin wave
 functions \cite{kokrawi}. They correspond to boosting static quark
 model wave functions to a collinear equal velocity frame as mentioned
 above \cite{hukoetho}. When explicit mass factors are scaled out
 of the baryon wave functions according to the HQET, one can set
 $H_2=H'_2$ in Eq.(\ref{quarkflow}), because of $CP$--invariance.
 After some straightforward algebraic manipulations involving the
 evaluation of the amplitude Eq.(\ref{quarkflow}) with the
 SU$(2)_{W}$ wave functions, one can calculate the nonleptonic
 transition amplitudes for the decays $1/2^+\rightarrow
 (1/2^+, 3/2^+)+(0^-,1^-)$.

In order to be able to discuss some general features of the
 solutions, we treat the decay $1/2^+\rightarrow 1/2^{+}+0^-$
 in more detail. Writing the amplitude $T_{B_1\rightarrow B_2+M}=\bar
 u_2(A+B\gamma_5)u_1$ one obtains the following amplitude expressions
\begin{eqnarray}
A&=&A^{fac}+\frac{1}{3}\frac{H_2}{M_1M_2\sqrt{M_3}}
                            \Big(-\frac{3}{4}Q_+(
M_1I_3-M_2 \hat I_3)\nonumber\\
& &+\frac{3}{4}M_1M_2M_3(I_3 - \hat I_3)\Big)\\
B&=&B^{fac}+\frac{1}{3}\frac{H_2}{4M_1M_2\sqrt{M_3}}
                               \Big (Q_+(M_1(I_3+2I_4)
+M_2(\hat I_3+2\hat I_4))\Big)\nonumber\\
& &+\frac{1}{3}\frac{H_3}{M_1M_2\sqrt{M_3}}3M_1M_2(M_1+M_2+M_3)I_5
\end{eqnarray}
 where the factorizing contributions $A^{fac}$ and $B^{fac}$
 (corresponding to diagrams Ia and Ib) are obtained from the
 current-induced form factors discussed in Section 5. We have
 defined $Q_+ = (M_1+M_2)^2-M_{3}^{2}$. The invariant flavour
 wave function contractions (Clebsch--Gordan coefficients)
 denoted by $I_i$ and $\hat I_i$ are defined in \cite{kokrawi}.
 $I_3$ and $I_4$ are associated with diagram IIa, $\hat I_3$ and
 $\hat I_4$ with diagram IIb, and $I_5$ with diagram III. Diagram
 III can be seen to contribute only to the p.c. amplitude $B$,
 whereas diagrams I and II contribute to both parity--conserving
 and parity--violating amplitudes. The parity--conserving and
 --violating amplitudes can be seen to be even and odd with
 respect to the generalized charge conjugation operation
 $(M_1;I_{3,4};I_5)$ $\rightarrow$ $(M_2;\hat I_{3,4};I_5)$
 as expected from the $CP$--conserving property of the nonleptonic
 Hamiltonian. For example, for the decay
 $\Lambda_{c}^{+} \rightarrow \Lambda \pi^+$ one finds
 $I_3=\hat I_3$ and thus the parity--violating amplitude $A$ in
 the symmetry limit $M_1=M_2$ vanishes, as remarked on earlier.
 With $M_1\gg M_2$ this statement no longer holds true.

The contributions of diagrams IIb and III are nonleading on
 the scale of the mass $M_1$ of the parent baryon. As a helicity
 analysis shows, they are nonleading because the contributions
 IIb and III are suppressed by helicity as a result of the
 $(V-A) \times (V-A)$ nature of the underlying quark transition
 \cite{kogold}. This implies that only the factorizing contribution
 Ia and the non--factorizing contribution IIa survive when $M_2/M_1
 \rightarrow 0$. This conclusion holds in general for all the
 ground state decay channels. These leading contributions
 can be seen to lead to an exclusive decay mode power behaviour
 $\Gamma \sim 1/M_1$ when $M_2$ and $M_3$ are kept fixed.
 The helicity suppressed contributions are down by an additional
 factor $(M_2/M_1)^2$. The same power behaviour holds true in
 nonleptonic meson decays. Compared to the inclusive nonleptonic rate
 $\Gamma_{FQD}^{nl} \sim M_{1}^{5}$ one infers that the exclusive
 branching ratio of a particular two--body channel decreases
 very rapidly as $M_1$ becomes large and $M_2$, $M_3$ are kept
 fixed. When both $M_1$ and $M_2$ become large with their ratios
 fixed, and $M_3$ kept fixed and small, one has again
 $\Gamma \sim (M_1)^5\cdot(M_2/M_1)^6$ with the helicity
 suppressed contributions diagrams IIb, III down by another factor
 $(M_2/M_1)^2$. We do not, however, see a mechanism that would
 suppress the non--factorizing contribution IIa relative to the
 factorizing contribution in this limit as is implicit in the
 analysis of \cite{bjang,marory}.

The flavour structure in the parity violating amplitude $A$
 has a remarkable property: there exists a one--to--one flavour
 correspondence with terms arising in the current algebra plus
 soft pion approach. This was first noticed empirically
 in the $\bigtriangleup C=0$, $\bigtriangleup Y=1$ \cite{koegu}
 and in the $\bigtriangleup C=0$, $\bigtriangleup Y=0$
 \cite{kokrawi2} transitions and was later proven in general
 \cite{kokrawi}. The correspondence between the quark model
 and current algebra approach works in the following way: the
 contributions proportional to $I_3$ and $\hat I_3$ have the
 flavour structure of the "equal time commutator" term when the
 symmetry limit $M_1=M_2$ is taken. The factorizing contribution
 $A^{fac}$ has the same interpretation in both schemes. In a
 similar vein, the nonfactorizing parity--conserving contributions
 can readily be interpreted as baryon pole contributions.

In the past few years, many new nonleptonic charm--baryon decays
 have been observed \cite{KOESI91}. By now about 25\% of the
 $\Lambda_{c}^{+}$ decay channels are accounted for; new nonleptonic decay
 modes of $\Xi_c$ and $\Omega_c$ have been seen recently. As for
 the bottom baryons, the experimental situation is meagre.
 Evidence for $\Lambda_b$ in the exclusive nonleptonic mode
 $\Lambda_b\rightarrow \Lambda J/\Psi$ was presented in \cite{UA1}
 which, however, was not confirmed by other
 collaborations \cite{pierre}. Low statistics evidence has been obtained
for the mode
 $\Lambda_b\rightarrow\Lambda_{c}^{+}\pi^-$ \cite{OPAL93}.
 Among the observed charm--baryon decay channels there are many two--body
 and quasi--two--body modes which can be compared with theoretical
 predictions. For the three-- and four--body decay modes only a few
 theoretical analyses have been carried out: Using a chiral
 Lagrangian, the authors of \cite{kps86} give predictions for decays of
 the type $\Lambda_{c}^{+} \ra 1/2^+ + 0^- + 0^-$,
 while a semiquantitative estimate of the decay channels
 $\Lambda_{c}^{+} \ra p \overline{K^0} \pi^0$ and $p
 \overline{K^0}\pi^+\pi^-$ has been made in \cite{fits93}.
Further theoretical work on the nonleptonic many--body decays is needed.
 In Table 10 we list some of the current--algebra and pole--model
 calculations for two--body modes
 and compare them to a quark model calculation with best fit
 values for the overlap parameters $H_2=H'_2$ and $H_3$ in
Eq.(\ref{quarkflow}). Compared to the quark model calculation of
\cite{kk92b} the calculations \cite{CHTS,xk,zencz} include long--distance
dynamics by considering the contributions of low--lying baryon intermediate
states. The calculations differ in the details of how coupling factors in
this approach are determined.
 In \cite{CHTS} the non--factorizing contribution
 has been evaluated by using the pole approximation, where the
 (parity--violating) s--wave amplitudes are dominated by the
 low-lying $1/2^-$ resonances, while the (parity--conserving)
 p--wave ones are governed by the ground state $1/2^+$ poles.
 The MIT bag model was employed to calculate the coupling constants,
 form factors and baryon matrix elements. The importance of
 including the $1/2^-$ pole terms to the s--wave contributions
 has also been emphasized by the authors of \cite{xk}. They used
 symmetry arguments to relate their couplings to those from
 semileptonic hyperon decays as well as the diquark model to determine
 the parity--conserving amplitudes. As for the parity--violating
 contribution, the authors of \cite{xk} claim that they are
 completely determined by the current algebra commutator term
 and the masses of the relevant $1/2^-$ resonances without
 introducing further new parameters. Zenczykowski \cite{zencz} sums
intermediate states contributing to the parity--violating amplitudes to
obtain an effective current--algebra expression. He uses broken SU(4) to
relate coupling factors in the charm sector to coupling factors in the
hyperon sector similar to the calculation \cite{xk}.

    All calculations \cite{CHTS,xk,kk92b,zencz} use the $N_c\to\infty$
approximation to determine the factorizing
contributions (``new factorization'').
The new factorization scheme is supported by the analysis of
the Cabibbo suppressed decay $\Lambda_{c}^{+}
 \ra p \phi$ which only gets contributions from the factorizing
 diagram I. As was first noticed in \cite{KOESI91,kk92b}, its
 measured rate can only be accounted for by dropping terms
 proportional to $1/N_c$.
We reiterate that the quark model and pole model or current algebra
approaches to nonleptonic charm baryon decays are not radically different
from one another because of the equality of spin--flavour
factors in both approaches \cite{kokrawi}.

 In the next few years the advent of new data will
 certainly constrain the current algebra and quark model
 calculations further.
 Almost all model calculations predict negative asymmetry parameter
 values close to their maximum value of $-1$ for the decays $\lam \ra
 \Lambda\pi^+$ and $\lam\ra p K^-$. They are thus in agreement with
 the measured asymmetry parameter in the decay
 $\lam\ra \Lambda\pi^+$ \cite{cleo,argus}.
 The decays $\lam\ra\Xi^0 K^+$ and $\Lambda_c^+\to\Sigma\pi$ are
particularly interesting: they obtain contributions
 only from the nonfactorizing diagrams IIa and III in Fig.18
 and thus give a measure of the nonfactorizing contributions to
 nonleptonic charm baryon decays.
 Their experimental observation proves that the nonfactorizing
(or W--exchange) contributions can certainly not be neglected as is
implicit in the analysis of \cite{bjang,marory}.
\newpage


\renewcommand{\arraystretch}{1.33}
\begin{tabular}{lllllll}
\multicolumn{7}{p{23.5cm}}{{\bf{Table 10.}} Current algebra and quark
 model predictions for nonleptonic charm baryon decays. The numbers
 cited are decay rates (in units $10^{11}s^{-1}$) and asymmetry
 parameters $\alpha_c$ (in parentheses)}\\
\hline\hline
 & Cheng and Tseng & Cheng and Tseng & Xu and Kamal & \.{Z}enczykowski
 & Quark Model & Experiment \\
 & Current Algebra \cite{CHTS} & Pole Model \cite{CHTS} & \cite{xk}
 & \cite{zencz} & \cite{kk92b}  & \cite{pdg} \\
\hline
$\lam\ra p\kbar$      & 1.82$(-0.90)$ & 0.63$(-0.49)$ &
 0.60$(0.51)$  & 0.99$(-0.90)$ & input$(-1.00)$ &
                                    $0.82\pm 0.12$ \\
$\lam\ra p\kbar(892)$ & & &
  & 1.19 & 1.54  & $0.45\pm 0.12$  \\
$\lam\ra \Delta^{++} K^-$ & & &
  &      & 1.35  & $0.34\pm 0.13$ \\
$\lam\ra p \phi$ & & &
  & 0.05 & 0.11  & $0.07\pm 0.05$ \\
$\lam\ra\Lambda\pi^+$ & 0.73$(-0.99)$ & 0.44$(-0.95)$ &
 0.81$(-0.67)$ & 0.31$(-0.86)$ & input$(-0.70)$ & $0.30\pm 0.05$
                                                       \\
& & & & & & $(-1.03\pm 0.29)$ \\
$\lam\ra\Lambda\rho^+$&               &               &
               & 0.27          & 9.54           & $<0.59$
\cite{9322}                                                      \\
$\lam\ra\sio\pi^+$    & 0.88$(-0.49)$ & 0.36$(0.78)$  &
 0.17$(0.92)$  & 0.23$(-0.76)$ & 0.16$(0.70)$ &$0.28\pm 0.10\pm 0.07$
                                                           \\
& & & & & & $(-0.43\pm 0.23\pm 0.20)$ \cite{procario}\\
$\lam\ra\sip\pi^0$    & 0.88$(-0.49)$ & 0.36$(0.78)$  &
 0.17$(0.91)$  & 0.23$(-0.76)$ & 0.16$(0.71)$ &$0.33\pm 0.08\pm 0.07$
\cite{9310}                                                          \\
$\lam\ra\sip\rho^0$   &               &               &
               & 0.28 & 1.56 &$<0.5$ \cite{9310}\\
$\lam\ra\sip\omega$   &               &               &
               & 0.18 & 2.01 &$0.88\pm 0.24\pm 0.19$ \cite{9310}\\
$\lam\ra\Xi^0 K^+$    &               &               &
 0.05$(0)$     & 0.04$(0)$     & 0.13$(0)$ &$0.13\pm 0.04$ \cite{9304}
                                                           \\
$\xip\ra\sip\kbar$    & 0.01$(0.43)$  & 0.19$(-0.09)$ &
 0.10$(0.24)$  & 0.15$(0.68)$  & 1.46$(-1.00)$  & \\
$\xip\ra\Xi^0\pi^+$   & 0.19$(-0.77)$ & 0.89$(-0.77)$ &
 0.76$(-0.81)$ & 0.16$(0.65)$  & 0.80$(-0.78)$  & \\
$\xio\ra\Lambda\kbar$ & 0.89$(-0.88)$ & 0.24$(-0.73)$ &
 0.33$(1.00)$  & 0.21$(-0.84)$ & 0.11$(-0.76)$  & \\
$\xio\ra\sio\kbar$    & 0.02$(0.85)$  & 0.12$(-0.59)$ &
 0.09$(-0.99)$ & 0.03$(-0.89)$ & 1.05$(-0.96)$  & \\
$\xio\ra\sip K^-$     &               &               &
 0.11$(0)$     & 0.04$(0)$     & 0.11$(0)$      & \\
$\xio\ra\Xi^0\pi^0$   & 1.12$(-0.78)$ & 0.25$(-0.54)$ &
 0.50$(0.92)$  & 0.15$(-0.99)$ & 0.03$(0.92)$   & \\
$\xio\ra\Xi^-\pi^+$   & 0.74$(-0.47)$ & 1.12$(-0.99)$ &
 1.55$(-0.38)$ & 0.46$(-0.78)$ & 0.93$(-0.38)$  & \\
$\omo\ra\Xi^0\kbar$   & 0.98$(0.44)$  & 0.13$(-0.93)$ &
               &               & 1.75$(0.51)$   & \\
 \hline\hline
\end{tabular}
\newpage

\newpage
\vspace*{1cm}\hspace*{3cm}
\section{Summary and Outlook}
\vspace*{2cm}

   In this review we have concentrated on the decay properties of ground
state and excited state charm and bottom baryons, focussing on recent
advances in the understanding of how QCD turns into a much simplified Heavy
Quark Effective Field Theory when the quarks become much heavier than the
QCD confinement scale. Present experiments are already proving the
usefulness of the concepts of HQET. One can expect a wealth of data
on heavy hadron physics in the future to be confronted with the predictions
of HQET.

   At present there are strong experimental programmes on heavy hadron
physics at high energy laboratories all over the world. ARGUS at DESY
has stopped running in 1993 after having produced a wealth of important
results on charm and bottom physics during its lifetime. CLEO is very much
alive and coping very well with CESR's present top performance at a design
luminosity of $3\times 10^{33}\,cm^{-2}\,s^{-1}$ with further improvements
lying ahead. LEP has begun to contribute to bottom physics in a significant
way. There will be two more years of running on the Z$^0$ peak and
there is still data on the tapes of previous runs waiting to be analyzed.
The hyperon beam experiment WA89 at CERN specializes on charm--strangeness
baryons and has seen first signs of the $\Xi_c'$ baryon. There have been
detector improvements and there are more runs coming up. At Fermilab the
collider mode detectors CDF and D0 have produced high statistics charm and
bottom hadron results and further detector and machine improvements are
being planned or have been installed. Then there are the photoproduction
and hadroproduction fixed target experiments E653, E672, E687, E691, E771,
E789 and E791 that have yielded some very accurate results on heavy hadron
physics in general and on some specific decay modes in particular.

   SLAC is tooling up with its approved B factory project which is expected
to start its
bottom physics program in 1999. The HERA--B project at DESY and the B factory
project at KEK are awaiting approval. While the primary objective of these
machines is to discover and study CP--violation in the bottom sector there
certainly will be ample fall--off for heavy hadron physics in general.

   All in all, we can expect an abundance of interesting new data on
charm and bottom baryons in the next few years. The field is
very much alive
and one can be sure that there will be plenty of  experimental
and theoretical activity in heavy baryon physics in the future.
As experience has
shown, real progress is achieved
when theoretical and experimental advances go hand in hand. In this sense
the theoretical heavy quark physics community is looking forward to a lot
of new experimental results on heavy quark physics in general and on heavy
baryon physics in particular.
\vspace*{1cm}

\noindent
{\bf Acknowledgements.}

\noindent
We would like to thank J.Landgraf, M.Lavelle and
H.W.Siebert for help and advice. Much of the material presented in this
review is drawn from work done in collaboration with S.Balk, P.Bialas,
F.E.Close, A.Czarnecki, C.A.Dominguez, F.Hussain, A.Ilakovac, U.Kilian,
M.Jezabek, B.K\"onig, P.Kroll,  J.H.K\"uhn, J.Landgraf, R.Migneron,
R.J.Phillips, A.Pilaftsis, K.Schilcher, D.J.Summers, G.Thompson, M.Tung,
Y.L.Wu and K.Zalewski.
Part of this work was done while J.G.K. was a visitor at the DESY theory
group. He would like to thank W.Buchm\"uller for hospitality and the DESY
directorate for support.
J.G.K. acknowledges partial support by the BMFT, Germany under contract
06MZ730. M.K. is supported by the DFG and D.P. is supported by the
Graduiertenkolleg ``Teilchenphysik'' in Mainz.

\end{document}